\documentclass[reprint,
superscriptaddress,
 amsmath,amssymb,
pra,
]{revtex4-2}

\usepackage{graphicx}

\usepackage{dcolumn}
\usepackage{bm}

\usepackage{float}
\usepackage{booktabs}
\usepackage{siunitx}
\usepackage{chemformula}
\usepackage{mhchem}
\usepackage{tabularx} 
\usepackage{textcomp}
\DeclareSIUnit\year{yr}

\begin{document}

\pagenumbering{Roman}

\title{Study of thorium in hypersonic gas jets: Ionization potentials of Th and Th$^+$}

\author{A. Claessens}
\email{This publication comprises part of the Ph.D. thesis of A. Claessens}
\affiliation{KU Leuven, Instituut voor Kern- en Stralingsfysica, 3001 Leuven, Belgium}
\author{F. Ivandikov}
\affiliation{KU Leuven, Instituut voor Kern- en Stralingsfysica, 3001 Leuven, Belgium}
\author{M. Brasseur}
\affiliation{Physique Atomique et Astrophysique, Universit\'e de Mons, B-7000 Mons, Belgium}
\author{A. Dragoun}
\email{Present adress: Division of Nuclear Chemistry, Department of Chemistry and Biochemistry, Faculty of Mathematics and Natural Sciences, University of Cologne, 50674 Cologne, Germany and
Institute of Neuroscience and Medicine - Nuclear Chemistry (INM-5), Forschungszentrum Jülich GmbH, 52428 Jülich, Germany
}
\affiliation{Johannes Gutenberg University Mainz, 55099 Mainz,
	Germany}
\affiliation{Helmholtz-Institut Mainz, 55099 Mainz,
	Germany}
\author{Ch. E. D\"ullmann}
\affiliation{Johannes Gutenberg University Mainz, 55099 Mainz,
	Germany}
\affiliation{Helmholtz-Institut Mainz, 55099 Mainz,
	Germany}
\affiliation{GSI Helmholtzzentrum für Schwerionenforschung GmbH, 64291 Darmstadt, Germany}
\author{R. Ferrer}
 \email{Corresponding author: rafael.ferrer@kuleuven.be}
\affiliation{KU Leuven, Instituut voor Kern- en Stralingsfysica, 3001 Leuven, Belgium}
\author{Yu. Kudryavtsev}
\affiliation{KU Leuven, Instituut voor Kern- en Stralingsfysica, 3001 Leuven, Belgium}

\author{P. Palmeri}
\affiliation{Physique Atomique et Astrophysique, Universit\'e de Mons, B-7000 Mons, Belgium}
\author{P. Quinet}
\affiliation{Physique Atomique et Astrophysique, Universit\'e de Mons, B-7000 Mons, Belgium}
\affiliation{IPNAS, Universit\'e de Li\`ege, Sart Tilman, B-4000 Li\`ege, Belgium}
\author{S. Raeder}
\affiliation{GSI Helmholtzzentrum für Schwerionenforschung GmbH, 64291 Darmstadt, Germany}

\author{D. Renisch}
\affiliation{Johannes Gutenberg University Mainz, 55099 Mainz,
Germany}
\affiliation{Helmholtz-Institut Mainz, 55099 Mainz,
	Germany}
\author{P. Van den Bergh}
\affiliation{KU Leuven, Instituut voor Kern- en Stralingsfysica, 3001 Leuven, Belgium}
\author{P. Van Duppen}
\affiliation{KU Leuven, Instituut voor Kern- en Stralingsfysica, 3001 Leuven, Belgium}

\date{\today}

\begin{abstract}
 Laser ionization spectroscopy was performed on both neutral and singly ionized \ch{^{232}Th} with the aim of identifying the nuclear-clock isomer in the singly charged ionic state of \ch{^{229}Th}. A search for an efficient laser ionization scheme of \ch{^{232}Th^+} was conducted in an argon-filled gas cell. This revealed a congested spectrum due to collisional quenching effects and the presence of several auto-ionizing states, one of which has a laser ionization efficiency of at least $1.2 \%$. Using a threshold approach, the second ionization potential was determined to be $\SI{12.300 \pm 0.009}{\eV}$. The subsequent study on atomic \ch{^{232}Th} validated the threshold approach. Conducting spectroscopy in a hypersonic gas jet, suppressed the gas-collision-induced quenching, revealing a Rydberg series that converges to the first ionization potential, determined to be  $\SI{6.306879 \pm 0.000014}{\electronvolt}$. The gas jet also cools down the thorium, allowing for high-resolution laser spectroscopy with a resolution of $\SI{240 \pm 30}{\mega \hertz}$. Using the Multiconfigurational Dirac-Hartree-Fock (MCDHF) method, the ionization potentials were computed, showing a relative difference of 0.06\% and 0.19\% between theory and our experimental values for the ionization potentials of \ch{Th} and \ch{Th^+} respectively. Further calculations using a pseudo-relativistic Hartree-Fock method reveal strong mixing in the used intermediate state at $\SI{26113.27}{\per \cm}$ of \ch{Th}. A dedicated fast-extraction gas cell with \ch{^{233}U} recoil sources was used to study \ch{^{229}Th^+} but no photo-ionization signal could be observed.
\end{abstract}

\maketitle

\section{Introduction}
The use of a nuclear transition in a next-generation optical clock can provide unprecedented accuracy to time measurements \cite{Peik2003}. The small electric and magnetic moments associated with atomic nuclei make the frequencies of nuclear resonances especially insensitive to external perturbations. The \ch{^{229}Th} nucleus has emerged as a viable candidate for such a nuclear clock owing to its laser-accessible isomeric state. Recently, the isomer has been populated by laser excitation in a VUV-transparent host material \cite{Tiedau2024,Elwell2024,Zhang2024}, constituting an important milestone towards the realization of the clock. One of its main foreseen strengths is its enhanced sensitivity towards temporal variations in the fundamental constants due to its near perfect cancellation of the Coulomb and strong interaction energies. The sensitivity is dependent on the mean-square charge radii and the electric quadrupole moments of both the ground and isomeric states of \ch{^{229}Th}. An accurate determination of these observables has been a key motivator for laser spectroscopy studies on thorium.  
\\
\\
The low excitation energy of the isomer strongly affects its decay mechanism and was historically the reason why it remained elusive for so long. The first direct observation of the isomer was achieved by the detection of its internal conversion decay \cite{VonDerWense2016}. This mechanism is the dominant decay path for the isomer in the neutral atomic state, in which it has a half-life of $\SI{7\pm1}{\micro \second}$ \cite{Seiferle2017}. This short half-life makes laser spectroscopy unfeasible on the atom. Laser spectroscopy studies of the isomer have therefore been performed on the doubly charged ion \cite{Thielking2018} and more recently on the triply charged ion \cite{Yamaguchi2024}. In these ionic states, the fast internal conversion decay is energetically forbidden and the isomer decays predominantly via a much slower M1 gamma transition that was first observed in \cite{Kraemer2023}. 
\\
\\
The isomeric state, though, has never been observed in the singly charged ion. Electronic bridge processes are possibly at play in this configuration \cite{Porsev2010,Porsev2010b,Karpeshin2018}. An upper bound for the half-life of the isomer in the singly charged state was found to be $\SI{10}{\milli \second}$ \cite{Seiferle2017}, significantly shorter than predicted by theory \cite{Karpeshin2018,Karpeshin2021}. The proximity of electronic transitions to the isomeric state influences the electronic bridge decay process. The density of electronic states has therefore been mapped in previous studies \cite{Herrera-Sancho2013, Meier2019} and an initial estimation of the second ionization potential (IP\textsubscript{2}) was inferred \cite{Herrera-Sancho2013}. 
\\
\\
Motivated by these observations, we initiated a program to find and study the isomeric state in the singly charged ion \ch{^{229(m)}Th^+} using the In-Gas Laser Ionization and spectroscopy (IGLIS) technique at KU Leuven \cite{Kudryavtsev2016a}. With the achievable resolution of the in-gas-jet method of IGLIS ($\sim \SI{200}{\MHz}$), simulations suggest that the isomeric state can be distinguished from the ground state based on the hyperfine structure (HFS) \cite{thesisVerlinde}.To pursue this goal, first preparatory studies had to be performed to obtain an efficient laser ionization scheme for \ch{Th^{+}}. An existing two-color three-step scheme reported in the literature \cite{Herrera-Sancho2013} was initially tested but was found to provide only around $0.1 \%$ laser ionization efficiency. These investigations, led to the extraction of the first two ionization limits IP\textsubscript{1} and IP\textsubscript{2} of Th, both with many-fold improved uncertainties which we report here. To carry out these studies, we used \ch{^{232}Th} produced via laser-assisted ablation in a gas cell, which facilitated an extensive investigation of the spectrum around the IP\textsubscript{1} and IP\textsubscript{2} by resonance ionization spectroscopy. The \ch{^{229(m)}Th^+} ions were obtained from a \ch{^{233}U} alpha recoil source, however, laser ionization of \ch{^{229(m)}Th^+} proved to be unsuccessful, as no photoions could be observed. Nevertheless, the results of characterization studies of the recoil sources and of laser spectroscopy on neutral \ch{^{229}Th} are reported. These experimental measurements were accompanied by atomic structure calculations using the Multiconfiguration Dirac-Hartree-Fock (MCDHF) method to compute the IPs. In the case of \ch{Th}, identification of the intermediate atomic state used in Sec.~\ref{sec:laserspecneutralTh} was performed by a pseudo-relativistic Hartree-Fock method with included core-polarization effects (HFR+CPOL). The radial parameters of this method were refined using least-squares to minimize the energy difference between calculated and known experimental values of lower-lying states.	
\begin{figure*}
	\includegraphics[width=\linewidth]{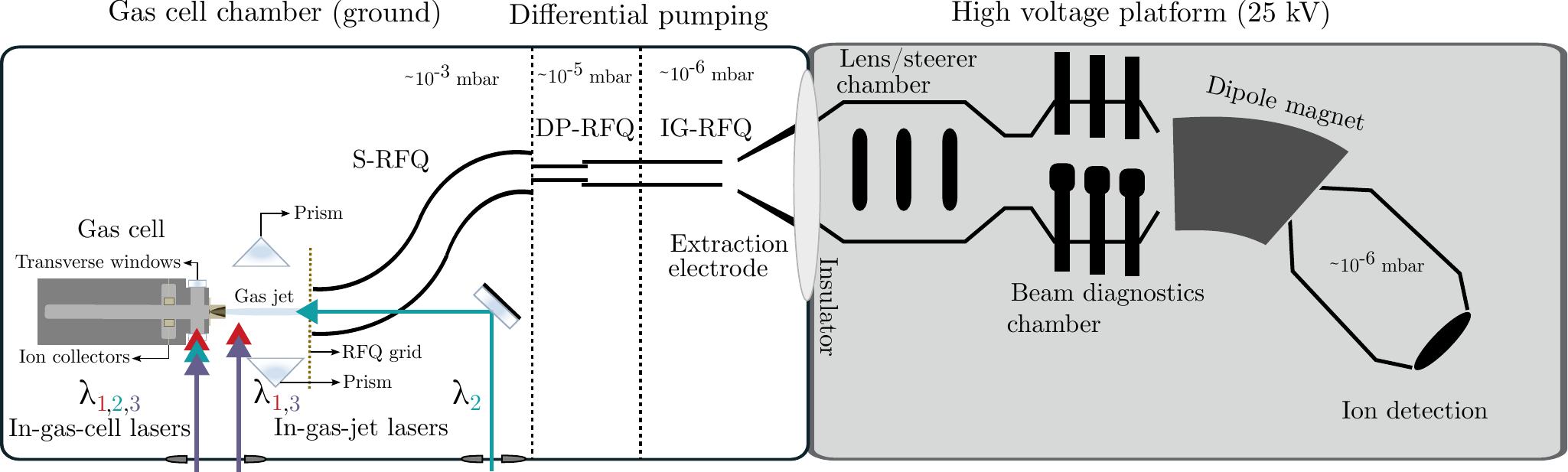}
	\caption{Schematic representation of the IGLIS jet laboratory at KU Leuven. The thorium isotopes are produced in the argon-filled gas cell and transported via gas flow through a de Laval nozzle into a hypersonic gas jet. The lasers can be overlapped in several different configurations. For low-resolution in-gas-cell spectroscopy, the laser can enter into the cell via transverse windows. For the high-resolution in-gas-jet spectroscopy, the lasers were overlapped transversely in a small portion of the gas jet or they can be reflected multiple times throught the gas jet by use of the prisms. Photoions are then captured by the RFQ system and extracted to the high voltage platform. After mass separation the ions are counted.}
	\label{fig:iglislab}
\end{figure*}
\section{Experimental setup} \label{sec:experimentalsetup}
Most laser spectroscopy studies of short-lived actinides have been performed using in-gas-cell techniques. A prime example is the isotope shift measurements conducted on the fission isomers of \ch{^{240{,}242}Am} \cite{Backe1998}. These measurements highlight the technique's high sensitivity, dealing with  production rates down to $\SI{10}{\per \s}$ and its applicability to short-lived species (e.g. \ch{^{240f}Am} has a half-life of $\SI{0.9}{\milli \s}$). While this was possible by the enhanced sensitivity of detecting fission fragments in the gas cell, detecting the decay channel of the thorium isomer is more cumbersome. Ion counting techniques can be used, requiring fast extraction of the isomer out of the gas cell volume and differential pumping (DP) as these detectors only operate in high vacuum. Furthermore, the spectroscopic broadening effects associated with a gas environment, i.e  Doppler and collisional broadening, lead to typical spectral resolutions of $\gtrsim$ 4 GHz. 
\\
\\
The use of supersonic gas jets in laser spectroscopy of short-lived nuclei was first implemented at the Leuven Isotope Separator On Line (LISOL) to drastically reduce the broadening effects associated with high-pressure gas environments \cite{Kudryavtsev2013,Raeder2016,Ferrer2017}. By using a convergent-divergent nozzle, a supersonic argon gas jet was created in which resonance ionization spectroscopy studies were performed on the neutron deficient actinium isotopes \ch{^{214,215}Ac} \cite{Ferrer2017}, improving the spectral resolution by one order of magnitude while preserving the overall efficiency and sensitivity of the gas cell results. A new generation of de Laval nozzles has recently been characterized offering a Mach number of more than $8$ (i.e. producing hypersonic gas jets) and bringing the resolution down to $\SI{200}{\MHz}$ for typical actinide transitions \cite{Ferrer2021,Lantis2024}. With such a hypersonic nozzle, the first in-gas-jet spectroscopy on \ch{^{254}No} has been performed using the JetRIS setup at GSI \cite{Lantis2024}. The laser spectroscopy studies on \ch{Th} and \ch{Th^+} reported in the present paper are performed with the same methodology in the off-line IGLIS laboratory at KU Leuven. The laboratory consists of a laser clean room and a gas jet setup and was specially designed to develop and optimize the in-gas-jet laser spectroscopy method. A detailed overview of the IGLIS laboratory is given in Ref.~\cite{Kudryavtsev2016a}. Only a short description of the main components relevant for the present work is given here. 
\\
\\
The front end of the jet setup (see Fig.~\ref{fig:iglislab}) houses the gas cell and the DP system. Two different gas cells are used in our experiments depending on the isotope we want to study. The gas cell or stagnation pressure $P_0$ is optimized per experiment. To achieve a short evacuation time ($ < \SI{5}{\milli \second}$) a dedicated fast-extraction gas cell was developed housing a pair of cylindrical \ch{^{233}U} recoil sources. The nuclear recoils are stopped in a high-purity argon buffer gas with which they get extracted via a low-P$_{0}$ de Laval nozzle \cite{Lantis2024,thesisClaessens} to hypersonic velocities, resulting in a well-collimated and low-temperature gas jet. The achievable spectral resolution in the gas jet can eventually be employed for separation and identification of the isomer from the ground-state based on its hyperfine splitting. In addition, a larger S\textsuperscript{3}-prototype gas cell is used for the production of \ch{^{232}Th} by laser ablation  (Sec.~\ref{sec:ablation}). Both gas cells can also be equipped with transverse optical viewports for in-gas-cell laser spectroscopy studies. As the performance of the nozzle is critically dependent on the background pressure environment in which the gas jet expands, an extensive DP is maintained. This specific nozzle requires a background pressure of $\SI{5e-3}{\milli \bar}$ (for $P_0 = \SI{80}{\milli\bar}$), which is achieved with an Edwards XDS 35i dry scroll pump backing three turbomolecular pumps, one per DP section. The radiofrequency quadrupole (RFQ) ion guide  system transporting the ions through the differential pumping sections consists of an S-shaped RFQ, a small DP RFQ, and the ion guide (IG) RFQ. 
\\
\\
The ions are extracted to a high-voltage platform, where an Einzel lens and a set of steerers are used for ion-beam manipulation. Beam diagnostics are possible with a Faraday cup and a microchannel plate (MCP) detector mounted with a phosphor screen as anode and a USB camera to record the beam spot images. To adjust the beam intensity, an attenuator with a $1\%$ transmission can be inserted. A dipole magnet with a resolving power $m/\Delta m=180$ at $\SI{25}{\kilo \volt}$ is used for mass separation, which can be further increased using mass slits \cite{thesisClaessens}. Ion detection is performed at the focal plane of the magnet using a channel electron multiplier (CEM).    
\\
\\
The IGLIS laser system comprises three wide-tunable, high-power, high-repetition-rate dye lasers (Sirah Credo), which are each optically pumped by Nd:YAG (Edgewave INNOSLAB). These are used for broadband excitation, scanning, as well as ionizing steps. The bandwidth of the dye lasers with a grating of $\SI{3000}{lines/\mm}$ is around $\SI{0.06}{\per \cm}$ at fundamental light of $\SI{530}{\nm}$. The pump lasers can be operated with $\SI{532}{\nm}$ ($\SI{355}{\nm}$) and an output power up to $\SI{90}{\watt}$ ($\SI{36}{\watt}$) at $\SI{10}{\kilo \hertz}$ pulse repetition rate. A higher power pump laser is also available and can provide up to $\SI{165}{\watt}$ at $\SI{532}{\nm}$  and $\SI{72}{\watt}$ at $\SI{355}{\nm}$ at $\SI{15}{\kilo \hertz}$. During the studies presented in this paper, the repetition rate was set to $\SI{7}{\kilo \hertz}$ as this provides the highest energy per pulse. 
\\
\\
For higher resolution laser spectroscopy studies a narrow-bandwidth laser system is employed which consists of a pulsed dye amplifier (PDA) seeded by a single-mode continous-wave laser (usually an external-cavity diode laser) and with up to three amplification stages. To ensure that the amplified light remains single-mode in nature, a single-longitudinal mode Nd:YAG (SLM) laser is used for optical pumping. This laser provides up to $\SI{60}{\watt}$ at $\SI{532}{\nm}$ and $\SI{15}{\kilo \hertz}$ pulse repetition rate. The characterization  and optimal specification of a prototype version of this laser is discussed in \cite{Verlinde2020}. The PDA system provides Fourier-limited laser pulses of $\SI{2.8}{\ns}$ corresponding to a bandwidth of $\SI{157}{\MHz}$ in fundamental light of $\SI{760}{\nm}$ created by a solution of styryl 8 laser dye in dimethyl sulfoxide. The wavelength is measured by a Highfinesse WS7/60 wavemeter, which is periodically calibrated using a frequency-stabilized He-Ne laser.
\begin{figure}
	\centering
	\includegraphics[width = \linewidth]{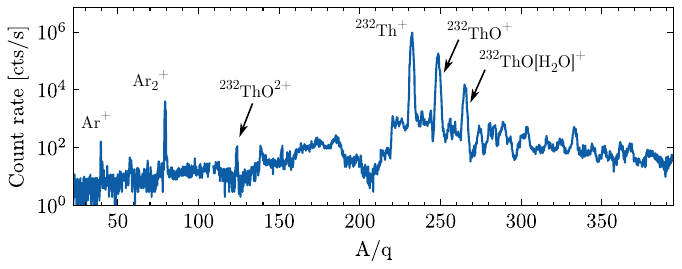}
	\caption{Mass spectrum when ablating thorium with identified molecular sidebands. This corresponds to an energy fluence of about $\SI{0.15}{\joule / \cm \squared}$. This spectrum was taken after several weeks of conditioning. About $\SI{1000000}{ions/\s}$ of $\ch{^{232}Th^+}$ and $\SI{200000}{ions/\s}$ of $\ch{^{232}ThO^+}$ are produced.}
	\label{fig:massSpectra_thoriumAblation}
\end{figure}
\section{Development of a laser ionization scheme for T\MakeLowercase{h}$^+$}
Efficient laser ionization is key to studying the isomer in \ch{^{229}Th^+}. Previous studies of the electronic structure of thorium, cf. \cite{Herrera-Sancho2012,Herrera-Sancho2013,Meier2019}, used two-photon laser-induced fluorescence to probe the electronic states in the range of $7 - \SI{10}{\eV}$ to investigate the possibility of exciting the isomer via an inverse electronic bridge process. Laser ionization of \ch{^{232}Th^+} was observed in these studies by a two-color, three-step scheme when the second laser reached states above $\SI{61000}{\per \cm}$. This allowed the authors of these studies to locate the second ionization potential IP\textsubscript{2} in a range between $\SI{11.9}{\eV}$ and $\SI{12.3}{\eV}$ \cite{Herrera-Sancho2013}. Reproducing this ionization scheme in our laboratory resulted in  a laser ionization efficiency of $ \lesssim 0.1 \%$. 
\\
\\
Motivated by the need for a more efficient ionization mechanism for Th$^{+}$, a three-color, three-step laser scheme was investigated by means of laser-assisted ablation of a \ch{^{232}Th} target located inside the gas cell. Preliminary results using the in-gas-cell method were already published in \cite{Claessens2023}, showing a dense spectrum of resonances around IP\textsubscript{2}. Building upon this work, the full uncertainty range of IP\textsubscript{2} was covered to extract several autoionizing (AI) states and therefore, a higher ionization efficiency laser scheme as well as a more accurate value of the IP\textsubscript{2}, as reported in the following sections.
\subsection{Laser-assisted ablation of Th$^+$} \label{sec:ablation}
Singly charged \ch{^{232}Th} ions were obtained by laser ablation of a metallic thorium foil (Goodfellows, $99.5 \%$ purity, $\SI{25}{\mm} \times \SI{25}{\mm}$ and $\SI{64}{\micro \meter}$ thick). This foil was installed in a gas cell that serves as the prototype model to be used in online experiments at the S\textsuperscript{3} Low Energy Branch (S\textsuperscript{3}-LEB) \cite{Kudryavtsev2016a} and through which a constant high-purity argon gas flow (grade 6.0, cleaned by monoTorr getter-based purifier) was passed. The foil was illuminated by the second harmonic of a Nd:YAG laser (Quanta-Ray/ Spectra Physics INDI series Laser) at $\SI{532}{\nm}$ and $\SI{20}{\hertz}$ pulse repetition rate and focused on a $\SI{3}{\mm \squared}$ area. With an average energy of $\approx\SI{5}{\milli \joule /pulse}$ and a buffer gas pressure of $\SI{15}{\milli \bar}$, a 10$^6$ ions$/\si{\s}$ beam of $\ch{^{232}Th^+}$ could be mass-separated, while only providing a background beam of several tens of $\,\si{ions/\s}$ in the \ch{^{232}Th^{2+}} mass region. The mass spectrum under these conditions is shown in Fig.~\ref{fig:massSpectra_thoriumAblation}. Under optimal conditions, it was possible to keep the beam stability within $10\%$ (difference between highest and lowest count rate over average count rate). On average, the stability was about $20 \%$. The most stable configuration of ablation was found when using a scanning mirror which would alternate between two different points on the foil every $\SI{0.5}{\s}$. 
\subsection{Search for atomic levels in the gas cell configuration}
\begin{figure}
	\centering	
	\includegraphics[width =\linewidth]{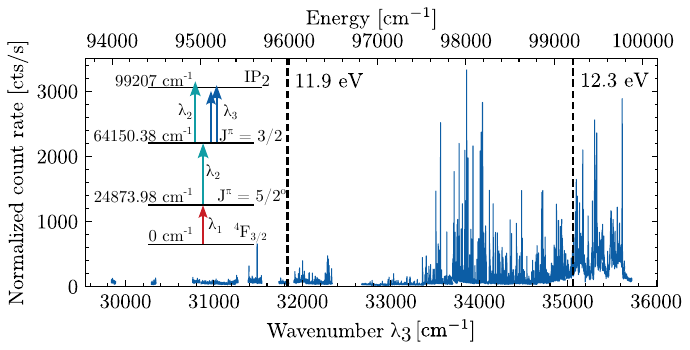}
	\caption{Level search conducted on \ch{Th^+} using the third laser step ($\lambda_3$) in the gas cell ($P_0 = \SI{15}{\milli \bar}$). The first ($\lambda_1$) and second step ($\lambda_2$) were fixed at a wavelength of $\SI{402.026}{\nm}$ and $\SI{254.606}{\nm}$ respectively. All spectra were normalized to an ion beam intensity of $500\,000 \, \text{cts/s}$ of \ch{Th^+}. As this search consists of about $40$ individual scans conducted at varying laser conditions, peak height does not indicate transition strength. The literature range of IP\textsubscript{2} is denoted by the dashed lines.}
	\label{fig:fulllevelsearch}
\end{figure}

Three dye lasers, each pumped by one Nd:YAG laser (see end of Sec.~\ref{sec:experimentalsetup})  were used to ionize \ch{^{232}Th^+} inside the gas cell. The three laser steps illuminated the full transverse entrance windows of the gas cell (see Fig.~\ref{fig:iglislab}). The first and second lasers were fixed to the known transitions while the third-step laser was used to scan the region around the IP\textsubscript{2}. The first-step transition corresponds to the strongest resonance starting from the ground state (\ch{$6d^27s$ \, $^4$F$_{3/2}$ -> $6d7s$($^3$D)$7p$} (J$^\pi = 5/2^\text{o}$) at an energy of $\SI{24873.98386 \pm 0.00014}{\per \cm}$ \cite{Redman2014,NIST_ASD_ThI}. The second excited state is situated at a measured energy of $\SI{64150.38 \pm 0.06}{\per \cm}$ and has an unknown configuration with a total angular momentum of $J = 3/2$. Starting from this state, a region of about $\SI{6000}{\per \cm }$ was covered during the third-step scans as shown in Fig.~\ref{fig:fulllevelsearch}. 
\\
\\
The shot-by-shot variance of the ablation ion source was averaged out by measuring the count rate at each set wavenumber seven times for $\SI{1}{\s}$. To minimize the variation in background, the power in the first and second step was reduced in order to suppress laser-induced background. Such background consists of two-color, three-step laser-ions and non-resonant background. The main contribution to the background originated from the laser ablation source itself followed by non-resonant background related to the third laser step. The ion beam intensities of \ch{Th^+} were regularly recorded and used to normalize the background between individual scans. Overall, more than $700$ different peaks were observed, differing in morphology and background, that enabled the extraction of the IP\textsubscript{2}.
\subsection{Ionization threshold} \label{sec:IP2threshold}
\begin{figure}
	\centering
	\includegraphics[width =\linewidth]{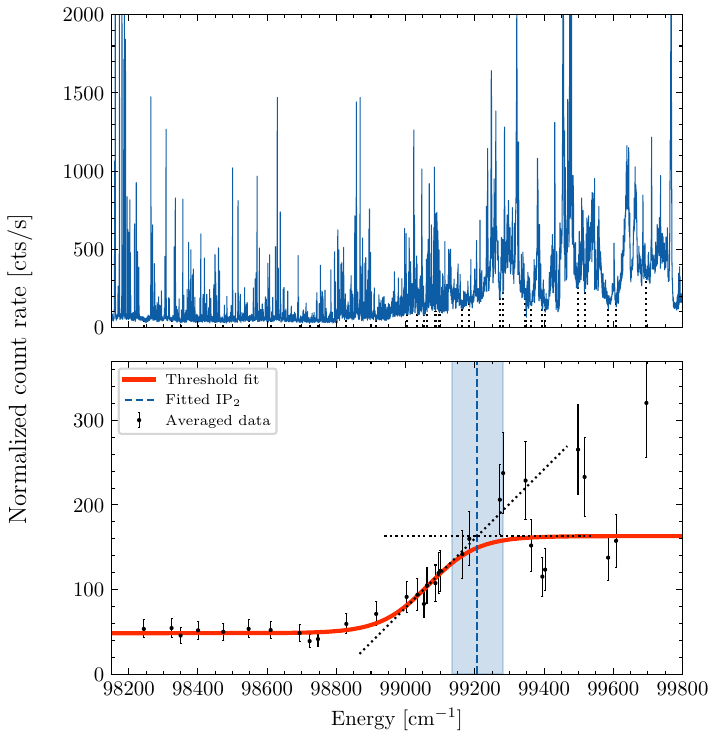}
	\caption{Top: Part of the normalized level search shown in Fig~\ref{fig:fulllevelsearch}. The selected background points are denoted by the dashed lines. Bottom: The threshold S-curve (Eq.~\ref{eq:sigmoid-Scurve})  fitted to the background points. The dotted lines represent the upper asymptote and tangent of the inflection point. The found IP\textsubscript{2} is the intersection of those lines and its uncertainty is denoted by the blue shaded area.}
	\label{fig:thresholdIP2_averaged}
\end{figure}

The raise in count rate between the peaks and the difference in peak width observed around $\SI{35000}{\per \cm}$ (cf. Fig.~\ref{fig:fulllevelsearch}) indicate a change in ionization behavior, i.e. the crossing of the IP\textsubscript{2}. The states shown in Table~\ref{tab:AIs_ThII} are substantially wider than the typical peaks found at lower energies which have a FWHM of $0.2 - \SI{0.3}{\per \cm}$. The photoionization threshold is determined by fitting a sigmoid function through selected points in the regions between the peaks. These points are taken as averages of $\SI{0.5}{\per \cm}$-wide regions absent of resonances. In the high-energy portion of the spectra, the points are chosen to be in the valleys of the ionizing structures. The photoionization threshold $E_{\text{IP}_2}$ is given by the intercept of the top asymptote with the tangent of the inflection point of the sigmoid function. The threshold is represented by a dashed line in Fig.~\ref{fig:thresholdIP2_averaged}. The sigmoid function $y$ and the photoionization threshold $E_{\text{IP}_2}$ are given by
\begin{equation} \label{eq:sigmoid-Scurve}
	y = \frac{A_2-A_1}{1+ e^{\frac{x-x_0}{\beta}}} + A_1, \qquad E_{\text{IP}_2} = 2\beta + x_0\, ,
\end{equation}
where $x_0$ is the inflection point of the S-curve and $\beta$ is a slope parameter. The arms of the curve are determined by $A_1$ and $A_2$ for the top and bottom asymptotes respectively. This analytic method was chosen to better capture the transition between the Rydberg states, which merge together to form the S-shape, and the continuum states with their autoionizing structures forming the high-energy plateau, and is based on the work on astatine \cite{Rothe2013,thesisRothe}. The photoionization threshold for Th$^+$ is therefore found to be $E_{\text{IP}_2} = \SI{99207 \pm 73}{\per \cm}$ or $\SI{12.300 \pm 0.009}{\electronvolt}$. 
\subsection{Laser ionization efficiency}
Laser ionization efficiencies of around $0.5 \%$ were obtained in runs that led to the data shown in Fig.~\ref{fig:fulllevelsearch}. Most of the resonances found are located below the IP\textsubscript{2}, indicating that ionization occurs via four or more steps. The most intense of these peaks were checked and confirmed to be dependent on all three laser steps (i.e. three-color). Most likely, collisional de-excitation and subsequent laser ionization are involved, a phenomenon that has been observed in atomic Th and is discussed in Sec.~\ref{sec:gascell_quenching}. Despite this, some of these transitions exhibit ionization efficiencies of several percent and could, for instance, eventually be used for radioactive ion beam (RIB) production in a gas cell-based laser ion source. Two examples are the transition at $\SI{33837.3}{\per \cm}$ with $1.9(3)\%$ (from \cite{Claessens2023}) and a transition at $\SI{34113.6}{\per \cm}$ with $3.4(5)\%$. The highest achievable laser powers for these transitions were used: $\SI{500}{\milli \watt}$, $\SI{180}{\milli \watt}$ and $2100-\SI{2200}{\milli \watt}$ of laser power for steps 1, 2 and 3, respectively, at $\SI{7}{\kilo \hertz}$ pulse repetition rate.
\begin{figure}
	\centering
	\includegraphics[width =\linewidth]{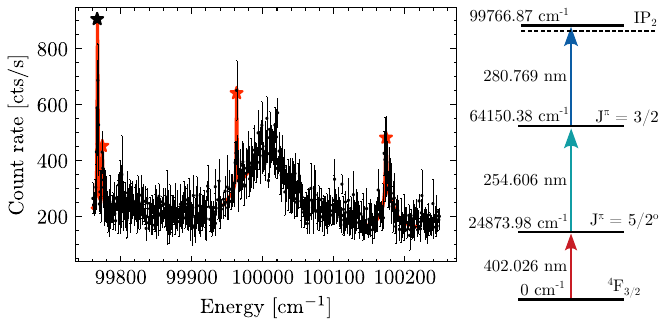}
	\caption{Left: Spectrum of the AI states with the best fits of four different AI states plotted in red. The $\SI{99766.87}{\per \cm}$ AI state is marked by the black star, the others with a red star. This scan was performed to try and identify a stronger AI state than the state at $\SI{99766.87}{\per \cm}$ and therefore larger stepsizes were used ($1 - \SI{2}{\per \cm}$). Right: The proposed laser ionization scheme using the strongest AI state.}
	\label{fig:gc_scan_aboveIP2}
\end{figure}
The strongest states found above IP\textsubscript{2} are shown in the left panel of Fig.~\ref{fig:gc_scan_aboveIP2} and are tabulated in Table~\ref{tab:AIs_ThII}. The strongest AI resonance is found at an energy of $\SI{99766.87}{\per \cm}$. Owing to the angular momentum of the state below, the found AI states are assigned to have odd parity with either $J = 1/2, 3/2$ or $5/2$. The efficiency for ionization via the $\SI{99766.87}{\per \cm}$ state was measured to be at least $1.2\%$ in an environment of $\SI{15}{\milli \bar}$. Relative strengths were measured only at an argon pressure of $\SI{80}{\milli \bar}$. The higher stagnation pressure affects the output of the ablation source and increases quenching effects, both diminishing the photoion intensities. Using these results, a new laser ionization scheme is proposed and shown in the right panel of Fig.~\ref{fig:gc_scan_aboveIP2}. Because of collisional de-excitation (quenching), an accurate efficiency for the states is difficult to ascertain. Attempts at ionizing Th$^+$ in the gas jet, where collisional effects are minimized due to the four orders of magnitude lower gas pressure, proved, however, unfruitful to assign a reliable ionization efficiency. A photoionization signal was observed, showing a clear difference between under and above IP\textsubscript{2}, but for unknown reasons the count rate varied enormously from laser shot to laser shot. This instability could be related to the multi-mode structure of the laser light and the narrow resonances of thorium in the gas jet. Spectral fluctuations over the different laser modes might thus drastically affect the count rate. Two other possible issues might also limit the performance of the in-gas-jet method. On the one hand, the technical difficulties of overlapping three lasers with the gas jet, and on the other hand, performing in-gas-jet spectroscopy on ions. For the former, technical constraints in overlapping the three lasers at the gas jet, meant that the second step ($\lambda_2$ in Fig.~\ref{fig:iglislab}) which has the lowest available power, was aligned in an anti-propagating geometry. The other two remaining lasers ($\lambda_{1,3}$) could easily be overlapped in a transverse geometry to the jet, either by a beam splitter or a D-shaped mirror. The passage of the laser though the nozzle throat and the gas cell could somehow affect the stability of the Th$^+$ signal extracted from the gas cell. As for the latter, the presence of the electric fields from the S-RFQ on the gas jet region could affect the Th$^+$ ion beam. Shielding the S-RFQ with a grid however did not eliminate the fluctuations.

\begin{table}
	\centering
	\caption{Energy, FWHM and relative intensities of the found AI states.}
	\begin{tabular}{@{}lll@{}}
		\toprule
		\multicolumn{3}{c}{Auto ionizing states}            \\ \midrule
		\multicolumn{1}{c}{Energy [$\si{\per \cm }$]}    & \multicolumn{1}{c}{FWHM [$\si{\per \cm}$]}  & \multicolumn{1}{c}{relative intensity}  \\ \midrule
		99766.87(22)   &   2.55(12)   & 1    \\
		99774.1(5)   &   2.59(14)   & 0.3    \\
		99963.1(8)   &   1.38(30)   & 0.5   \\
		100173.5(4)  &   2.6(15)   &  0.4   \\ \bottomrule
	\end{tabular}
	\label{tab:AIs_ThII}
\end{table}
\section{Attempt at Laser spectroscopy of $^{229}$T\MakeLowercase{h}$^+$}  \label{sec:recoilsources}
\begin{figure}[H]
	\includegraphics[width =\linewidth]{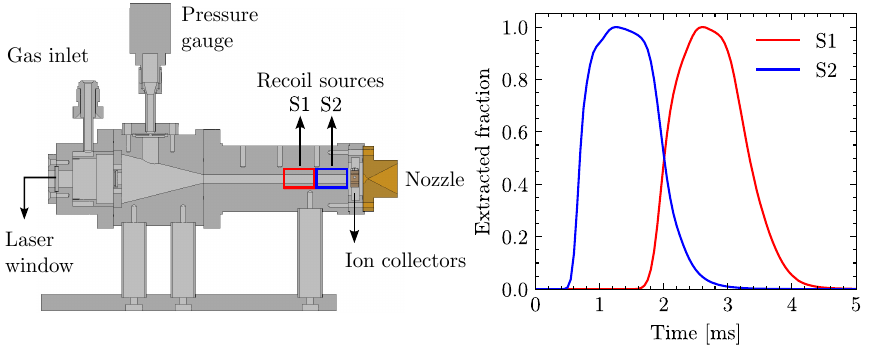}
	\caption{Left: Schematic of the fast gas cell and the positions of the two recoil sources. Right: the extraction time out of the fast gas cell of the recoil ions computed using COMSOL Multiphysics. The red(blue) bounding box in the schematic and the red(blue) extraction time curves correspond to the S1(S2) recoil source.}
	\label{fig:fastgascell}
\end{figure}

Without access to an accelerator facility, the most viable approach to populate the isomeric state in \ch{^{229}Th^+} is using an alpha recoil source of \ch{^{233}U}. In the alpha decay, approximately $2\%$ \cite{VonDerWense2016} of the thorium recoils end up in the isomeric state. Fast extraction of thorium is essential for studying the isomer in the singly charged ionic form (see Introduction). Therefore, a dedicated fast-extraction gas cell was designed and gas-cell evacuation times were simulated using the COMSOL Multiphysics software. The gas cell consists of a cone-shaped volume extending to a narrow channel in which two cylindrical sources are mounted. Simulations suggest extraction times of about $\SI{1}{\ms}$ for the first cylinder (S1) and $\SI{2.5}{\ms}$ for the second (S2) \cite{thesisVerlinde}.  A schematic of the gas cell and the simulated evacuation times are shown in the left and right panels of  Fig.~\ref{fig:fastgascell} respectively. 
Each cylindrical source contains a $\SI{20}{\mm} \times \SI{15.7}{\mm}$ titanium foil ($\SI{10}{\micro \meter}$ thick) onto which \ch{^{233}U} was deposited by molecular plating \cite{Vascon2012}. One source is loaded with a measured activity of $\SI{30.8}{\kilo \becquerel}$ and the other with $\SI{9.2}{\kilo \becquerel}$, corresponding to $\SI{86.4}{\micro \gram}$ and $\SI{25.8}{\micro \gram}$ of \ch{^{233}U} respectively. The exact chemical composition of the layers is not known but is expected to be \ch{UO2} mixed with some hydroxide and carbonate species. The isotopic composition is  87.87\% \ch{^{233}U}, 11.21\% \ch{^{238}U}, 0.83\% \ch{^{234}U} and $<0.1\%$ for each isotope of \ch{^{232{,}235{,}236}U}. The production process and characterization of  the sources is discussed in \cite{Haas2020}. Recoil efficiencies were determined at JGU Mainz by adding a tracer of \ch{^{243}Am} in similarly produced foils. Recoiling \ch{^{239}Np} daughters were caught on an aluminium foil in vacuum and quantified by gamma spectroscopy. Efficiencies of $3.2(8) \%$ ($\SI{30.8}{\kilo \becquerel}$) and $6.9(20) \%$ ($\SI{9.2}{\kilo \becquerel}$) were inferred \cite{reportSourcesMainz}.
\begin{figure} \label{fig:massspectrumrecoil}
	\includegraphics[width = \linewidth]{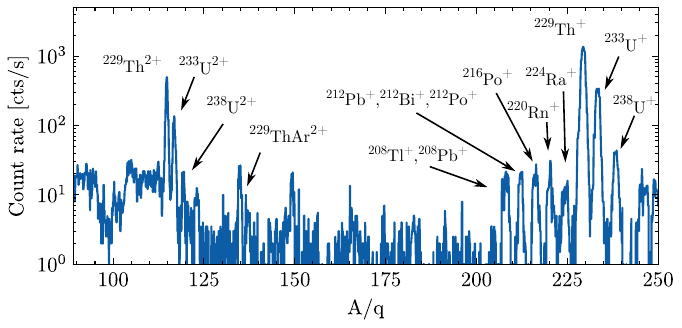}
	\caption{Mass spectrum of the recoil sources in optimal conditions. The composition of the used uranium is discussed in the text and shows the in-growth of daughters of the \ch{^{232}U} isotopic contaminant.}
\end{figure}
The sources were loaded into the fast gas cell such that the high activity source ($\SI{30.8}{\kilo \becquerel}$) was inserted first (in position S1 cf. Fig.~\ref{fig:fastgascell}), leaving the $\SI{9.2}{\kilo \becquerel}$ source closer to the exit hole (position S2).
\\
\\
Initial characterization of the sources was published in \cite{Claessens2023}. Count rates of $\SI{680}{counts/\s}$ for \ch{^{229}Th^+} and $\SI{210}{counts/\s}$ for \ch{^{229}Th^{2+}} were determined, indicating an extraction efficiency of $85\%$. More recent mass spectra (see Fig~\ref{fig:massspectrumrecoil}) using $P_0 = \SI{38}{\milli \bar}$ and a free gas jet (diameter $\SI{1.5}{\mm}$) showed higher intensities of up to $\SI{1360}{counts/\s}$ for \ch{^{229}Th^+} and $\SI{500}{counts/\s}$ for \ch{^{229}Th^{2+}}. The increase in count rate by a factor of two is rather puzzling as this would correspond to an unphysical transport efficiency. This might indicate a large uncertainty regarding the determination of the recoil efficiencies. The reason for the overall improvement in count rate is also not well understood but conditioning of the recoil sources under prolonged ($> \SI{1}{\year}$) argon atmosphere or the cumulative effect of slight beam tuning improvements are both considered.  
\\
\\
Using the ionization scheme reported above with a measured laser ionization efficiency of 1.2\% and testing different laser geometry arrangements, several attempts to laser ionization of \ch{^{229}Th^+} were carried out. Unfortunately, no photoion signal was observed, unlike for \ch{^{232}Th^+}.  A plausible explanation could be that most of \ch{^{229}Th^+} ions end up in long-lived excited electronic states after recoiling into the argon buffer gas. Either these states are populated directly during recombination in the gas or via the de-excitation pathway. Such states would be unavailable for laser excitation, reducing the efficiency of the developed laser ionization scheme. Any laser-ions could then easily be obscured by the background of \ch{Th^{2+}} ions coming from the recoil sources. Attempts at reducing this ion background using e.g. a xenon admixture to the argon buffer gas were unsuccessful.

\section{Laser spectroscopy of T\MakeLowercase{h}$^0$} \label{sec:laserspecneutralTh}
Practical difficulties, such as the overlapping of three laser beams with the gas jet and the presence of electric fields in the jet region made it impossible to study Th$^+$ in the gas jet. In fact, the spectral congestion observed in the scans of Th$^+$ while searching for atomic levels in the gas cell  obscured any clear indication of a Rydberg series. Such a series would provide a significant improvement on the precision of the  IP\textsubscript{2}. To try to understand, on the one hand, systematic effects in the determination of the ionization potential via the threshold method and, on the other hand, the quenching mechanisms in the gas cell medium, we initiated a series of measurements using neutral Th, for which precise experimental data are available in the literature, e.g. IP\textsubscript{1} of $\SI{50867\pm 2}{\per \cm}$ \cite{Kohler1997}. Furthermore, these measurements would result in the  first determination of the IP\textsubscript{1} of an element via Rydberg series in the gas jet.

\subsection{Search for atomic levels around IP\textsubscript{1}}
Laser ionization schemes of atomic Th are well-known in the literature \cite{Liu2020,Raeder2011}. Both three- and two-step schemes are known. A two-step scheme was chosen for our studies using the \ch{$6d^27s^2$ ^3F_{2} -> }(J$^\pi = 2^\text{o}$) ground-state transition at $\SI{382.947}{\nm}$ from \cite{Tomita2018} for the first step. This light can easily be produced by both our broadband laser and narrowband PDA. Unfortunately, this choice of first step allows ionization via a one-color, two-step process, resulting in a permanent background. The laser ionization scheme is shown in the inset of Fig.~\ref{fig:spectrum_gascellvsgasjetshield_long}. The second-step laser was scanned just below the ionization limit IP\textsubscript{1} from a wavelength of $\SI{403.9}{\nm}$ to $\SI{404.9}{\nm}$ with a typical step size of $\SI{0.11}{\per \cm}$ and a laser bandwidth of $\sim\SI{0.06}{\per \cm}$. The laser power was kept high at $\SI{250}{\milli\watt}$, inducing enough power broadening to compensate for the large step size. Both lasers were operated at $\SI{7}{\kilo \hertz}$. Laser-assisted ablation provided the thorium atoms and the ion collector electrodes in the gas cell were used to remove any non-laser related background ions. 
\\
\\
Similarly to the Th$^+$ case, a search for atomic levels in the vicinity of the ionization threshold did not reveal a clear Rydberg series when performing in-gas-cell laser spectroscopy, as illustrated in the top panel of Fig.~\ref{fig:spectrum_gascellvsgasjetshield_long}. When a similar region was scanned in the gas jet, a structure of converging Rydberg states appeared below the known IP\textsubscript{1}. To suppress any Stark splitting and broadening originating from the RFQ electric fields, a wire grid was installed at the end of the $\SI{60}{\mm}$ gas jet at a distance of $\SI{10}{\mm}$ from the RFQ entrance (see Fig.~\ref{fig:iglislab}). The wire grid shielding not only improves the resolution but also separates the laser excitation from the field ionization region (see Sec.~\ref{sec:Stark}). This further simplified the spectrum as short-lived, low-$n$ valence states are allowed to decay during hypersonic travel (approximate travel time of $\SI{110}{\micro \second}$). The resulting in-gas-jet spectrum (see bottom panel in Fig.~\ref{fig:spectrum_gascellvsgasjetshield_long}) shows clear Rydberg states just below the IP\textsubscript{1} as the transitions originating from collisional de-excitation are almost fully suppressed in the gas jet.
\subsection{In-gas-jet spectroscopy of Rydberg series}\label{section:RydbergSeries}
\begin{figure*}
	\includegraphics[width=\linewidth]{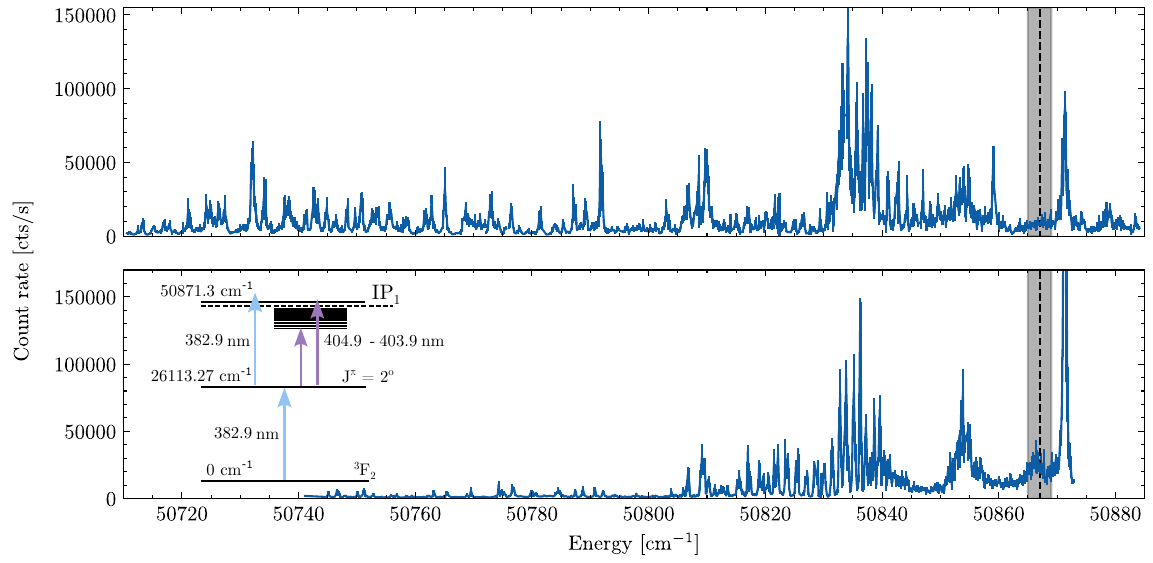}
	\caption{Top: Spectrum obtained by scanning the 2nd step with both steps transversely illuminating the gas cell. Bottom: The same range with the lasers transversely intersecting the gas jet. The gas jet itself is shielded from the electric fields of the RFQ using a wire grid. The value of IP\textsubscript{1} from \cite{Kohler1997} is denoted by the dashed line with shaded $2\sigma$ uncertainty and the ionization scheme is given in the inset. A clear change in the in-gas-jet spectrum is observed when approaching the ionization threshold. The auto-ionizing state at $\SI{50871.3}{\per \cm}$ (right-most peak) goes up to 700 000 cts/s in the gas jet spectrum.}
	\label{fig:spectrum_gascellvsgasjetshield_long}
\end{figure*}
\begin{figure}
	\centering
	\includegraphics[width = \linewidth]{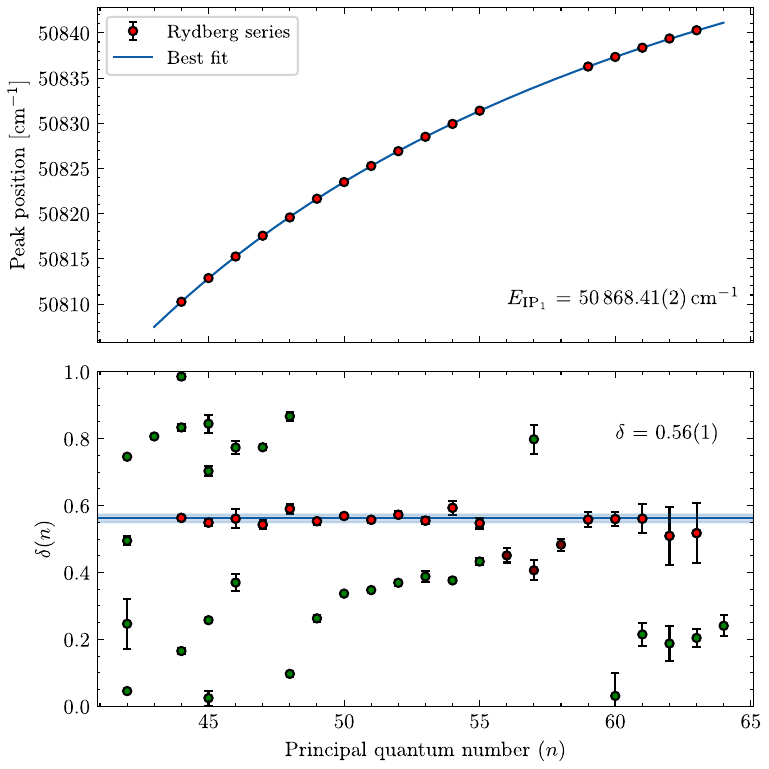}
	\caption{Top: Convergence of the identified Rydberg series (red) to $E_{\text{IP}_1}$. Bottom: the quantum defect $\delta$ with respect to the assigned principal quantum number $n$ for all 49 peaks. The identified series is denoted in red with three of its members ($n=56,57,58$, in maroon) being perturbed by an interfering state at $n=57$. These three peaks were not taken into consideration for the convergence. Other partial series and miscellaneous peaks are colored in green.}
	\label{fig:Rybergconvergencefit_annotated}
\end{figure}

To identify different Rydberg series in a spectrum, the quantum defects are analyzed. In the region $50800-\SI{50845}{\per \cm}$, 49 peaks were observed and their positions were determined by a Voigt fit. They are listed in Table.~\ref{tab:Rydbergstates}. An effective quantum number $n^\ast$ and a quantum defect $\delta$ was assigned to each peak by
\begin{equation*}
	n^\ast = \sqrt{	\frac{R_\mu}{E_{\text{guess}} - E_{n}}} \text{ and } \delta \equiv -n^\ast \pmod 1,
\end{equation*}
for which the literature IP\textsubscript{1} value was taken as initial guess $E_{\text{guess}}$. The reduced Rydberg constant $R_\mu$ is $\SI{109737.31568160 \pm 0.00000021}{\per \cm}$ \cite{CODATA2018}.  The quantum defect defined in this manner is equal to the real quantum defect up to an integer value. A series is identified by repeated constant quantum defects over a long chain of principal quantum numbers $n = n^\ast + \delta$. Such a chain can be perturbed by interloper states with the same parity and angular momentum. Seventeen unperturbed members of a single Rydberg series were identified and fitted using the Rydberg-Ritz formula,
\begin{equation}
	E_n = E_{\operatorname{IP}} -  \frac{R_{\mu}}{ {(n - \delta )}^2}.
\end{equation}
The resulting fit, shown in Fig.~\ref{fig:Rybergconvergencefit_annotated}, converged to a quantum defect of $\delta = 0.56(1)$ and the IP\textsubscript{1} of \ch{Th} was determined to be $50868.41(2)_\text{stat.}(11)_\text{sys.} \, \si{\per \cm}$. The statistical uncertainty of $\SI{0.02}{\per \cm}$ is completely determined by the large number of Rydberg states in the chain. The stepsize of the scan ($\SI{0.11}{\per \cm}$) is assigned as a systematic uncertainty and dominates the error budget. The full list of peaks and their associated quantum defects is given in Table.~\ref{tab:Rydbergstates}
\\
\\
The threshold method used in Sec.~\ref{sec:IP2threshold} is also applied on the atomic spectra to obtain a value for IP\textsubscript{1}. In the spectrum of the gas cell (cf. top of Fig~\ref{fig:spectrum_gascellvsgasjetshield_long}), a value of $\SI{50847 \pm 34}{\per \cm}$ was determined which agrees well with the found $\SI{50868.41 \pm 0.11}{\per \cm}$ Rydberg-Ritz value in the gas jet. This highlights the systematic deviation associated with the threshold method. These deviations have been recorded for several species in the PhD thesis of S. Rothe (App. C of \cite{thesisRothe}). In the case of \ch{Th}, the threshold approach gives respectively $\SI{21}{\per \cm}$ lower values than the Rydberg-Ritz method. These deviations are well within the $\SI{73}{\per \cm}$ statistical uncertainty of the threshold approach for the case of \ch{Th^+}. 

\begin{table}
	\centering
		\caption{Table of the found resonances in the Rydberg region. Energy and quantum defect are given. For the states belonging to the series, the principal quantum number is provided. States denoted by $\ast$ are believed to be part of the series but are disturbed by an interfering valence state and are not used in the Ritz-Rydberg convergence fit.}
	\begin{tabular}{@{}llllll@{}}
		\toprule
		\multicolumn{1}{c}{$n$}   & \multicolumn{1}{c}{Energy [$\si{\per \centi \meter }$]}    &\multicolumn{1}{c}{$\delta$} &	\multicolumn{1}{c}{$n$} & \multicolumn{1}{l}{Energy [$\si{\per \centi \meter }$]}    &\multicolumn{1}{c}{$\delta$}   \\ \midrule
		& $50803.93(2)$ & $0.75(1)$  & 		 & $50823.92(1)$ & $0.34(1)$ \\ 
		& $50804.71(4)$ & $0.49(1)$  &		$51$ & $50825.28(1)$ & $0.56(1)$ \\ 
		& $50805.46(23)$ & $0.25(8)$ & 		 & $50825.64(2)$ & $0.35(1)$ \\ 
		& $50806.07(2)$ & $0.05(1)$ & 		$52$ & $50826.92(1)$ & $0.57(1)$ \\ 
		& $50806.77(2)$ & $0.81(1)$ & 			 & $50827.24(2)$ & $0.37(1)$ \\ 
		& $50809.10(2)$ & $0.99(1)$ & 		$53$ & $50828.51(1)$ & $0.56(1)$ \\ 
		& $50809.52(3)$ & $0.83(1)$ &			 & $50828.77(3)$ & $0.39(2)$ \\ 
		$44$ & $50810.25(2)$ & $0.56(1)$ & 		$54$ & $50829.94(3)$ & $0.59(2)$ \\ 
		& $50811.30(2)$ & $0.17(1)$ & 			 & $50830.25(1)$ & $0.38(1)$ \\ 
		& $50812.13(7)$ & $0.84(3)$ & 		$55$ & $50831.40(2)$ & $0.55(2)$ \\ 
		& $50812.49(4)$ & $0.70(2)$ & 			 & $50831.56(2)$ & $0.43(1)$ \\ 
		$45$ & $50812.87(3)$ & $0.55(1)$ & 		$56^\ast$	 & $50832.85(3)$ & $0.45(2)$ \\ 
		& $50813.59(2)$ & $0.26(1)$ & 			 & $50833.67(5)$ & $0.80(4)$ \\ 
		& $50814.16(5)$ & $0.02(2)$ & 		$57^\ast$	 & $50834.15(4)$ & $0.41(3)$ \\ 
		& $50814.76(4)$ & $0.77(1)$ & 		$58^\ast$	 & $50835.24(2)$ & $0.48(2)$ \\ 
		$46$ & $50815.26(6)$ & $0.56(3)$ & 		$59$ & $50836.28(2)$ & $0.56(2)$ \\ 
		& $50815.71(6)$ & $0.37(2)$ & 		$60$ & $50837.35(2)$ & $0.56(2)$ \\ 
		& $50817.05(2)$ & $0.77(1)$ & 			 & $50837.90(7)$ & $0.03(7)$ \\ 
		$47$ & $50817.56(3)$ & $0.54(1)$ &		$61$ & $50838.37(4)$ & $0.56(4)$ \\ 
		& $50819.01(3)$ & $0.87(1)$ & 			 & $50838.71(3)$ & $0.21(4)$ \\ 
		$48$ & $50819.59(3)$ & $0.59(2)$ & 		$62$ & $50839.39(8)$ & $0.51(9)$ \\ 
		& $50820.59(2)$ & $0.10(1)$ & 			 & $50839.69(5)$ & $0.19(5)$ \\ 
		$49$ & $50821.66(2)$ & $0.55(1)$ & 		$63$ & $50840.30(8)$ & $0.52(9)$ \\ 
		& $50822.21(2)$ & $0.26(1)$ & 			 & $50840.58(2)$ & $0.20(3)$ \\ 
		$50$ & $50823.50(1)$ & $0.57(1)$ & 			 & $50841.42(3)$ & $0.24(3)$ \\ 
		\bottomrule 
		&&&$\infty$ & \multicolumn{2}{l}{ $50868.41(2)_{\text{stat.}}(11)_{\text{sys.}}$} \\ 
		\bottomrule
	\end{tabular}
	\label{tab:Rydbergstates}
\end{table}
\subsection{Field ionization and time-of-flight} \label{sec:Stark}
\begin{figure}
	\centering
	\includegraphics[width = \linewidth]{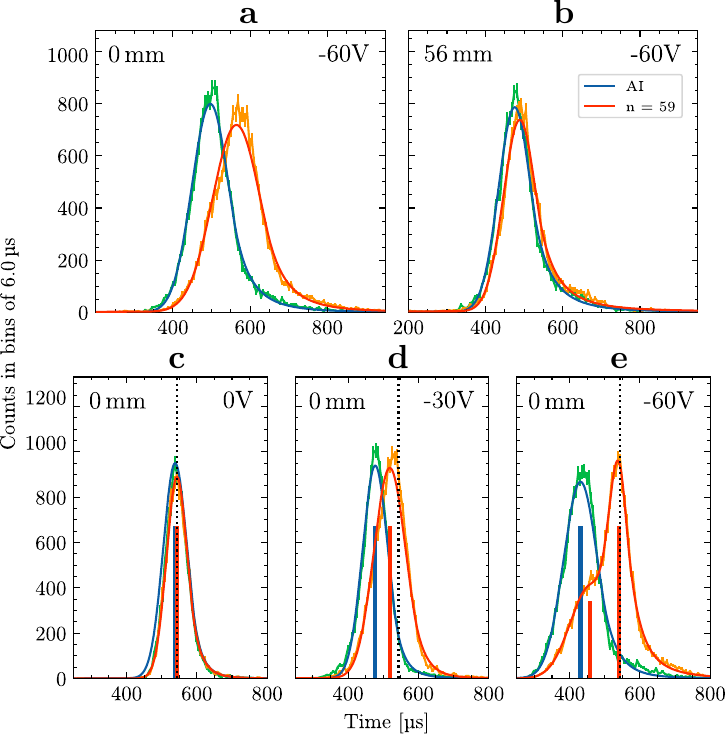}
	\caption{Top: Time profiles of the different photoions close (a) and far (b) from the nozzle exit plane without shielding grid in front of the S-RFQ. A voltage of $\SI{-60}{\volt}$ was applied on the first S-RFQ electrodes, revealing different time structures for the ions originating from the AI and from the $n = 59$ Rydberg state. Time is measured relative to the TDC trigger at $\SI{0}{\micro \second}$. Bottom: Time of flight when the shielding grid is installed. Three different voltages are applied on the grid. The dashed line indicates the arrival time of the ions when no voltage is applied. At a voltage of $\SI{-60}{\volt}$ (e), two components can be distinguished, the fast component originates from the one-color, two-step background and the slow components are the Rydberg atoms.}
	\label{fig:TOF_Rydberg}
\end{figure}

Rydberg states are especially sensitive to external electrical fields.
Without the shielding wire grid in front of the S-RFQ, the Rydberg states split due to the trapping RF field. The RFQ ion guide operates at a frequency of $\SI{1}{\mega \hertz}$ and a typical peak-to-peak amplitude of $\SI{340}{\volt}$. The splitting is observed as a broadening of the Rydberg states to a FWHM of about $\SI{0.5}{\per \cm}$. Ignoring the Stark splitting, a Rydberg state with effective quantum number $n^{\ast}$ is ionized by an electric field $F$ when \cite{Lethokhov}
\begin{equation}
	F \geq \frac{3.214 \times 10^8}{n^{\ast 4}} \si{\volt \per \cm}.
\end{equation}
The lowest identified Rydberg state has a $n^{\ast} = 43.44$ and would require at least $F = \SI{90}{\volt \per \cm}$. The grounded shielding grid ensures that electric fields of this strength are present only between it and the RFQ, localizing the field ionization to a $\SI{10}{\mm}$ region.
\\
\\
To fully conclude where and by which mechanism ionization occurs, time-of-flight measurements were performed in the gas jet. By applying a DC voltage on the shielding grid and on the RFQ electrodes, ions in the gas jet are subjected to an acceleration on top of their initial gas-jet velocity. The atoms, however, remain unaffected by the field. The arrival times of the ions originating from the most intense Rydberg state ($n =59$) and those from the found AI state at $\SI{50871.3}{\per \cm}$ were compared. Photoions from the AI state will result in the fastest possible travel time as they are ionized promptly ($< \si{\nano \second}$). By overlapping the two lasers beams in a well localized small region of the gas jet, one can then establish the ionization region and ionization mechanism of the Rydberg atoms. Using a time-to-digital converter (TDC) module (512 channels, \SI{4}{\nano \second} maximum resolution), two sets of time-of-flight measurements were performed with the lasers at a repetition rate of $\SI{500}{ \hertz}$.  The TDC was triggered into the falling edge of the TTL pulse firing the step-one laser.
\\
\\
In the first measurements, the shielding grid was removed and an offset DC bias was applied on the entrance electrodes of the S-RFQ. To ensure transport through the RFQ system, the voltage on the remaining segmented rod electrodes was increased so that the dragging field could be kept constant. The lasers were overlapped with the gas jet in two different positions along the gas jet axis. One close to the nozzle exit plane, which we assigned as the origin (i.e., $\SI{0}{\mm}$) and one near the end of the gas jet at a distance of $\SI{56}{\mm}$. When no voltage was applied the Rydberg and AI photoions were indistinguishable from each other. In the case of $\SI{-60}{\volt}$ bias, the time profiles separate (see Fig.~\ref{fig:TOF_Rydberg}) showing that the Rydberg atoms are not promptly ionized in the gas jet. The photoions originating from the ionization through the AI state are almost unaffected by the position of the ionization region along the jet as the voltage imparts a large acceleration on the ions. The slower and wider time structure exhibited by the Rydberg atoms at $\SI{0}{\mm}$ and $\SI{-60}{\volt}$, points to a more dispersed ionization region in the gas jet. 
\\
\\
In a second set of measurements, the shielding grid was reinstalled. An identical bias voltage was applied on the grid and on the entrance segments of the RFQ electrodes. The DC dragging fields were adjusted accordingly. The lasers were both again overlapped in a transverse geometry in the gas jet close to the nozzle exit plane. Three different measurements are shown in Fig.~\ref{fig:TOF_Rydberg}. The results reveal the existence of two time components contributing to the time structure when probing the Rydberg transition. A fast component seen at $\SI{-60}{\volt}$, nearly following the autoionized atoms, was identified as being related to one-color, two-step background. The slow component overlaps perfectly with the initial time structure at $\SI{0}{\volt}$ once it is resolved from the fast component. This confirms that most Rydberg atoms are ionized in a well-defined region, unaffected by the DC gradient, i.e., between the grid and S-RFQ. Hence, field ionization by the RFQ field is the main mechanism by which the Rydberg atoms are ionized in the gas jet. As the field-free travel times are around $\SI{120}{\micro\second}$, one can thus assume that, for lower $n$, a significant fraction of the Rydberg states are de-exciting during their hypersonic travel. To use field ionization of Rydberg states as an efficient ionization mechanism in the gas jet, it might be preferable to remove the shielding of the S-RFQ or to introduce an extra field. These avenues are currently being explored. 
\subsection{Dark states} \label{sec:gascell_quenching}
\begin{figure}
		\centering
		\includegraphics[width = 0.5 \textwidth]{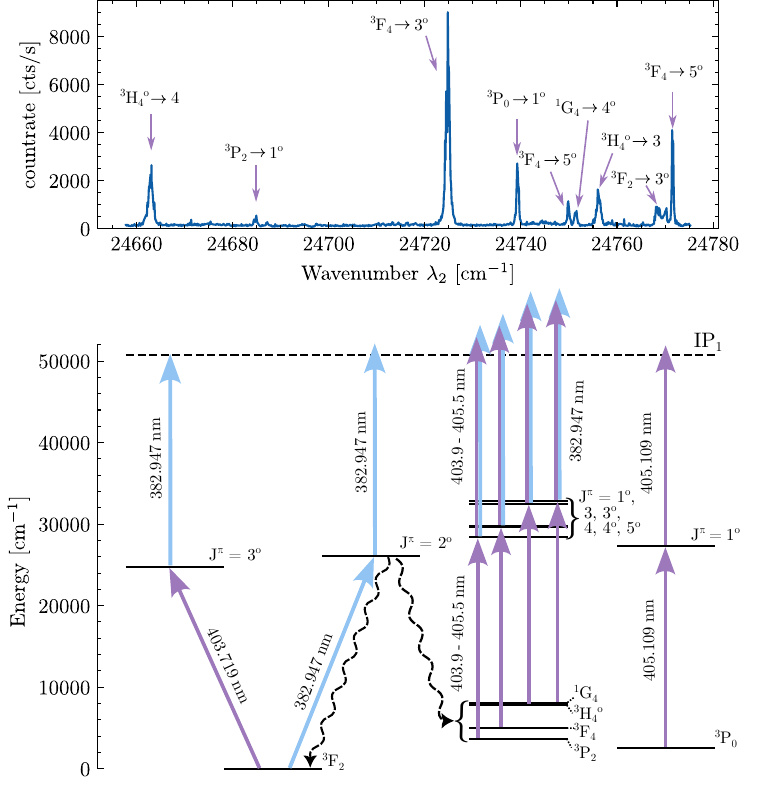}
		\caption{Four different identified processes in $\SI{100}{\per \cm}$. From left to right on the scheme: A reversal of the order of the laser steps, a one-color, two-step transition coming from the ground state, gas-induced quenching (denoted by the wiggly lines) from the excited state to lower lying states that are ionized by a two-step transition and part of the population residing in an excited state that gets ionized via a one-color, two-step process. Identification of these states is purely based on the exact wavelengths of the transitions and on the laser steps that were involved.}
		\label{fig:quenching_levelscheme}
\end{figure}

The appearance of many transitions below the IP\textsubscript{1} when scanning in the gas cell greatly complicated the spectrum making it impossible to identify any Rydberg series. To better understand the underlying origin, the population of the metastable states was investigated. By overlapping the laser steps in the gas cell through a 5 mm entrance window, each atom interacts with successive laser pulses about five times due to the low velocity of the gas. By delaying the first step in time such that it arrived $\SI{150}{\nano \second}$ after the second step, short-lived states were allowed to decay revealing any possible dark states. Due to the repetition rate of $\SI{7}{\kilo \hertz}$, the decay time was limited to about  $\sim \SI{140}{\micro \second}$.
\\
\\
The resulting spectrum and associated level scheme are shown in Fig.~\ref{fig:quenching_levelscheme}. The power of the lasers was set to $\SI{0.5}{\milli \watt}$ for the first step and $\SI{260}{\milli \watt}$ for the second step. The scheme was composed using known transitions from spectrometer measurements \cite{Redman2014,Zalubas1976} and with help from the NIST Database \cite{NIST_ASD_ThI}. An interesting observation is the one-color, two-step transition (\ch{^3P_0 -> '1' ^o}) that is observed starting from the excited \ch{^3P_0} state. This is the lowest known excited state in thorium at an energy of $\SI{2558.06}{\per \cm}$. The observation of this transition (without the first step) indicates that a sizable population is present in this state, probably populated after the ablation process. It is reasonable to assume that the thorium recoils could also end up in similar states. The high density of levels in a three or four valence electron system can significantly affect the laser ionization efficiency in gas cells. This is for example seen in Fig.~\ref{fig:spectrum_gascellvsgasjetshield_long}, where the intensity of the AI state is significantly reduced in the gas cell. Complicated spectra originating from many populated lower states was also encountered in spectroscopic studies of protactinium \cite{Naubereit2018}.
\subsection{Narrowband spectroscopy of \ch{^{229}Th}}
\begin{figure} \label{fig:narrowbandspectrum229Thneutral}
	\includegraphics[width=\linewidth]{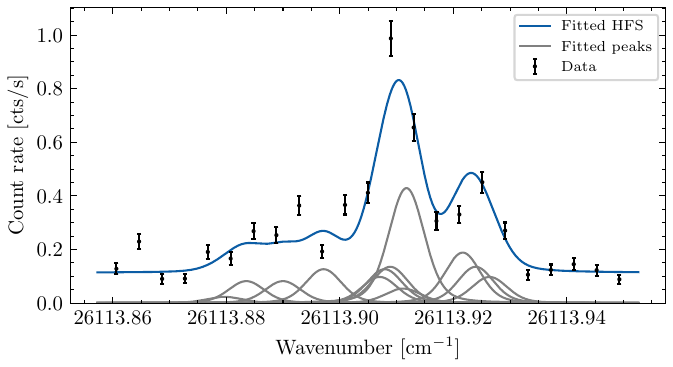}
	\caption{Hyperfine spectrum of neutral \ch{^{229}Th} measured in the gas jet. The HFS consists of $13$ individual peaks. Because of the low statistics, the Racah intensities as well as the $A$ and $B$ HFS parameters were kept fixed. The literature values of $A$ and $B$ are found in \cite{Sonnenschein2012}.}
\end{figure}

After the unsuccessful attempt at laser spectroscopy of \ch{^{229}Th^+}, a spectrum on neutral \ch{^{229}Th} (Fig.\ref{fig:narrowbandspectrum229Thneutral}) was taken with a narrowband laser system in the gas jet using the \ch{^{233}U} recoil sources. Although the gas cell is optimized for the fast extraction of ionic species, a low signal of laser-ionized \ch{^{229}Th} could be observed with a rate of about $\SI{1}{count/\second}$, well above the dark count noise of the CEM employed for ion detection. This was achieved by blocking the ions from the alpha decay using the collector electrodes in the gas cell. The \ch{$6d^27s^2$ ^3F_{2} -> }(J$^\pi = 2^\text{o}$) ground-state transition at $\SI{382.947}{\nm}$ was used as the excitation step and the $\SI{403.909}{\nm}$ transition to an AI state for ionization (see scheme in Fig.~\ref{fig:spectrum_gascellvsgasjetshield_long}). Both laser steps were overlapped in a transverse geometry to the gas jet using a beamsplitter. To increase the duty cycle of the laser steps, a repetition rate of $\SI{10}{\kilo \hertz}$ was used together with two prisms. These prisms reflect the lasers up to 8 times back into the gas jet (see Fig.~\ref{fig:iglislab}).
\\
\\
Each point was measured for $\SI{300}{\second}$ with a laser power of $\SI{7}{\milli\watt}$ in the first step and $\SI{1.4}{\watt}$ in the second step.  A hyperfine spectrum consisting of Voigt peaks and fixed hyperfine constants was fitted to the narrowband spectrum. Only the centroid positions and Gaussian contribution to the FWHM were left as free variables. The hyperfine parameters were taken from \cite{Sonnenschein2012}. A FWHM of $\SI{238 \pm 30}{ \mega \hertz}$ could be obtained, corresponding to a jet temperature $T \sim \SI{7}{\kelvin}$ and demonstrating the foreseen performance of the gas jet environment as characterized in \cite{Lantis2024}.  This resolution is sufficient to discern the isomeric state from the ground state  in \ch{^{229}Th^+}\cite{thesisVerlinde}. The centroid was determined to be $\SI{26113.910 \pm 0.003}{\per \cm}$.

\section{Atomic structure calculations}
\subsection{$LS$ composition of the lowest levels in \ch{Th}}
A first set of atomic structure calculations was carried out in order to classify the lowest energy levels of \ch{Th} paying particular attention to the odd-parity level at 26113.27 cm$^{-1}$ ($J$ = 2) which is the intermediate level used in the two-step scheme from the 6d$^2$7s$^2$ $^3$F$_2$ ground state chosen in our laser spectroscopy study of neutral thorium described in Sec.~\ref{sec:laserspecneutralTh}. The pseudo-relativistic Hartree-Fock (HFR) method originally developed by \cite{Cowan1981} and modified to include core-polarization effects, giving rise to the so-called HFR+CPOL approach, as described by \cite{Quinet1999,Quinet2002} was used in the present work. The 6d$^2$7s$^2$, 6d$^3$7s, 6d$^4$, 5f6d$^2$7p even configurations and the 5f6d7s$^2$, 6d$^2$7s7p, 6d$^3$7p, 6d7s$^2$7p, 5f6d$^2$7s odd configurations were explicitly included in the physical model while the core-polarization corrections were estimated using a dipole polarizability, $\alpha$$_d$, equal to 10.26 a$_0^3$ (taken from \cite{Fraga1976} for a \ch{Th^{4+}} ionic core) and a cut-off radius, $r_c$, equal to 1.88 a$_0$ (which corresponds to the mean value of $r$ for the outermost core orbital (6p), as obtained from our HFR calculations). A least-squares adjustment of the radial parameters was then performed in order to minimize the differences between the HFR+CPOL energy levels and the available experimental values in both parities taken from the NIST compilation \cite{NIST_ASD_ThI}. In a first step, only the levels below 20000 cm$^{-1}$ were considered in the fitting process in each of the parities because, beyond this limit, it was extremely complicated to make a trustworthy correspondence between calculated and experimental levels.\\
 
As mentioned above, the odd level at 26113.27 cm$^{-1}$ ($J$ = 2) used as an intermediate state in the two-step scheme was particularly difficult to find in the HFR+CPOL calculations because it is located in a region where many levels have extremely large mixtures, so that different computed levels were likely to correspond to this specific state. In order to identify the corresponding level in the HFR+CPOL calculations, we looked for the one that gave both a Land\'e factor in good agreement with the experimental value taken from NIST \cite{NIST_ASD_ThI}, i.e. $g$ = 0.980, and whose calculated transition probabilities best reproduced the intensities observed for the main lines depopulating the level at 26113.27 cm$^{-1}$, also taken from the NIST database. Among the nine levels appearing in the energy range between 24000 and 28000 cm$^{-1}$ in the $J$ = 2 Hamiltonian matrix of odd parity, a final fitted level appearing at 26283 cm$^{-1}$ could then be identified in calculations as corresponding to the experimental level at 26113.27 cm$^{-1}$ when the latter was included in the fit in a second step. Indeed, only this theoretical level gave a calculated Land\'e factor ($g$ = 0.944) in good agreement with the experimental value and transition probabilities ($gA$) that could qualitatively reproduce the measured intensities ($I$) for the lines depopulating the level at 26113.27 cm$^{-1}$ towards lower even-parity levels at 0 cm$^{-1}$ ($gA$ = 7.38 $\times$ 10$^7$ s$^{-1}$, $I$ = 14000), at 3687.99 cm$^{-1}$ ($gA$ = 4.92 $\times$ 10$^7$ s$^{-1}$, $I$ = 4000) and at 6362.40 cm$^{-1}$ ($gA$ = 1.57 $\times$ 10$^7$ s$^{-1}$, $I$ = 600). Finally, this led us to find that the level of interest at 26113.27 cm$^{-1}$ is in fact extremely mixed since, according to our calculations, the first three main $LS$-coupling components are 8.8\% 6d$^2$7s7p ($^1$D)$^3$F + 8.1\% 6d$^3$7p ($^4$F)$^3$F + 7.9 \% 5f6d$^2$7s ($^1$D)$^3$D. More broadly, it is worth pointing out that, for this specific level, the (slightly) dominant configuration seems to be 6d$^2$7s7p, with the HFR+CPOL calculations actually giving 35.8\% 6d$^2$7s7p + 31.8\% 5f6d$^2$7s + 24.2\% 6d$^3$7p + 5.1\% 6d7s$^2$7p + 3.1\% 5f6d7s$^2$ when summing all the basis state contributions for each electronic configuration. This designation is also justified by the fact that the most intense transition depopulating the level at 26113.27 cm$^{-1}$ is, according to \cite{NIST_ASD_ThI}, the transition to the 6d$^2$7s$^2$ $^3$F$_2$ ground level, which our calculations confirm since the HFR+CPOL electric dipole matrix element $\langle$7s$||$$r$$||$7p$\rangle$ between the 6d$^2$7s$^2$ and 6d$^2$7s7p configurations is the largest one. For the levels below 20000 cm$^{-1}$, the mean deviations were found to be equal to 105 and 142 cm$^{-1}$ for the even and odd parities, respectively. The final fitted parameters are given in Table~\ref{parameters_ThI} while the calculated energy levels are compared to experimental data in Table~\ref{energylevels_ThI}. In the latter table, the $LS$-coupling compositions are also listed showing very strong mixtures in many cases.\\
 \begin{table}[h!]
\centering
\caption{\label{parameters_ThI} Final fitted radial parameters (in cm$^{-1}$) in \ch{Th} and their corresponding HFR+CPOL values. All the remaining Slater and spin-orbit parameters have been respectively fixed to 70\% and 100\% of their HFR+CPOL values.}
\begin{ruledtabular}
\begin{tabular}{ccccc}
Configuration & Parameter   & HFR & Fit     & Fit/HFR \\[1mm] \hline
\textbf{Even Parity} & & & & \\
6d$^2$7s$^2$ & E$_{av}$                               & 8826.9   & 7366.4   & 0.83 \\
       & F$^2$(6d,6d)       & 34346.3  & 20380.4  & 0.59 \\
       & F$^4$(6d,6d)     & 22698.6  & 13755.8  & 0.61 \\
       & $\alpha$                            & 0.0      & 0.6      &        \\
       & $\beta$                              & 0.0      & 137.1     &        \\
       & $\zeta$(6d)                         & 1672.7   & 1319.2   & 0.79 \\
6d$^3$7s  & E$_{av}$                             & 17782.2  & 18440.5  & 1.04 \\
       & F$^2$(6d,6d)        & 32583.0  & 21167.1  & 0.65 \\
       & F$^4$(6d,6d)       & 21398.6  & 14470.1  & 0.68 \\
       & $\alpha$                             & 0.0      & 5.3        &        \\
       & $\beta$                              & 0.0      & -993.0    &        \\
       & $\zeta$(6d)                          & 1528.9   & 1316.2   & 0.86 \\
       & G$^2$(6d,7s)       & 18890.3  & 12586.1  & 0.67 \\
       &                                   &          &          &        \\
       & R$^2$(6d,7s;6d,6d) & -22709.4 & -13056.4 & 0.57 \\
& & & & \\
\textbf{Odd Parity} & & & & \\
5f6d7s$^2$          & E$_{av}$                & 20477.1  & 16202.8  & 0.79  \\
                   & $\zeta$ (5f)          & 1276.1   & 987.3    & 0.77  \\
                   & $\zeta$ (6d)          & 1376.7   & 1155.4   & 0.84  \\
                   & F$^2$(5f,6d)       & 26379.3  & 21254.9  & 0.81  \\
                   & F$^4$(5f,6d)       & 14609.7  & 20438.9  & 1.40  \\
                   & G$^1$(5f,6d)       & 19521.7  & 13884.8  & 0.71  \\
                   & G$^3$(5f,6d)       & 13462.9  & 13432.1  & 1.00  \\
                   & G$^5$(5f,6d)       & 9833.1   & 6556.9   & 0.67  \\
6d$^2$7s7p            & E$_{av}$                & 29067.4  & 33661.0  & 1.16  \\
                   & F$^2$(6d,6d)       & 35501.9  & 29172.7  & 0.82  \\
                   & F$^4$(6d,6d)       & 23559.7  & 12837.9  & 0.55  \\
                   & $\alpha$              & 0        & 584.0    &        \\
                   & $\beta$               & 0        & -816.5   &        \\
                   & $\zeta$ (6d)          & 1770.4   & 1435.6   & 0.81  \\
                   & $\zeta$ (7p)          & 2677.8   & 3367.4   & 1.26  \\
                   & F$^2$(6d,7p)       & 14586.0  & 9289.6   & 0.64  \\
                   & G$^2$(6d,7s)       & 19142.1  & 11009.6  & 0.58  \\
                   & G$^1$(6d,7p)       & 8712.7   & 7835.7   & 0.90  \\
                   & G$^3$(6d,7p)       & 6321.7   & 6834.2   & 1.08  \\
                   & G$^1$(7s,7p)       & 21370.0  & 17073.8  & 0.80  \\
6d$^3$7p              & E$_{av}$                & 38534.6  & 38346.4  & 1.00 \\
6d7s$^2$7p            & E$_{av}$                & 23127.4  & 23833.7  & 1.03  \\
5f6d$^2$7s                    & E$_{av}$                & 31643.7  & 30962.7  & 0.99  \\
                   & F$^2$(6d,6d)       & 29656.7  & 31959.1  & 1.08  \\
                   & F$^4$(6d,6d)       & 19319.4  & 13959.7  & 0.72  \\
                   & $\alpha$             & 0        & 526.7    &        \\
                   & $\beta$               & 0        & 842.8    &        \\
                   & $\zeta$ (5f)          & 1243.1   & 999.0    & 0.80  \\
                   & $\zeta$ (6d)          & 1235.5   & 881.4    & 0.71  \\
                   & F$^2$(5f,6d)       & 24815.4  & 21334.1  & 0.86  \\
                   & F$^4$(5f,6d)       & 13681.7  & 22014.6  & 1.61  \\
\end{tabular}
\end{ruledtabular}
\end{table}
\addtocounter{table}{-1}
\begin{table}[h!]
\centering
\caption{Continued.}
\begin{ruledtabular}
\begin{tabular}{ccccc}
Configuration & Parameter   & HFR & Fit     & Fit/HFR \\[1mm] \hline
5f6d$^2$7s         & G$^1$(5f,6d)       & 18854.4  & 13325.5  & 0.71  \\
                   & G$^3$(5f,6d)       & 12780.9  & 12011.8  & 0.94  \\
                   & G$^5$(5f,6d)       & 9273.4   & 12518.5  & 1.35  \\
                   & G$^3$(5f,7s)       & 4374.3   & 6768.5   & 1.55  \\
                   & G$^2$(6d,7s)       & 19055.9  & 12662.6  & 0.66 \\ 
                   & & & & \\
                   & R$^1$(5f,7s;5f,7p) & -14537.0 & -10068.4 & 0.69  \\
                   & R$^2$(5f,7s;5f,7p) & -3553.0  & -3834.6  & 1.08  \\
                   & R$^2$(5f,7s;5f,6d) & -10488.1 & -5795.8  & 0.55  \\
                   & R$^3$(5f,7s;5f,6d) & -3387.0  & -4198.4  & 1.24  \\
                   & R$^2$(6d,7s;6d,6d) & -22174.1 & -13684.5 & 0.62  \\
                   & R$^1$(6d,7p;5f,6d) & 10243.6  & 6493.4   & 0.63  \\
                   & R$^3$(6d,7p;5f,6d) & 4594.3   & 2912.4   & 0.63  \\
                   & R$^2$(6d,7p;5f,6d) & 2474.7   & 1568.6   & 0.63  \\
                   & R$^4$(6d,7p;5f,6d) & 1937.7   & 1228.6   & 0.63  \\
\end{tabular}
\end{ruledtabular}
\end{table}

\begin{table*}[h!]
\centering
\caption{\label{energylevels_ThI} Lowest energy levels of \ch{Th} with their respective $LS$ compositions (only the first three components are given), as computed in the present work using the semi-empirical HFR+CPOL approach. The experimental values are taken from the NIST database \cite{NIST_ASD_ThI}. All energies are given in cm$^{-1}$.}
\resizebox{16.5cm}{!}{
\begin{ruledtabular}
\begin{tabular}{cccccccccc}
E$_{exp}$  & E$_{cal}$ & $\Delta E$ &  J & \multicolumn{6}{c}{LS composition}   \\ \hline
\textbf{Even Parity} & & & & & & & & & \\

0.00     & 0     & 0 & 2 & 81.8$\%$ & 6d$^2$7s$^2$ ($^3$F) $^3$F                   & 11.1$\%$ & 6d$^2$7s$^2$ ($^1$D) $^1$D & 2.1$\%$  & 6d$^3$7s  ($^2$F) $^3$F    \\
2558.06  & 2439  & 119  & 0 & 68.8$\%$ & 6d$^2$7s$^2$ ($^3$P) $^3$P & 18.9$\%$ & 6d$^3$7s    ($^2$P) $^3$P & 5.0$\%$    & 6d$^4$      ($^3$P) $^3$P  \\
2869.26  & 2775  & 94   & 3 & 93.4$\%$ & 6d$^2$7s$^2$ ($^3$F) $^3$F & 3.4$\%$  & 6d$^3$7s  ($^2$F) $^3$F    & 2.7$\%$  & 6d$^4$    ($^3$F) $^3$F    \\
3687.99  & 3741  & -53  & 2 & 45.0$\%$   & 6d$^2$7s$^2$ ($^3$P) $^3$P                   & 18.7$\%$ & 6d$^2$7s$^2$ ($^1$D) $^1$D & 15.8$\%$ & 6d$^3$7s  ($^2$P) $^3$P    \\
3865.47  & 3666  & 199  & 1 & 67.9$\%$ & 6d$^2$7s$^2$ ($^3$P) $^3$P                   & 22.7$\%$ & 6d$^3$7s  ($^2$P) $^3$P    & 4.5$\%$  & 6d$^4$    ($^3$P) $^3$P    \\
4961.66  & 4794  & 168  & 4 & 80.5$\%$ & 6d$^2$7s$^2$ ($^3$F) $^3$F                   & 11.2$\%$ & 6d$^2$7s$^2$ ($^1$G) $^1$G & 3.9$\%$  & 6d$^3$7s  ($^2$F) $^3$F    \\
5563.14  & 5448  & 115  & 1 & 88.6$\%$ & 6d$^3$7s  ($^4$F) $^5$F                      & 4.8$\%$  & 6d$^3$7s  ($^2$D) $^3$D    & 2.4$\%$  & 6d$^3$7s  ($^2$D) $^3$D    \\
6362.40  & 6299  & 63   & 2 & 89.8$\%$ & 6d$^3$7s  ($^4$F) $^5$F                      & 3.7$\%$  & 6d$^3$7s  ($^2$D) $^3$D    & 1.5$\%$  & 6d$^3$7s  ($^2$D) $^3$D    \\
7280.12  & 7452  & -172 & 2 & 51.2$\%$ & 6d$^2$7s$^2$ ($^1$D) $^1$D                   & 19.0$\%$   & 6d$^2$7s$^2$ ($^3$P) $^3$P & 9.3$\%$  & 6d$^3$7s  ($^2$P) $^3$P    \\
7502.29  & 7502  & 0    & 3 & 93.4$\%$ & 6d$^3$7s  ($^4$F) $^5$F                      & 1.8$\%$  & 6d$^3$7s  ($^2$D) $^3$D    & 0.9$\%$  & 6d$^3$7s  ($^2$G) $^3$G    \\
8111.00  & 8146  & -35  & 4 & 66.4$\%$ & 6d$^2$7s$^2$ ($^1$G) $^1$G                   & 11.5$\%$ & 6d$^3$7s  ($^2$G) $^1$G    & 8.7$\%$  & 6d$^2$7s$^2$ ($^3$F) $^3$F \\
8800.25  & 8847  & -46  & 4 & 84.1$\%$ & 6d$^3$7s  ($^4$F) $^5$F                      & 5.2$\%$  & 6d$^2$7s$^2$ ($^1$G) $^1$G & 5.0$\%$    & 6d$^3$7s  ($^2$G) $^3$G    \\
9804.81  & 9921  & -116 & 5 & 79.9$\%$ & 6d$^3$7s  ($^4$F) $^5$F                      & 15.9$\%$ & 6d$^3$7s  ($^2$G) $^3$G    & 1.2$\%$  & 6d$^3$7s  ($^2$H) $^3$H    \\
11601.03 & 11503 & 98   & 1 & 89.3$\%$ & 6d$^3$7s  ($^4$P) $^5$P                      & 4.8$\%$  & 6d$^2$7s$^2$ ($^3$P) $^3$P & 2.5$\%$  & 6d$^3$7s  ($^2$P) $^3$P    \\
11802.93 & 11798 & 5    & 2 & 75.3$\%$ & 6d$^3$7s  ($^4$P) $^5$P                      & 11.0$\%$   & 6d$^2$7s$^2$ ($^3$P) $^3$P & 3.4$\%$  & 6d$^3$7s  ($^2$P) $^3$P    \\
12847.97 & 12822 & 26   & 3 & 73.2$\%$ & 6d$^3$7s  ($^2$G) $^3$G                      & 7.5$\%$  & 6d$^3$7s  ($^4$P) $^5$P    & 7.3$\%$  & 6d$^3$7s  ($^4$F) $^3$F    \\
13088.56 & 13131 & -42  & 3 & 85.4$\%$ & 6d$^3$7s  ($^4$P) $^5$P                      & 7.8$\%$  & 6d$^3$7s  ($^2$G) $^3$G    & 2.5$\%$  & 6d$^3$7s  ($^2$D) $^3$D    \\
13297.43 & 13193 & 104  & 4 & 55.0$\%$   & 6d$^3$7s  ($^2$G) $^3$G                      & 30.5$\%$ & 6d$^3$7s  ($^2$H) $^3$H    & 4.9$\%$  & 6d$^3$7s  ($^4$F) $^5$F    \\
13847.77 & 13946 & -98  & 2 & 54.3$\%$ & 6d$^3$7s  ($^4$F) $^3$F                      & 9.7$\%$  & 6d$^3$7s  ($^2$D) $^3$D    & 8.2$\%$  & 6d$^3$7s  ($^4$P) $^5$P    \\
13962.52 & 14231 & -268 & 1 & 36.5$\%$ & 6d$^3$7s  ($^2$D) $^3$D                      & 13.3$\%$ & 6d$^3$7s  ($^4$P) $^3$P    & 11.9$\%$ & 6d$^3$7s  ($^2$P) $^1$P    \\
14204.26 & 14244 & -39  & 5 & 42.7$\%$ & 6d$^3$7s  ($^2$H) $^3$H                      &35.7$\%$ & 6d$^3$7s  ($^2$G) $^3$G     &  13.6$\%$ & 6d$^3$7s  ($^4$F) $^5$F    \\
14226.82 & 14250  & -23  & 0 & 24.5$\%$ & 6d$^3$7s  ($^2$P) $^3$P    & 24.4$\%$ & 6d$^3$7s  ($^4$P) $^3$P   & 17.8$\%$ & 6d$^2$7s$^2$ ($^1$S) $^1$S \\
15493.22 & 15412 & 81   & 4 & 52.3$\%$ & 6d$^3$7s  ($^2$H) $^3$H                      & 24.0$\%$   & 6d$^3$7s  ($^2$G) $^3$G    & 8.0$\%$    & 6d$^3$7s  ($^2$G) $^1$G    \\
15863.89 & 16043 & -179 & 2 & 47.1$\%$ & 6d$^3$7s  ($^2$D) $^3$D                      & 18.0$\%$   & 6d$^3$7s  ($^4$F) $^3$F    & 6.7$\%$  & 6d$^3$7s  ($^2$D) $^3$D    \\
15970.09 & 16050 & -80  & 3 & 38.2$\%$ & 6d$^3$7s  ($^2$D) $^3$D & 35.3$\%$ & 6d$^3$7s  ($^4$F) $^3$F                         & 7.0$\%$    & 6d$^3$7s  ($^2$G) $^3$G    \\
16351.94 & 16484 & -132 & 0 & 57.5$\%$ & 6d$^2$7s$^2$ ($^1$S) $^1$S      			& 11.1$\%$ & 6d$^3$7s  ($^4$P) $^3$P   & 7.0$\%$    & 6d$^4$   ($^3$P) $^3$P \\ 
16554.24 & 16614 & -59  & 6 & 94.8$\%$ & 6d$^3$7s  ($^2$H) $^3$H                      & 2.8$\%$  & 6d$^4$    ($^3$H) $^3$H    & 0.8$\%$  & 5f6d$^2$7p ($^3$F) $^3$H  \\
17073.81 & 16861 & 213  & 1 & 32.3$\%$ & 6d$^3$7s  ($^4$P) $^3$P                      & 22.7$\%$ & 6d$^3$7s  ($^2$D) $^3$D    & 15.5$\%$ & 6d$^4$    ($^3$P) $^3$P    \\
17166.10 & 17193 & -27  & 5 & 49.9$\%$ & 6d$^3$7s  ($^2$H) $^3$H                      & 37.5$\%$ & 6d$^3$7s  ($^2$G) $^3$G    & 4.1$\%$  & 6d$^3$7s  ($^4$F) $^5$F    \\
17398.40 & 17384 & 15   & 3 & 41.7$\%$ & 6d$^3$7s  ($^2$D) $^3$D                      & 32.9$\%$ & 6d$^3$7s  ($^4$F) $^3$F    & 5.7$\%$  & 6d$^4$    ($^3$F) $^3$F    \\
17959.89 & 17874 & 86   & 4 & 58.0$\%$   & 6d$^3$7s  ($^4$F) $^3$F                      & 16.1$\%$ & 6d$^3$7s  ($^2$G) $^1$G    & 9.3$\%$  & 6d$^4$    ($^3$F) $^3$F    \\
18549.40 & 18440 & 109  & 2 & 59.9$\%$ & 6d$^3$7s  ($^4$P) $^3$P                      & 18.3$\%$ & 6d$^4$    ($^3$P) $^3$P    & 6.6$\%$  & 6d$^3$7s  ($^2$F) $^3$F    \\
18574.61 & 18689 & -114 & 1 & 47.3$\%$ & 6d$^3$7s  ($^2$P) $^3$P                      & 23.5$\%$ & 6d$^3$7s  ($^2$P) $^1$P    & 13.0$\%$   & 6d$^2$7s$^2$ ($^3$P) $^3$P \\
19273.28 & 18913 & 360  & 2 & 26.5$\%$ & 6d$^3$7s  ($^2$D) $^1$D                      & 23.9$\%$ & 6d$^3$7s  ($^2$P) $^3$P    & 8.3$\%$  & 6d$^3$7s  ($^2$D) $^3$D    \\
19532.42 & 19361 & 171  & 4 & 56.3$\%$ & 6d$^3$7s  ($^2$F) $^3$F                      & 10.5$\%$ & 6d$^3$7s  ($^2$G) $^1$G    & 7.4$\%$  & 6d$^4$    ($^3$F) $^3$F    \\
19713.03 & 19991 & -278 & 3 & 70.8$\%$ & 6d$^3$7s  ($^2$F) $^3$F                      & 11.8$\%$ & 6d$^3$7s  ($^4$F) $^3$F    & 3.7$\%$  & 6d$^3$7s  ($^2$D) $^3$D    \\
\end{tabular}
\end{ruledtabular}}
\end{table*}
\addtocounter{table}{-1}
\begin{table*}[h!]
\centering
\caption{Continued.}
\resizebox{16.5cm}{!}{
\begin{ruledtabular}
\begin{tabular}{cccccccccc}
E$_{exp}$  & E$_{cal}$ & $\Delta E$ &  J & \multicolumn{6}{c}{LS composition}   \\ \hline
\textbf{Odd Parity} & & & & & & & & & \\
7795.27  & 7832  & -37  & 4 & 63.2$\%$ & 5f6d7s$^2$ ($^2$F) $^3$H & 28.8$\%$ & 5f6d7s$^2$ ($^2$F) $^1$G  & 3.6$\%$  & 5f6d$^2$7s ($^1$D) $^3$H \\
8243.60  & 8245  & -1   & 2 & 52.3$\%$ & 5f6d7s$^2$ ($^2$F) $^3$F & 28.0$\%$   & 5f6d7s$^2$ ($^2$F) $^1$D  & 8.1$\%$  & 6d7s$^2$7p ($^2$D) $^3$F \\
10414.14 & 10346 & 68   & 4 & 52.6$\%$ & 5f6d7s$^2$ ($^2$F) $^1$G & 29.0$\%$   & 5f6d7s$^2$ ($^2$F) $^3$H  & 11.0$\%$   & 5f6d7s$^2$ ($^2$F) $^3$F \\
10526.54 & 10508 & 19   & 3 & 70.7$\%$ & 5f6d7s$^2$ ($^2$F) $^3$G & 9.4$\%$  & 5f6d7s$^2$ ($^2$F) $^3$F  & 7.7$\%$  & 5f6d7s$^2$ ($^2$F) $^1$F \\
10783.15 & 10623 & 160  & 2 & 34.6$\%$ & 6d7s$^2$7p ($^2$D) $^3$F & 13.1$\%$ & 6d7s$^2$7p ($^2$D) $^1$D  & 11.8$\%$ & 6d$^2$7s7p ($^1$D) $^3$F \\
11197.03 & 11210 & -13  & 5 & 91.4$\%$ & 5f6d7s$^2$ ($^2$F) $^3$H & 6.7$\%$  & 5f6d$^2$7s ($^1$D) $^3$H  & 1.0$\%$    & 5f6d$^2$7s ($^3$F) $^3$H \\
11241.73 & 11389 & -147 & 3 & 76.7$\%$ & 5f6d7s$^2$ ($^2$F) $^3$F & 9.2$\%$  & 5f6d7s$^2$ ($^2$F) $^3$G  & 5.8$\%$  & 6d7s$^2$7p ($^2$D) $^3$F \\
11877.84 & 11841 & 36   & 1 & 39.4$\%$ & 5f6d7s$^2$ ($^2$F) $^3$D & 25.9$\%$ & 6d7s$^2$7p ($^2$D) $^3$D  & 5.3$\%$  & 6d7s$^2$7p ($^2$D) $^1$P \\
12114.37 & 12124 & -10  & 2 & 36.3$\%$ & 5f6d7s$^2$ ($^2$F) $^1$D & 32.1$\%$ & 5f6d7s$^2$ ($^2$F) $^3$F  & 9.1$\%$  & 5f6d7s$^2$ ($^2$F) $^3$D \\
13175.11 & 13101 & 74   & 4 & 79.0$\%$   & 5f6d7s$^2$ ($^2$F) $^3$G & 9.1$\%$  & 5f6d7s$^2$ ($^2$F) $^3$F  & 5.1$\%$  & 6d$^2$7s7p ($^3$F) $^3$G \\
13945.31 & 14134 & -189 & 3 & 26.0$\%$   & 5f6d7s$^2$ ($^2$F) $^3$D & 13.7$\%$ & 6d7s$^2$7p ($^2$D) $^3$F  & 10.8$\%$ & 5f6d7s$^2$ ($^2$F) $^1$F \\
14032.08 & 14042 & -10  & 2 & 39.1$\%$ & 5f6d7s$^2$ ($^2$F) $^3$D & 11.7$\%$ & 6d7s$^2$7p ($^2$D) $^3$D  & 10.8$\%$ & 6d7s$^2$7p ($^2$D) $^3$P \\
14206.92 & 14165 & 42   & 4 & 66.5$\%$ & 5f6d7s$^2$ ($^2$F) $^3$F & 14.3$\%$ & 5f6d7s$^2$ ($^2$F) $^1$G  & 6.6$\%$  & 5f6d7s$^2$ ($^2$F) $^3$G \\
14243.99 & 14167 & 77   & 1 & 25.0$\%$   & 5f6d7s$^2$ ($^2$F) $^3$D  & 22.5$\%$ & 6d7s$^2$7p ($^2$D) $^3$P &  21.0$\%$   & 5f6d7s$^2$ ($^2$F) $^3$P \\
14247.31 & 14407 & -160 & 0 & 38.9$\%$ & 5f6d7s$^2$ ($^2$F) $^3$P & 30.3$\%$ & 6d7s$^2$7p ($^2$D) $^3$P  & 12.6$\%$ & 5f6d$^2$7s ($^1$D) $^3$P \\
14465.22 & 14907 & -442 & 2 & 14.6$\%$ & 5f6d7s$^2$ ($^2$F) $^3$D  & 13.1$\%$ & 5f6d7s$^2$ ($^2$F) $^1$D &  8.5$\%$ & 6d7s$^2$7p ($^2$D) $^3$P \\
14481.87 & 14306 & 176  & 6 & 90.5$\%$ & 5f6d7s$^2$ ($^2$F) $^3$H & 7.6$\%$  & 5f6d$^2$7s ($^1$D) $^3$H  & 1.2$\%$  & 5f6d$^2$7s ($^3$F) $^3$H \\
15166.90 & 15062 & 105 & 3 & 24.2$\%$ & 6d7s$^2$7p ($^3$D) $^3$F  & 23.7$\%$ & 5f6d7s$^2$ ($^2$F) $^3$D &  12.5$\%$ & 6d$^2$7s7p ($^1$D) $^3$F \\
15490.08 & 15476 & 14   & 5 & 84.2$\%$ & 5f6d7s$^2$ ($^2$F) $^3$G & 6.1$\%$  & 6d$^2$7s7p ($^3$F) $^3$G  & 4.0$\%$    & 5f6d$^2$7s ($^1$D) $^3$G \\
15618.98 & 15882 & -263 & 3 & 71.1$\%$ & 5f6d$^2$7s ($^3$F) $^5$H  & 19.8$\%$ & 5f6d$^2$7s ($^3$F) $^3$G &  3.0$\%$ & 5f6d$^2$7s ($^1$D) $^3$G \\
15736.97 & 15481 & 256  & 1 & 38.4$\%$ & 6d7s$^2$7p ($^2$D) $^3$D & 16.2$\%$ & 5f6d7s$^2$ ($^2$F) $^3$D  & 9.9$\%$  & 5f6d7s$^2$ ($^2$F) $^3$P \\
16217.48 & 16531 & -313 & 2 & 35.7$\%$ & 6d$^2$7s7p ($^3$F) $^5$G  & 9.0$\%$ & 5f6d7s$^2$ ($^2$F) $^3$P &  7.5$\%$ & 6d7s$^2$7p ($^2$D) $^3$P \\
16346.65 & 16494 & -148 & 4 & 72.3$\%$ & 5f6d$^2$7s ($^3$F) $^5$F & 11.9$\%$ & 5f6d$^2$27s ($^3$F) $^5$H & 6.0$\%$    & 5f6d$^2$7s ($^3$F) $^3$G \\
16783.85 & 16829 & -45  & 4 & 47.5$\%$ & 5f6d$^2$7s ($^3$F) $^5$H & 22.4$\%$ & 5f6d$^2$7s ($^3$F) $^5$I  & 19.0$\%$   & 5f6d$^2$7s ($^3$F) $^3$G \\
17224.30 & 17121 & 103 & 2 & 11.9$\%$ & 6d7s$^2$7p ($^2$D) $^3$D  & 11.3$\%$ & 6d$^2$7s7p ($^3$F) $^5$G &  9.4$\%$ & 6d7s$^2$7p ($^2$D) $^1$D \\
17354.64 & 17169 & 186  & 1 & 39.8$\%$ & 6d$^2$7s7p ($^3$F) $^5$F & 34.8$\%$ & 5f6d$^2$7s ($^3$F) $^5$F  & 7.0$\%$    & 5f6d$^2$7s ($^3$P) $^5$F \\
17411.22 & 17983 & -572 & 3 & 34.8$\%$ & 6d$^2$7s7p ($^3$F) $^5$G  & 18.8$\%$ & 5f6d$^2$7s ($^3$F) $^5$G &  7.2$\%$ & 6d$^3$7p ($^4$F) $^5$G \\
17501.18 & 17617 & -116 & 5 & 69.7$\%$ & 5f6d$^2$7s ($^3$F) $^5$I & 9.3$\%$  & 5f6d$^2$7s ($^3$F) $^3$G  & 7.1$\%$  & 5f6d$^2$7s ($^3$F) $^5$H \\
17847.08 & 17959 & -112 & 2 & 24.7$\%$ & 6d7s$^2$7p ($^2$D) $^3$D  & 17.3$\%$ & 5f6d7s$^2$ ($^2$F) $^3$P &  9.3$\%$ & 6d7s$^2$7p ($^2$D) $^3$D \\
18011.38 & 18002 & 9    & 5 & 34.1$\%$ & 5f6d$^2$7s ($^3$F) $^5$H  & 27.2$\%$ & 5f6d$^2$7s ($^3$F) $^3$G &  23.7$\%$ & 5f6d$^2$7s ($^3$F) $^5$I \\
18053.62 & 18758 & -704 & 4 & 32.3$\%$ & 6d7s$^2$7p ($^2$D) $^3$F  & 23.4$\%$ & 6d$^2$7s7p ($^1$D) $^3$F &  18.6$\%$ & 6d$^2$7s7p ($^3$F) $^5$G \\
18069.07 & 18014 & 55 & 3 & 36.6$\%$ & 5f6d7s$^2$ ($^2$F) $^1$F  & 22.0$\%$ & 5f6d7s$^2$ ($^2$F) $^3$D &  9.2$\%$ & 6d7s$^2$7p ($^2$D) $^1$F \\
18382.83 & 18299 & 84   & 0 & 28.5$\%$ & 5f6d$^2$7s ($^3$F) $^5$D & 16.0$\%$   & 6d$^2$7s7p ($^3$P) $^5$D  & 14.3$\%$ & 5f6d$^2$7s ($^3$P) $^5$D \\
18614.34 & 18544 & 70 & 1 & 22.2$\%$ & 5f6d$^2$7s ($^3$F) $^5$D & 12.1$\%$   & 5f6d$^2$7s ($^3$F) $^3$S  & 1.0$\%$ & 6d$^3$7p ($^2$P) $^1$P \\
18809.89 & 18909 & -99 & 4 & 30.6$\%$ & 5f6d$^2$7s ($^3$F) $^5$H & 23.7$\%$ & 5f6d$^2$7s ($^3$F) $^3$G  & 10.0$\%$ & 5f6d$^2$7s ($^3$F) $^5$G \\
18930.29 & 18952 & -22 & 3 & 24.4$\%$ & 5f6d$^2$7s ($^3$F) $^5$F  & 20.0$\%$ & 6d$^2$7s7p ($^3$F) $^5$F &  17.4$\%$ & 5f6d$^2$7s ($^3$F) $^3$G \\
19039.15 & 18831 & 208 & 2 & 23.2$\%$ & 5f6d$^2$7s ($^3$F) $^5$G  & 22.6$\%$ & 6d$^3$7p ($^4$F) $^5$G &  21.8$\%$ & 6d$^2$7s7p ($^3$F) $^5$G \\
19227.34 & 19159 & 68   & 6 & 94.3$\%$ & 5f6d$^2$7s ($^3$F) $^5$I & 2.0$\%$    & 5f6d$^2$7s ($^3$F) $^3$I  & 1.4$\%$  & 5f6d$^2$7s ($^3$F) $^3$I \\
19503.14 & 19352 & 151 & 3 & 17.8$\%$ & 6d7s$^2$7p ($^2$D) $^3$D  & 13.4$\%$ & 6d$^2$7s7p ($^3$F) $^5$G &  6.4$\%$ & 5f6d$^2$7s ($^3$F) $^3$G \\
19516.98 & 19579 & -62 & 2 & 29.7$\%$ & 5f6d$^2$7s ($^3$F) $^5$D  & 19.8$\%$ & 5f6d$^2$7s ($^3$P) $^5$D &  9.0$\%$ & 6d$^2$7s7p ($^3$P) $^5$D \\
19588.36 & 19401 & 187 & 5 & 54.1$\%$ & 5f6d$^2$7s ($^3$F) $^5$H  & 22.8$\%$ & 5f6d$^2$7s ($^3$F) $^3$G &  9.0$\%$ & 5f6d$^2$7s ($^3$P) $^3$G \\
19817.18 & 19246 & 571 & 1 & 21.4$\%$ & 5f6d7s$^2$ ($^2$F) $^3$P  & 13.5$\%$ & 6d$^2$7s7p ($^3$P) $^5$P &  11.2$\%$ & 6d7s$^2$7p ($^2$D) $^3$P \\
19948.40 & 19827 & 121 & 4 & 21.9$\%$ & 5f6d$^2$7s ($^3$F) $^5$F  & 19.0$\%$ & 6d$^2$7s7p ($^3$F) $^5$G &  16.7$\%$ & 6d$^2$7s7p ($^3$F) $^5$F \\
19986.17 & 20166 & -180 & 6 & 94.8$\%$ & 5f6d$^2$7s ($^3$F) $^5$H & 1.1$\%$  & 5f6d$^2$7s ($^3$F) $^3$H  & 1.1$\%$  & 5f6d$^2$7s ($^3$F) $^3$H \\
\end{tabular}
\end{ruledtabular}}
\end{table*}

\subsection{IP$_1$ and IP$_2$}
A second set of atomic calculations was undertaken to calculate the ionization potentials (IP\textsubscript{1} and IP\textsubscript{2}) of thorium. In this case, the fully relativistic Multiconfiguration Dirac-Hartree-Fock (MCDHF) method \cite{Grant2007} was used as implemented in the latest version of the General Relativistic Atomic Structure Program (GRASP), namely GRASP2018 \cite{Froese2019}. In order to estimate the ground-state levels in \ch{Th}, \ch{Th^{+}} and \ch{Th^{2+}}, as accurately as possible, we began the calculations by considering the following multireferences (MR): 6d$^2$7s$^2$, 6d$^3$7s, 6d$^4$, 5f$^2$6d7s, 5f$^2$6d$^2$, 5f$^2$7s$^2$, 5f6d$^2$7p, 5f6d7s7p, 5f7s$^2$7p for \ch{Th}, 6d$^2$7s, 6d7s$^2$, 6d$^3$, 5f$^2$7s, 5f$^2$6d, 5f6d7p, 5f7s7p for \ch{Th^+}, and 5f6d, 5f7s, 6d7p, 7s7p for \ch{Th^{2+}}.\\
\\
For \ch{Th} and \ch{Th^+}, we considered the same core orbitals optimized on the ground-state level of \ch{Th} (6d$^2$7s$^2$ $^3$F$_2$) in order to enforce the relaxation of the core orbitals in \ch{Th^+}. The 7s,7p,6d and 5f valence orbitals were then optimized on the ground levels separately for \ch{Th} (6d$^2$7s$^2$ $^3$F$_2$) and \ch{Th^+} (6d$^2$7s $^4$F$_{3/2}$) with the Optimal Level (OL) method \cite{Grant2007}, where 254 and 52 Configuration State Functions (CSF)s were generated from each MR, respectively. For \ch{Th^{2+}}, we started with the optimization of all orbitals using the ground-state level only (5f6d $^3$H$_4$) and the 7s,7p,6d and 5f orbitals were then optimized on the ground-state level with the OL method where the 6 MR CSFs were generated.\\
\\
In the next computational steps, valence-valence interactions were treated through single and double substitutions from the reference orbitals (7s, 7p, 6d, 5f) to correlation orbitals belonging to different active sets (AS) listed in Table~\ref{MCDHFIPs}. These sets were chosen in such a way that more and more orbitals with increasing $n$-values were considered until convergence was achieved on both IP\textsubscript{1} and IP\textsubscript{2}. For each AS, the new correlation orbitals were optimized on the ground-state level by freezing the orbitals already optimized in the previous AS. The results obtained in our calculations are given in Table~\ref{MCDHFIPs} where we can note that the last AS, i.e. $\{$10(s--h)$\}$ allowed us to obtain theoretical results in very good agreement with the experimental ionization potentials measured in the present work, i.e. a relative difference between calculations and experiment of 0.06\% and 0.19\% for IP\textsubscript{1} and IP\textsubscript{2}, respectively. We actually observed that both IP\textsubscript{1} and IP\textsubscript{2} final results reproduce the experiment better when using the above optimization procedure for the core orbitals and, therefore, the core relaxation effects on the IPs.\\
\\
It is interesting to note that the values considered to be the best computed by \cite{Weigand2014}, namely those obtained using the Multiconfiguration Dirac-Hartree-Fock method based on a Dirac-Coulomb Hamiltonian with a perturbative treatment of the Breit interaction and a relativistic small-core pseudopotential (MCDHF/DC+B PP), were IP\textsubscript{1} = 50813 cm$^{-1}$ and IP\textsubscript{2} = 100093 cm$^{-1}$. These two IP values \cite{Weigand2014} show a slightly larger discrepancy with respect to the experimental values (0.11\% and 0.89\%, respectively) than our theoretical results.
\begin{table}[H]
	\centering
		\caption{Ionization potentials (in cm$^{-1}$) of \ch{Th} (IP\textsubscript{1}) and \ch{Th^+} (IP\textsubscript{2}) as computed in the present work using the MCDHF method. Each active set (AS) is denoted by $\{$$nl$, $n'l'$,...$\}$ where $n$, $n'$, ... represent the maximum values of the principal quantum number for the different orbitals. Experimental values, also deduced from the present work, are given for comparison along with the theoretical MCDHF/DC+B PP values from \cite{Weigand2014}.}
	\begin{tabular}{@{}lll@{}}
		\toprule
        Active set & IP\textsubscript{1} & IP\textsubscript{2} \\
            \midrule
        MR               &  44676.0 & 99716.8 \\
        $\{$7s,7p,6d,5f,5g$\}$ &  49674.7 & 97330.5 \\
        $\{$7s,7p,6(d--h)$\}$  &  49863.9 & 98782.8 \\
        $\{$7(s--h)$\}$        &  50026.4 & 98899.4 \\
        $\{$8(s--h)$\}$        &  50817.6 & 99349.5 \\
        $\{$9(s--h)$\}$        &  50884.0 & 99384.1 \\
        $\{$10(s--h)$\}$       &  50896.2 & 99390.8 \\
        Experiment             &  50868.41(11) &  99207(73) \\
        MCDHF/DC+B PP \cite{Weigand2014} & 50813 & 100093 \\ 
		\bottomrule
	\end{tabular}
	\label{MCDHFIPs}
\end{table}

\section{Conclusion and discussion}
A level search in the vicinity of the IP\textsubscript{2} of thorium was conducted and presented with the initial aim of developing an efficient laser ionization scheme of \ch{Th^+} to be used for the study of \ch{^{229m}Th^{+}}. The \ch{Th^+} level search was performed with \ch{^{232}Th} in an argon-filled gas cell where a natural metallic \ch{Th} foil was laser-ablated. It produced a complicated, convoluted spectrum influenced by gas-induced quenching effects. Differences in background and peak width of the observed resonances, allowed for the use of a threshold approach to extract IP\textsubscript{2}. A value of $\SI{12.300 \pm 0.009}{\electronvolt}$  ($\SI{99207 \pm 73}{\per \cm}$) was found for IP\textsubscript{2}, presenting an 22-fold improvement over the previous value. This improved value helped the identification of several auto-ionizing states of which the most efficient was found at $\SI{99766.87 \pm 0.22}{\per\cm}$ with an efficiency of at least $1.2\%$. Additionally, ionization efficiency of up to $3.4(5)\%$ was achieved by utilizing transitions originating from states populated by collisional de-excitation. In the latter case, such resonances could be used for gas-cell based laser ionization.
\\
\\
To evaluate the accuracy of the threshold approach, a similar level search around IP\textsubscript{1} was performed, both in-gas-cell and in-gas-jet. The nearly collision-free gas jet revealed a telltale Rydberg structure consisting of several (partial) Rydberg series. Rydberg-Ritz analysis of the most complete series converged to an IP\textsubscript{1} of $\SI{6.306879 \pm 0.000014}{\electronvolt}$ ($\SI{50868.41 \pm 0.11}{\per \cm}$), presenting a 9-fold reduction in uncertainty compared to the literature \cite{Kohler1997}. Hence the first demonstration of in-gas-jet laser ionization via Rydberg states could be reported. Time-of-flight measurements showed that these Rydberg states are field ionized in the S-RFQ. 
\\
\\ 
Atomic structure calculations by the HFR+CPOL method were performed using optimized radial parameters for \ch{Th}. This revealed that the $J =2$ state at $\SI{26113.27}{\per \cm}$ lies in an energy region with strongly-mixed states. The state was identified based on its Land\'e factor and transition probability. It is highly mixed and its first three main LS-coupling components are 8.8\% 6d$^2$7s7p ($^1$D)$^3$F + 8.1\% 6d$^3$7p ($^4$F)$^3$F + 7.9 \% 5f6d$^2$7s ($^1$D)$^3$D. As the code cannot reach high $n$ values, extrapolation from the Rydberg states to lower $n$ states was attempted but did not allow for the identification of the spectroscopic designation of the series. MCDHF calculations were conducted to obtain the ground states of \ch{Th}, \ch{Th^+} and \ch{Th^{2+}}. The final AS ($\{10(s-h)\}$) provides the best theoretical values for IP\textsubscript{1} and IP\textsubscript{2} with a relative difference of 0.06\% and 0.19\% versus experiment, respectively.
\\
\\ 
We presented a dedicated fast-extraction gas cell equipped with \ch{^{233}U} alpha-recoil sources to demonstrate the existence of the nuclear-clock isomer in the not-yet-observed singly charged state. Extraction times to the gas jet of around $\SI{1}{\ms}$ and $\SI{2.5}{\ms}$ can be expected based on COMSOL Multiphysics simulations for the two sources, which is below the current upper boundary for the singly charged isomeric half-life ($\SI{10}{\ms}$ \cite{VonDerWense2016}). Mass separation showed that most of the recoil ions end up in a singly charged state with count rates of $\SI{1360}{counts/\s}$ for \ch{^{229}Th^+} and $\SI{500}{counts/\s}$ for \ch{^{229}Th^{2+}}. Despite this, no laser ionization signal of even the ground state \ch{^{229}Th^+} was observed. The recoil process and subsequent charge state quenching might leave a substantial fraction of ions in (several) meta-stable states, making them unavailable for the used laser ionization scheme. The occupation of these dark states was observed in the gas jet for laser-ablated thorium atoms. Unstable photoion signal was encountered when laser ionizing \ch{^{232}Th^{+}} in the gas jet, possibly due to the multi-mode structure of the laser pulses, the electric field of the RFQ's or influence of the lasers entering the gas cell, further necessitating the conclusion of the project.
\\
\\
Future in-gas-jet work on the thorium isotopic chain is planned at S\textsuperscript{3}-LEB at GANIL. These studies could benefit from the
added flexibility of studying the ion should neutralization in the gas cell turn out to be inefficient. The efficiency in the IGLIS technique, and RIS techniques in general, can be limited by the ionization step. In forseseen cases such as nobelium or silver, for which no AI states are known, high laser powers are required to efficiently ionize to the continuum. By using a high-lying Rydberg state and subsequent field ionization in-gas-jet, cross sections might get significantly higher.	
\section{Acknowledgments}
We would like to thank M. Kaja and K. Wendt from JGU Mainz for their help with the Rydberg-Ritz analysis. This work has received funding from European Union’s Horizon 2020 research and innovation programme under Grant Agreement No. 861198-LISA-H2020-MSCA- ITN-2019, the Research Foundation Flanders (FWO, Belgium) BOF KU Leuven (C14/22/104) and the FWO and F.R.S.-FNRS under the Excellence of Science (EOS) program (40007501).

\bibliography{allpapers}

\begin{thebibliography}{50}%
\makeatletter
\providecommand \@ifxundefined [1]{%
 \@ifx{#1\undefined}
}%
\providecommand \@ifnum [1]{%
 \ifnum #1\expandafter \@firstoftwo
 \else \expandafter \@secondoftwo
 \fi
}%
\providecommand \@ifx [1]{%
 \ifx #1\expandafter \@firstoftwo
 \else \expandafter \@secondoftwo
 \fi
}%
\providecommand \natexlab [1]{#1}%
\providecommand \enquote  [1]{``#1''}%
\providecommand \bibnamefont  [1]{#1}%
\providecommand \bibfnamefont [1]{#1}%
\providecommand \citenamefont [1]{#1}%
\providecommand \href@noop [0]{\@secondoftwo}%
\providecommand \href [0]{\begingroup \@sanitize@url \@href}%
\providecommand \@href[1]{\@@startlink{#1}\@@href}%
\providecommand \@@href[1]{\endgroup#1\@@endlink}%
\providecommand \@sanitize@url [0]{\catcode `\\12\catcode `\$12\catcode
  `\&12\catcode `\#12\catcode `\^12\catcode `\_12\catcode `\%12\relax}%
\providecommand \@@startlink[1]{}%
\providecommand \@@endlink[0]{}%
\providecommand \url  [0]{\begingroup\@sanitize@url \@url }%
\providecommand \@url [1]{\endgroup\@href {#1}{\urlprefix }}%
\providecommand \urlprefix  [0]{URL }%
\providecommand \Eprint [0]{\href }%
\providecommand \doibase [0]{https://doi.org/}%
\providecommand \selectlanguage [0]{\@gobble}%
\providecommand \bibinfo  [0]{\@secondoftwo}%
\providecommand \bibfield  [0]{\@secondoftwo}%
\providecommand \translation [1]{[#1]}%
\providecommand \BibitemOpen [0]{}%
\providecommand \bibitemStop [0]{}%
\providecommand \bibitemNoStop [0]{.\EOS\space}%
\providecommand \EOS [0]{\spacefactor3000\relax}%
\providecommand \BibitemShut  [1]{\csname bibitem#1\endcsname}%
\let\auto@bib@innerbib\@empty
\bibitem [{\citenamefont {Peik}\ and\ \citenamefont {Tamm}(2003)}]{Peik2003}%
  \BibitemOpen
  \bibfield  {author} {\bibinfo {author} {\bibfnamefont {E.}~\bibnamefont
  {Peik}}\ and\ \bibinfo {author} {\bibfnamefont {C.}~\bibnamefont {Tamm}},\
  }\bibfield  {title} {\bibinfo {title} {{Nuclear laser spectroscopy of the 3.5
  eV transition in Th-229}},\ }\href
  {https://doi.org/10.1209/epl/i2003-00210-x} {\bibfield  {journal} {\bibinfo
  {journal} {Europhysics Letters}\ }\textbf {\bibinfo {volume} {61}},\ \bibinfo
  {pages} {181} (\bibinfo {year} {2003})}\BibitemShut {NoStop}%
\bibitem [{\citenamefont {Tiedau}\ \emph {et~al.}(2024)\citenamefont {Tiedau},
  \citenamefont {Okhapkin}, \citenamefont {Zhang}, \citenamefont {Thielking},
  \citenamefont {Zitzer},\ and\ \citenamefont {Peik}}]{Tiedau2024}%
  \BibitemOpen
  \bibfield  {author} {\bibinfo {author} {\bibfnamefont {J.}~\bibnamefont
  {Tiedau}}, \bibinfo {author} {\bibfnamefont {M.~V.}\ \bibnamefont
  {Okhapkin}}, \bibinfo {author} {\bibfnamefont {K.}~\bibnamefont {Zhang}},
  \bibinfo {author} {\bibfnamefont {J.}~\bibnamefont {Thielking}}, \bibinfo
  {author} {\bibfnamefont {G.}~\bibnamefont {Zitzer}},\ and\ \bibinfo {author}
  {\bibfnamefont {E.}~\bibnamefont {Peik}},\ }\bibfield  {title} {\bibinfo
  {title} {{Laser Excitation of the Th-229 Nucleus}},\ }\href
  {https://doi.org/10.1103/PhysRevLett.132.182501} {\bibfield  {journal}
  {\bibinfo  {journal} {Physical Review Letters}\ }\textbf {\bibinfo {volume}
  {132}},\ \bibinfo {pages} {182501} (\bibinfo {year} {2024})}\BibitemShut
  {NoStop}%
\bibitem [{\citenamefont {Elwell}\ \emph {et~al.}(2024)\citenamefont {Elwell},
  \citenamefont {Schneider}, \citenamefont {Jeet}, \citenamefont {Terhune},
  \citenamefont {Morgan}, \citenamefont {Alexandrova}, \citenamefont {Tan},
  \citenamefont {Derevianko},\ and\ \citenamefont {Hudson}}]{Elwell2024}%
  \BibitemOpen
  \bibfield  {author} {\bibinfo {author} {\bibfnamefont {R.}~\bibnamefont
  {Elwell}}, \bibinfo {author} {\bibfnamefont {C.}~\bibnamefont {Schneider}},
  \bibinfo {author} {\bibfnamefont {J.}~\bibnamefont {Jeet}}, \bibinfo {author}
  {\bibfnamefont {J.~E.~S.}\ \bibnamefont {Terhune}}, \bibinfo {author}
  {\bibfnamefont {H.~W.~T.}\ \bibnamefont {Morgan}}, \bibinfo {author}
  {\bibfnamefont {A.~N.}\ \bibnamefont {Alexandrova}}, \bibinfo {author}
  {\bibfnamefont {H.~B.~T.}\ \bibnamefont {Tan}}, \bibinfo {author}
  {\bibfnamefont {A.}~\bibnamefont {Derevianko}},\ and\ \bibinfo {author}
  {\bibfnamefont {E.~R.}\ \bibnamefont {Hudson}},\ }\bibfield  {title}
  {\bibinfo {title} {{Laser excitation of the $^{229}$Th nuclear isomeric
  transition in a solid-state host}},\ }\href
  {https://doi.org/10.1103/PhysRevLett.133.013201} {\bibfield  {journal}
  {\bibinfo  {journal} {Physical Review Letters}\ }\textbf {\bibinfo {volume}
  {133}},\ \bibinfo {pages} {13201} (\bibinfo {year} {2024})},\ \Eprint
  {https://arxiv.org/abs/2404.12311} {arXiv:2404.12311} \BibitemShut {NoStop}%
\bibitem [{\citenamefont {Zhang}\ \emph {et~al.}(2024)\citenamefont {Zhang},
  \citenamefont {Ooi}, \citenamefont {Higgins}, \citenamefont {Doyle},
  \citenamefont {von~der Wense}, \citenamefont {Beeks}, \citenamefont
  {Leitner}, \citenamefont {Kazakov}, \citenamefont {Li}, \citenamefont
  {Thirolf}, \citenamefont {Schumm},\ and\ \citenamefont {Ye}}]{Zhang2024}%
  \BibitemOpen
  \bibfield  {author} {\bibinfo {author} {\bibfnamefont {C.}~\bibnamefont
  {Zhang}}, \bibinfo {author} {\bibfnamefont {T.}~\bibnamefont {Ooi}}, \bibinfo
  {author} {\bibfnamefont {J.~S.}\ \bibnamefont {Higgins}}, \bibinfo {author}
  {\bibfnamefont {J.~F.}\ \bibnamefont {Doyle}}, \bibinfo {author}
  {\bibfnamefont {L.}~\bibnamefont {von~der Wense}}, \bibinfo {author}
  {\bibfnamefont {K.}~\bibnamefont {Beeks}}, \bibinfo {author} {\bibfnamefont
  {A.}~\bibnamefont {Leitner}}, \bibinfo {author} {\bibfnamefont
  {G.}~\bibnamefont {Kazakov}}, \bibinfo {author} {\bibfnamefont
  {P.}~\bibnamefont {Li}}, \bibinfo {author} {\bibfnamefont {P.~G.}\
  \bibnamefont {Thirolf}}, \bibinfo {author} {\bibfnamefont {T.}~\bibnamefont
  {Schumm}},\ and\ \bibinfo {author} {\bibfnamefont {J.}~\bibnamefont {Ye}},\
  }\bibfield  {title} {\bibinfo {title} {{Dawn of a nuclear clock: frequency
  ratio of the $^{229m}$Th isomeric transition and the $^{87}$Sr atomic
  clock}},\ }\bibfield  {journal} {\bibinfo  {journal} {Nature}\ }\textbf
  {\bibinfo {volume} {633}},\ \href
  {https://doi.org/10.1038/s41586-024-07839-6} {10.1038/s41586-024-07839-6}
  (\bibinfo {year} {2024}),\ \Eprint {https://arxiv.org/abs/2406.18719}
  {arXiv:2406.18719} \BibitemShut {NoStop}%
\bibitem [{\citenamefont {{Von Der Wense}}\ \emph {et~al.}(2016)\citenamefont
  {{Von Der Wense}}, \citenamefont {Seiferle}, \citenamefont {Laatiaoui},
  \citenamefont {Neumayr}, \citenamefont {Maier}, \citenamefont {Wirth},
  \citenamefont {Mokry}, \citenamefont {Runke}, \citenamefont {Eberhardt},
  \citenamefont {D{\"{u}}llmann}, \citenamefont {Trautmann},\ and\
  \citenamefont {Thirolf}}]{VonDerWense2016}%
  \BibitemOpen
  \bibfield  {author} {\bibinfo {author} {\bibfnamefont {L.}~\bibnamefont {{Von
  Der Wense}}}, \bibinfo {author} {\bibfnamefont {B.}~\bibnamefont {Seiferle}},
  \bibinfo {author} {\bibfnamefont {M.}~\bibnamefont {Laatiaoui}}, \bibinfo
  {author} {\bibfnamefont {J.~B.}\ \bibnamefont {Neumayr}}, \bibinfo {author}
  {\bibfnamefont {H.~J.}\ \bibnamefont {Maier}}, \bibinfo {author}
  {\bibfnamefont {H.~F.}\ \bibnamefont {Wirth}}, \bibinfo {author}
  {\bibfnamefont {{\relax Ch}.}~\bibnamefont {Mokry}}, \bibinfo {author}
  {\bibfnamefont {J.}~\bibnamefont {Runke}}, \bibinfo {author} {\bibfnamefont
  {K.}~\bibnamefont {Eberhardt}}, \bibinfo {author} {\bibfnamefont {{\relax
  Ch}.~E.}\ \bibnamefont {D{\"{u}}llmann}}, \bibinfo {author} {\bibfnamefont
  {N.~G.}\ \bibnamefont {Trautmann}},\ and\ \bibinfo {author} {\bibfnamefont
  {P.~G.}\ \bibnamefont {Thirolf}},\ }\bibfield  {title} {\bibinfo {title}
  {{Direct detection of the \ch{^{229}Th} nuclear clock transition}},\ }\href
  {https://doi.org/10.1038/nature17669} {\bibfield  {journal} {\bibinfo
  {journal} {Nature}\ }\textbf {\bibinfo {volume} {533}},\ \bibinfo {pages}
  {47} (\bibinfo {year} {2016})},\ \Eprint {https://arxiv.org/abs/1710.11398}
  {arXiv:1710.11398} \BibitemShut {NoStop}%
\bibitem [{\citenamefont {Seiferle}\ \emph {et~al.}(2017)\citenamefont
  {Seiferle}, \citenamefont {von~der Wense},\ and\ \citenamefont
  {Thirolf}}]{Seiferle2017}%
  \BibitemOpen
  \bibfield  {author} {\bibinfo {author} {\bibfnamefont {B.}~\bibnamefont
  {Seiferle}}, \bibinfo {author} {\bibfnamefont {L.}~\bibnamefont {von~der
  Wense}},\ and\ \bibinfo {author} {\bibfnamefont {P.~G.}\ \bibnamefont
  {Thirolf}},\ }\bibfield  {title} {\bibinfo {title} {Lifetime measurement of
  the \ch{^{229}Th} nuclear isomer},\ }\href
  {https://doi.org/10.1103/PhysRevLett.118.042501} {\bibfield  {journal}
  {\bibinfo  {journal} {Physical Review Letters}\ }\textbf {\bibinfo {volume}
  {118}},\ \bibinfo {pages} {042501} (\bibinfo {year} {2017})},\ \Eprint
  {https://arxiv.org/abs/arXiv:1801.05205v1} {arXiv:arXiv:1801.05205v1}
  \BibitemShut {NoStop}%
\bibitem [{\citenamefont {Thielking}\ \emph {et~al.}(2018)\citenamefont
  {Thielking}, \citenamefont {Okhapkin}, \citenamefont {G{\l}owacki},
  \citenamefont {Meier}, \citenamefont {{Von Der Wense}}, \citenamefont
  {Seiferle}, \citenamefont {D{\"{u}}llmann}, \citenamefont {Thirolf},\ and\
  \citenamefont {Peik}}]{Thielking2018}%
  \BibitemOpen
  \bibfield  {author} {\bibinfo {author} {\bibfnamefont {J.}~\bibnamefont
  {Thielking}}, \bibinfo {author} {\bibfnamefont {M.~V.}\ \bibnamefont
  {Okhapkin}}, \bibinfo {author} {\bibfnamefont {P.}~\bibnamefont
  {G{\l}owacki}}, \bibinfo {author} {\bibfnamefont {D.~M.}\ \bibnamefont
  {Meier}}, \bibinfo {author} {\bibfnamefont {L.}~\bibnamefont {{Von Der
  Wense}}}, \bibinfo {author} {\bibfnamefont {B.}~\bibnamefont {Seiferle}},
  \bibinfo {author} {\bibfnamefont {{\relax Ch}.~E.}\ \bibnamefont
  {D{\"{u}}llmann}}, \bibinfo {author} {\bibfnamefont {P.~G.}\ \bibnamefont
  {Thirolf}},\ and\ \bibinfo {author} {\bibfnamefont {E.}~\bibnamefont
  {Peik}},\ }\bibfield  {title} {\bibinfo {title} {{Laser spectroscopic
  characterization of the nuclear-clock isomer \ch{^{229m}Th} }},\ }\href
  {https://doi.org/10.1038/s41586-018-0011-8} {\bibfield  {journal} {\bibinfo
  {journal} {Nature}\ }\textbf {\bibinfo {volume} {556}},\ \bibinfo {pages}
  {321} (\bibinfo {year} {2018})},\ \Eprint {https://arxiv.org/abs/1709.05325}
  {arXiv:1709.05325} \BibitemShut {NoStop}%
\bibitem [{\citenamefont {Yamaguchi}\ \emph {et~al.}(2024)\citenamefont
  {Yamaguchi}, \citenamefont {Shigekawa}, \citenamefont {Haba}, \citenamefont
  {Kikunaga}, \citenamefont {Shirasaki}, \citenamefont {Wada},\ and\
  \citenamefont {Katori}}]{Yamaguchi2024}%
  \BibitemOpen
  \bibfield  {author} {\bibinfo {author} {\bibfnamefont {A.}~\bibnamefont
  {Yamaguchi}}, \bibinfo {author} {\bibfnamefont {Y.}~\bibnamefont
  {Shigekawa}}, \bibinfo {author} {\bibfnamefont {H.}~\bibnamefont {Haba}},
  \bibinfo {author} {\bibfnamefont {H.}~\bibnamefont {Kikunaga}}, \bibinfo
  {author} {\bibfnamefont {K.}~\bibnamefont {Shirasaki}}, \bibinfo {author}
  {\bibfnamefont {M.}~\bibnamefont {Wada}},\ and\ \bibinfo {author}
  {\bibfnamefont {H.}~\bibnamefont {Katori}},\ }\bibfield  {title} {\bibinfo
  {title} {{Laser spectroscopy of triply charged 229Th isomer for a nuclear
  clock}},\ }\bibfield  {journal} {\bibinfo  {journal} {Nature}\ }\href
  {https://doi.org/10.1038/s41586-024-07296-1} {10.1038/s41586-024-07296-1}
  (\bibinfo {year} {2024})\BibitemShut {NoStop}%
\bibitem [{\citenamefont {Kraemer}\ \emph {et~al.}(2023)\citenamefont
  {Kraemer}, \citenamefont {Moens}, \citenamefont {Athanasakis-Kaklamanakis},
  \citenamefont {Bara}, \citenamefont {Beeks}, \citenamefont {Chhetri},
  \citenamefont {Chrysalidis}, \citenamefont {Claessens}, \citenamefont
  {Cocolios}, \citenamefont {Correia}, \citenamefont {Witte}, \citenamefont
  {Ferrer}, \citenamefont {Geldhof}, \citenamefont {Heinke}, \citenamefont
  {Hosseini}, \citenamefont {Huyse}, \citenamefont {K{\"{o}}ster},
  \citenamefont {Kudryavtsev}, \citenamefont {Laatiaoui}, \citenamefont {Lica},
  \citenamefont {Magchiels}, \citenamefont {Manea}, \citenamefont {Merckling},
  \citenamefont {Pereira}, \citenamefont {Raeder}, \citenamefont {Schumm},
  \citenamefont {Sels}, \citenamefont {Thirolf}, \citenamefont {Tunhuma},
  \citenamefont {{Van Den Bergh}}, \citenamefont {{Van Duppen}}, \citenamefont
  {Vantomme}, \citenamefont {Verlinde}, \citenamefont {Villarreal},\ and\
  \citenamefont {Wahl}}]{Kraemer2023}%
  \BibitemOpen
  \bibfield  {author} {\bibinfo {author} {\bibfnamefont {S.}~\bibnamefont
  {Kraemer}}, \bibinfo {author} {\bibfnamefont {J.}~\bibnamefont {Moens}},
  \bibinfo {author} {\bibfnamefont {M.}~\bibnamefont
  {Athanasakis-Kaklamanakis}}, \bibinfo {author} {\bibfnamefont
  {S.}~\bibnamefont {Bara}}, \bibinfo {author} {\bibfnamefont {K.}~\bibnamefont
  {Beeks}}, \bibinfo {author} {\bibfnamefont {P.}~\bibnamefont {Chhetri}},
  \bibinfo {author} {\bibfnamefont {K.}~\bibnamefont {Chrysalidis}}, \bibinfo
  {author} {\bibfnamefont {A.}~\bibnamefont {Claessens}}, \bibinfo {author}
  {\bibfnamefont {T.~E.}\ \bibnamefont {Cocolios}}, \bibinfo {author}
  {\bibfnamefont {J.~G.}\ \bibnamefont {Correia}}, \bibinfo {author}
  {\bibfnamefont {H.~D.}\ \bibnamefont {Witte}}, \bibinfo {author}
  {\bibfnamefont {R.}~\bibnamefont {Ferrer}}, \bibinfo {author} {\bibfnamefont
  {S.}~\bibnamefont {Geldhof}}, \bibinfo {author} {\bibfnamefont
  {R.}~\bibnamefont {Heinke}}, \bibinfo {author} {\bibfnamefont
  {N.}~\bibnamefont {Hosseini}}, \bibinfo {author} {\bibfnamefont
  {M.}~\bibnamefont {Huyse}}, \bibinfo {author} {\bibfnamefont
  {U.}~\bibnamefont {K{\"{o}}ster}}, \bibinfo {author} {\bibfnamefont
  {Y.}~\bibnamefont {Kudryavtsev}}, \bibinfo {author} {\bibfnamefont
  {M.}~\bibnamefont {Laatiaoui}}, \bibinfo {author} {\bibfnamefont
  {R.}~\bibnamefont {Lica}}, \bibinfo {author} {\bibfnamefont {G.}~\bibnamefont
  {Magchiels}}, \bibinfo {author} {\bibfnamefont {V.}~\bibnamefont {Manea}},
  \bibinfo {author} {\bibfnamefont {C.}~\bibnamefont {Merckling}}, \bibinfo
  {author} {\bibfnamefont {L.~M.}\ \bibnamefont {Pereira}}, \bibinfo {author}
  {\bibfnamefont {S.}~\bibnamefont {Raeder}}, \bibinfo {author} {\bibfnamefont
  {T.}~\bibnamefont {Schumm}}, \bibinfo {author} {\bibfnamefont
  {S.}~\bibnamefont {Sels}}, \bibinfo {author} {\bibfnamefont {P.~G.}\
  \bibnamefont {Thirolf}}, \bibinfo {author} {\bibfnamefont {S.~M.}\
  \bibnamefont {Tunhuma}}, \bibinfo {author} {\bibfnamefont {P.}~\bibnamefont
  {{Van Den Bergh}}}, \bibinfo {author} {\bibfnamefont {P.}~\bibnamefont {{Van
  Duppen}}}, \bibinfo {author} {\bibfnamefont {A.}~\bibnamefont {Vantomme}},
  \bibinfo {author} {\bibfnamefont {M.}~\bibnamefont {Verlinde}}, \bibinfo
  {author} {\bibfnamefont {R.}~\bibnamefont {Villarreal}},\ and\ \bibinfo
  {author} {\bibfnamefont {U.}~\bibnamefont {Wahl}},\ }\bibfield  {title}
  {\bibinfo {title} {{Observation of the radiative decay of the \ch{^{229}Th}
  nuclear clock isomer}},\ }\href {https://doi.org/10.1038/s41586-023-05894-z}
  {\bibfield  {journal} {\bibinfo  {journal} {Nature}\ }\textbf {\bibinfo
  {volume} {617}},\ \bibinfo {pages} {706} (\bibinfo {year} {2023})},\ \Eprint
  {https://arxiv.org/abs/2209.10276} {arXiv:2209.10276} \BibitemShut {NoStop}%
\bibitem [{\citenamefont {Porsev}\ \emph {et~al.}(2010)\citenamefont {Porsev},
  \citenamefont {Flambaum}, \citenamefont {Peik},\ and\ \citenamefont
  {Tamm}}]{Porsev2010}%
  \BibitemOpen
  \bibfield  {author} {\bibinfo {author} {\bibfnamefont {S.~G.}\ \bibnamefont
  {Porsev}}, \bibinfo {author} {\bibfnamefont {V.~V.}\ \bibnamefont
  {Flambaum}}, \bibinfo {author} {\bibfnamefont {E.}~\bibnamefont {Peik}},\
  and\ \bibinfo {author} {\bibfnamefont {C.}~\bibnamefont {Tamm}},\ }\bibfield
  {title} {\bibinfo {title} {{Excitation of the isomeric \ch{^{229m}Th} nuclear
  state via an electronic bridge process in \ch{^{229}Th^+} }},\ }\href
  {https://doi.org/10.1103/PhysRevLett.105.182501} {\bibfield  {journal}
  {\bibinfo  {journal} {Physical Review Letters}\ }\textbf {\bibinfo {volume}
  {105}},\ \bibinfo {pages} {1} (\bibinfo {year} {2010})}\BibitemShut {NoStop}%
\bibitem [{\citenamefont {Porsev}\ and\ \citenamefont
  {Flambaum}(2010)}]{Porsev2010b}%
  \BibitemOpen
  \bibfield  {author} {\bibinfo {author} {\bibfnamefont {S.~G.}\ \bibnamefont
  {Porsev}}\ and\ \bibinfo {author} {\bibfnamefont {V.~V.}\ \bibnamefont
  {Flambaum}},\ }\bibfield  {title} {\bibinfo {title} {{Electronic bridge
  process in $^{229}$Th$^+$}},\ }\href
  {https://doi.org/10.1103/PhysRevA.81.042516} {\bibfield  {journal} {\bibinfo
  {journal} {Physical Review A - Atomic, Molecular, and Optical Physics}\
  }\textbf {\bibinfo {volume} {81}},\ \bibinfo {pages} {1} (\bibinfo {year}
  {2010})}\BibitemShut {NoStop}%
\bibitem [{\citenamefont {Karpeshin}\ and\ \citenamefont
  {Trzhaskovskaya}(2018)}]{Karpeshin2018}%
  \BibitemOpen
  \bibfield  {author} {\bibinfo {author} {\bibfnamefont {F.~F.}\ \bibnamefont
  {Karpeshin}}\ and\ \bibinfo {author} {\bibfnamefont {M.~B.}\ \bibnamefont
  {Trzhaskovskaya}},\ }\bibfield  {title} {\bibinfo {title} {{Impact of the
  ionization of the atomic shell on the lifetime of the\ch{^{229m}Th}}},\
  }\href {https://doi.org/10.1016/j.nuclphysa.2017.10.003} {\bibfield
  {journal} {\bibinfo  {journal} {Nuclear Physics A}\ }\textbf {\bibinfo
  {volume} {969}},\ \bibinfo {pages} {173} (\bibinfo {year} {2018})},\ \Eprint
  {https://arxiv.org/abs/1701.05340} {arXiv:1701.05340} \BibitemShut {NoStop}%
\bibitem [{\citenamefont {Karpeshin}\ and\ \citenamefont
  {Trzhaskovskaya}(2021)}]{Karpeshin2021}%
  \BibitemOpen
  \bibfield  {author} {\bibinfo {author} {\bibfnamefont {F.~F.}\ \bibnamefont
  {Karpeshin}}\ and\ \bibinfo {author} {\bibfnamefont {M.~B.}\ \bibnamefont
  {Trzhaskovskaya}},\ }\bibfield  {title} {\bibinfo {title} {{A proposed
  solution for the lifetime puzzle of the\ch{^{229m}Th^+} isomer}},\ }\href
  {https://doi.org/10.1016/j.nuclphysa.2021.122173} {\bibfield  {journal}
  {\bibinfo  {journal} {Nuclear Physics A}\ }\textbf {\bibinfo {volume}
  {1010}},\ \bibinfo {pages} {122173} (\bibinfo {year} {2021})}\BibitemShut
  {NoStop}%
\bibitem [{\citenamefont {Herrera-Sancho}\ \emph {et~al.}(2013)\citenamefont
  {Herrera-Sancho}, \citenamefont {Nemitz}, \citenamefont {Okhapkin},\ and\
  \citenamefont {Peik}}]{Herrera-Sancho2013}%
  \BibitemOpen
  \bibfield  {author} {\bibinfo {author} {\bibfnamefont {O.~A.}\ \bibnamefont
  {Herrera-Sancho}}, \bibinfo {author} {\bibfnamefont {N.}~\bibnamefont
  {Nemitz}}, \bibinfo {author} {\bibfnamefont {M.~V.}\ \bibnamefont
  {Okhapkin}},\ and\ \bibinfo {author} {\bibfnamefont {E.}~\bibnamefont
  {Peik}},\ }\bibfield  {title} {\bibinfo {title} {{Energy levels of Th$^+$
  between 7.3 and 8.3 eV}},\ }\href
  {https://doi.org/10.1103/PhysRevA.88.012512} {\bibfield  {journal} {\bibinfo
  {journal} {Physical Review A - Atomic, Molecular, and Optical Physics}\
  }\textbf {\bibinfo {volume} {88}},\ \bibinfo {pages} {1} (\bibinfo {year}
  {2013})}\BibitemShut {NoStop}%
\bibitem [{\citenamefont {Meier}\ \emph {et~al.}(2019)\citenamefont {Meier},
  \citenamefont {Thielking}, \citenamefont {G{\l}owacki}, \citenamefont
  {Okhapkin}, \citenamefont {M{\"{u}}ller}, \citenamefont {Surzhykov},\ and\
  \citenamefont {Peik}}]{Meier2019}%
  \BibitemOpen
  \bibfield  {author} {\bibinfo {author} {\bibfnamefont {D.~M.}\ \bibnamefont
  {Meier}}, \bibinfo {author} {\bibfnamefont {J.}~\bibnamefont {Thielking}},
  \bibinfo {author} {\bibfnamefont {P.}~\bibnamefont {G{\l}owacki}}, \bibinfo
  {author} {\bibfnamefont {M.~V.}\ \bibnamefont {Okhapkin}}, \bibinfo {author}
  {\bibfnamefont {R.~A.}\ \bibnamefont {M{\"{u}}ller}}, \bibinfo {author}
  {\bibfnamefont {A.}~\bibnamefont {Surzhykov}},\ and\ \bibinfo {author}
  {\bibfnamefont {E.}~\bibnamefont {Peik}},\ }\bibfield  {title} {\bibinfo
  {title} {{Electronic level structure of Th$^+$ in the range of the
  \ch{^{229m}Th} isomer energy}},\ }\href
  {https://doi.org/10.1103/PhysRevA.99.052514} {\bibfield  {journal} {\bibinfo
  {journal} {Physical Review A}\ }\textbf {\bibinfo {volume} {99}},\ \bibinfo
  {pages} {1} (\bibinfo {year} {2019})}\BibitemShut {NoStop}%
\bibitem [{\citenamefont {Kudryavtsev}\ \emph {et~al.}(2016)\citenamefont
  {Kudryavtsev}, \citenamefont {Creemers}, \citenamefont {Ferrer},
  \citenamefont {Granados}, \citenamefont {Gaffney}, \citenamefont {Huyse},
  \citenamefont {Mogilevskiy}, \citenamefont {Raeder}, \citenamefont {Sels},
  \citenamefont {{Van den Bergh}}, \citenamefont {{Van Duppen}},\ and\
  \citenamefont {Zadvornaya}}]{Kudryavtsev2016a}%
  \BibitemOpen
  \bibfield  {author} {\bibinfo {author} {\bibfnamefont {Y.}~\bibnamefont
  {Kudryavtsev}}, \bibinfo {author} {\bibfnamefont {P.}~\bibnamefont
  {Creemers}}, \bibinfo {author} {\bibfnamefont {R.}~\bibnamefont {Ferrer}},
  \bibinfo {author} {\bibfnamefont {C.}~\bibnamefont {Granados}}, \bibinfo
  {author} {\bibfnamefont {L.}~\bibnamefont {Gaffney}}, \bibinfo {author}
  {\bibfnamefont {M.}~\bibnamefont {Huyse}}, \bibinfo {author} {\bibfnamefont
  {E.}~\bibnamefont {Mogilevskiy}}, \bibinfo {author} {\bibfnamefont
  {S.}~\bibnamefont {Raeder}}, \bibinfo {author} {\bibfnamefont
  {S.}~\bibnamefont {Sels}}, \bibinfo {author} {\bibfnamefont {P.}~\bibnamefont
  {{Van den Bergh}}}, \bibinfo {author} {\bibfnamefont {P.}~\bibnamefont {{Van
  Duppen}}},\ and\ \bibinfo {author} {\bibfnamefont {A.}~\bibnamefont
  {Zadvornaya}},\ }\bibfield  {title} {\bibinfo {title} {{A new in-gas-laser
  ionization and spectroscopy laboratory for off-line studies at KU Leuven}},\
  }\href {https://doi.org/10.1016/j.nimb.2016.02.040} {\bibfield  {journal}
  {\bibinfo  {journal} {Nuclear Instruments and Methods in Physics Research
  Section B: Beam Interactions with Materials and Atoms}\ }\textbf {\bibinfo
  {volume} {376}},\ \bibinfo {pages} {345} (\bibinfo {year}
  {2016})}\BibitemShut {NoStop}%
\bibitem [{\citenamefont {Verlinde}(2021)}]{thesisVerlinde}%
  \BibitemOpen
  \bibfield  {author} {\bibinfo {author} {\bibfnamefont {M.}~\bibnamefont
  {Verlinde}},\ }\emph {\bibinfo {title} {Towards the In-Gas-Jet Laser
  Ionization Spectroscopy of the \ch{^{229}Th} Isomer}},\ \href@noop {} {Ph.D.
  thesis},\ \bibinfo  {school} {KU Leuven} (\bibinfo {year} {2021})\BibitemShut
  {NoStop}%
\bibitem [{\citenamefont {Backe}\ \emph {et~al.}(1998)\citenamefont {Backe},
  \citenamefont {Hies}, \citenamefont {Kunz}, \citenamefont {Lauth},
  \citenamefont {Curtze}, \citenamefont {Schwamb}, \citenamefont {Sewtz},
  \citenamefont {Theobald}, \citenamefont {Zahn}, \citenamefont {Eberhardt},
  \citenamefont {Trautmann}, \citenamefont {Habs}, \citenamefont {Repnow},\
  and\ \citenamefont {Fricke}}]{Backe1998}%
  \BibitemOpen
  \bibfield  {author} {\bibinfo {author} {\bibfnamefont {H.}~\bibnamefont
  {Backe}}, \bibinfo {author} {\bibfnamefont {M.}~\bibnamefont {Hies}},
  \bibinfo {author} {\bibfnamefont {H.}~\bibnamefont {Kunz}}, \bibinfo {author}
  {\bibfnamefont {W.}~\bibnamefont {Lauth}}, \bibinfo {author} {\bibfnamefont
  {O.}~\bibnamefont {Curtze}}, \bibinfo {author} {\bibfnamefont
  {P.}~\bibnamefont {Schwamb}}, \bibinfo {author} {\bibfnamefont
  {M.}~\bibnamefont {Sewtz}}, \bibinfo {author} {\bibfnamefont
  {W.}~\bibnamefont {Theobald}}, \bibinfo {author} {\bibfnamefont
  {R.}~\bibnamefont {Zahn}}, \bibinfo {author} {\bibfnamefont {K.}~\bibnamefont
  {Eberhardt}}, \bibinfo {author} {\bibfnamefont {N.}~\bibnamefont
  {Trautmann}}, \bibinfo {author} {\bibfnamefont {D.}~\bibnamefont {Habs}},
  \bibinfo {author} {\bibfnamefont {R.}~\bibnamefont {Repnow}},\ and\ \bibinfo
  {author} {\bibfnamefont {B.}~\bibnamefont {Fricke}},\ }\bibfield  {title}
  {\bibinfo {title} {{Isotope Shift Measurements for Superdeformed Fission
  Isomeric States}},\ }\href {https://doi.org/10.1103/physrevlett.80.920}
  {\bibfield  {journal} {\bibinfo  {journal} {Physical Review Letters}\
  }\textbf {\bibinfo {volume} {80}},\ \bibinfo {pages} {920} (\bibinfo {year}
  {1998})}\BibitemShut {NoStop}%
\bibitem [{\citenamefont {Kudryavtsev}\ \emph {et~al.}(2013)\citenamefont
  {Kudryavtsev}, \citenamefont {Ferrer}, \citenamefont {Huyse}, \citenamefont
  {{Van den Bergh}},\ and\ \citenamefont {{Van Duppen}}}]{Kudryavtsev2013}%
  \BibitemOpen
  \bibfield  {author} {\bibinfo {author} {\bibfnamefont {Y.}~\bibnamefont
  {Kudryavtsev}}, \bibinfo {author} {\bibfnamefont {R.}~\bibnamefont {Ferrer}},
  \bibinfo {author} {\bibfnamefont {M.}~\bibnamefont {Huyse}}, \bibinfo
  {author} {\bibfnamefont {P.}~\bibnamefont {{Van den Bergh}}},\ and\ \bibinfo
  {author} {\bibfnamefont {P.}~\bibnamefont {{Van Duppen}}},\ }\bibfield
  {title} {\bibinfo {title} {{The in-gas-jet laser ion source: Resonance
  ionization spectroscopy of radioactive atoms in supersonic gas jets}},\
  }\href {https://doi.org/10.1016/j.nimb.2012.12.008} {\bibfield  {journal}
  {\bibinfo  {journal} {Nuclear Instruments and Methods in Physics Research
  Section B: Beam Interactions with Materials and Atoms}\ }\textbf {\bibinfo
  {volume} {297}},\ \bibinfo {pages} {7} (\bibinfo {year} {2013})},\ \Eprint
  {https://arxiv.org/abs/1211.6649} {arXiv:1211.6649} \BibitemShut {NoStop}%
\bibitem [{\citenamefont {Raeder}\ \emph {et~al.}(2016)\citenamefont {Raeder},
  \citenamefont {Bastin}, \citenamefont {Block}, \citenamefont {Creemers},
  \citenamefont {Delahaye}, \citenamefont {Ferrer}, \citenamefont
  {Fl{\'{e}}chard}, \citenamefont {Franchoo}, \citenamefont {Ghys},
  \citenamefont {Gaffney}, \citenamefont {Granados}, \citenamefont {Heinke},
  \citenamefont {Hijazi}, \citenamefont {Huyse}, \citenamefont {Kron},
  \citenamefont {Kudryavtsev}, \citenamefont {Laatiaoui}, \citenamefont
  {Lecesne}, \citenamefont {Luton}, \citenamefont {Moore}, \citenamefont
  {Martinez}, \citenamefont {Mogilevskiy}, \citenamefont {Naubereit},
  \citenamefont {Piot}, \citenamefont {Rothe}, \citenamefont {Savajols},
  \citenamefont {Sels}, \citenamefont {Sonnenschein}, \citenamefont {Traykov},
  \citenamefont {{Van Beveren}}, \citenamefont {{Van Den Bergh}}, \citenamefont
  {{Van Duppen}}, \citenamefont {Wendt},\ and\ \citenamefont
  {Zadvornaya}}]{Raeder2016}%
  \BibitemOpen
  \bibfield  {author} {\bibinfo {author} {\bibfnamefont {S.}~\bibnamefont
  {Raeder}}, \bibinfo {author} {\bibfnamefont {B.}~\bibnamefont {Bastin}},
  \bibinfo {author} {\bibfnamefont {M.}~\bibnamefont {Block}}, \bibinfo
  {author} {\bibfnamefont {P.}~\bibnamefont {Creemers}}, \bibinfo {author}
  {\bibfnamefont {P.}~\bibnamefont {Delahaye}}, \bibinfo {author}
  {\bibfnamefont {R.}~\bibnamefont {Ferrer}}, \bibinfo {author} {\bibfnamefont
  {X.}~\bibnamefont {Fl{\'{e}}chard}}, \bibinfo {author} {\bibfnamefont
  {S.}~\bibnamefont {Franchoo}}, \bibinfo {author} {\bibfnamefont
  {L.}~\bibnamefont {Ghys}}, \bibinfo {author} {\bibfnamefont {L.~P.}\
  \bibnamefont {Gaffney}}, \bibinfo {author} {\bibfnamefont {C.}~\bibnamefont
  {Granados}}, \bibinfo {author} {\bibfnamefont {R.}~\bibnamefont {Heinke}},
  \bibinfo {author} {\bibfnamefont {L.}~\bibnamefont {Hijazi}}, \bibinfo
  {author} {\bibfnamefont {M.}~\bibnamefont {Huyse}}, \bibinfo {author}
  {\bibfnamefont {T.}~\bibnamefont {Kron}}, \bibinfo {author} {\bibfnamefont
  {Y.}~\bibnamefont {Kudryavtsev}}, \bibinfo {author} {\bibfnamefont
  {M.}~\bibnamefont {Laatiaoui}}, \bibinfo {author} {\bibfnamefont
  {N.}~\bibnamefont {Lecesne}}, \bibinfo {author} {\bibfnamefont
  {F.}~\bibnamefont {Luton}}, \bibinfo {author} {\bibfnamefont {I.~D.}\
  \bibnamefont {Moore}}, \bibinfo {author} {\bibfnamefont {Y.}~\bibnamefont
  {Martinez}}, \bibinfo {author} {\bibfnamefont {E.}~\bibnamefont
  {Mogilevskiy}}, \bibinfo {author} {\bibfnamefont {P.}~\bibnamefont
  {Naubereit}}, \bibinfo {author} {\bibfnamefont {J.}~\bibnamefont {Piot}},
  \bibinfo {author} {\bibfnamefont {S.}~\bibnamefont {Rothe}}, \bibinfo
  {author} {\bibfnamefont {H.}~\bibnamefont {Savajols}}, \bibinfo {author}
  {\bibfnamefont {S.}~\bibnamefont {Sels}}, \bibinfo {author} {\bibfnamefont
  {V.}~\bibnamefont {Sonnenschein}}, \bibinfo {author} {\bibfnamefont
  {E.}~\bibnamefont {Traykov}}, \bibinfo {author} {\bibfnamefont
  {C.}~\bibnamefont {{Van Beveren}}}, \bibinfo {author} {\bibfnamefont
  {P.}~\bibnamefont {{Van Den Bergh}}}, \bibinfo {author} {\bibfnamefont
  {P.}~\bibnamefont {{Van Duppen}}}, \bibinfo {author} {\bibfnamefont
  {K.}~\bibnamefont {Wendt}},\ and\ \bibinfo {author} {\bibfnamefont
  {A.}~\bibnamefont {Zadvornaya}},\ }\bibfield  {title} {\bibinfo {title}
  {{Developments towards in-gas-jet laser spectroscopy studies of actinium
  isotopes at LISOL}},\ }\href {https://doi.org/10.1016/j.nimb.2015.12.014}
  {\bibfield  {journal} {\bibinfo  {journal} {Nuclear Instruments and Methods
  in Physics Research, Section B: Beam Interactions with Materials and Atoms}\
  }\textbf {\bibinfo {volume} {376}},\ \bibinfo {pages} {382} (\bibinfo {year}
  {2016})}\BibitemShut {NoStop}%
\bibitem [{\citenamefont {Ferrer}\ \emph {et~al.}(2017)\citenamefont {Ferrer},
  \citenamefont {Barzakh}, \citenamefont {Bastin}, \citenamefont {Beerwerth},
  \citenamefont {Block}, \citenamefont {Creemers}, \citenamefont {Grawe},
  \citenamefont {de~Groote}, \citenamefont {Delahaye}, \citenamefont
  {Fl{\'{e}}chard}, \citenamefont {Franchoo}, \citenamefont {Fritzsche},
  \citenamefont {Gaffney}, \citenamefont {Ghys}, \citenamefont {Gins},
  \citenamefont {Granados}, \citenamefont {Heinke}, \citenamefont {Hijazi},
  \citenamefont {Huyse}, \citenamefont {Kron}, \citenamefont {Kudryavtsev},
  \citenamefont {Laatiaoui}, \citenamefont {Lecesne}, \citenamefont {Loiselet},
  \citenamefont {Lutton}, \citenamefont {Moore}, \citenamefont
  {Mart{\'{i}}nez}, \citenamefont {Mogilevskiy}, \citenamefont {Naubereit},
  \citenamefont {Piot}, \citenamefont {Raeder}, \citenamefont {Rothe},
  \citenamefont {Savajols}, \citenamefont {Sels}, \citenamefont {Sonnenschein},
  \citenamefont {Thomas}, \citenamefont {Traykov}, \citenamefont {{Van
  Beveren}}, \citenamefont {{Van den Bergh}}, \citenamefont {{Van Duppen}},
  \citenamefont {Wendt},\ and\ \citenamefont {Zadvornaya}}]{Ferrer2017}%
  \BibitemOpen
  \bibfield  {author} {\bibinfo {author} {\bibfnamefont {R.}~\bibnamefont
  {Ferrer}}, \bibinfo {author} {\bibfnamefont {A.}~\bibnamefont {Barzakh}},
  \bibinfo {author} {\bibfnamefont {B.}~\bibnamefont {Bastin}}, \bibinfo
  {author} {\bibfnamefont {R.}~\bibnamefont {Beerwerth}}, \bibinfo {author}
  {\bibfnamefont {M.}~\bibnamefont {Block}}, \bibinfo {author} {\bibfnamefont
  {P.}~\bibnamefont {Creemers}}, \bibinfo {author} {\bibfnamefont
  {H.}~\bibnamefont {Grawe}}, \bibinfo {author} {\bibfnamefont
  {R.}~\bibnamefont {de~Groote}}, \bibinfo {author} {\bibfnamefont
  {P.}~\bibnamefont {Delahaye}}, \bibinfo {author} {\bibfnamefont
  {X.}~\bibnamefont {Fl{\'{e}}chard}}, \bibinfo {author} {\bibfnamefont
  {S.}~\bibnamefont {Franchoo}}, \bibinfo {author} {\bibfnamefont
  {S.}~\bibnamefont {Fritzsche}}, \bibinfo {author} {\bibfnamefont {L.~P.}\
  \bibnamefont {Gaffney}}, \bibinfo {author} {\bibfnamefont {L.}~\bibnamefont
  {Ghys}}, \bibinfo {author} {\bibfnamefont {W.}~\bibnamefont {Gins}}, \bibinfo
  {author} {\bibfnamefont {C.}~\bibnamefont {Granados}}, \bibinfo {author}
  {\bibfnamefont {R.}~\bibnamefont {Heinke}}, \bibinfo {author} {\bibfnamefont
  {L.}~\bibnamefont {Hijazi}}, \bibinfo {author} {\bibfnamefont
  {M.}~\bibnamefont {Huyse}}, \bibinfo {author} {\bibfnamefont
  {T.}~\bibnamefont {Kron}}, \bibinfo {author} {\bibfnamefont {Y.}~\bibnamefont
  {Kudryavtsev}}, \bibinfo {author} {\bibfnamefont {M.}~\bibnamefont
  {Laatiaoui}}, \bibinfo {author} {\bibfnamefont {N.}~\bibnamefont {Lecesne}},
  \bibinfo {author} {\bibfnamefont {M.}~\bibnamefont {Loiselet}}, \bibinfo
  {author} {\bibfnamefont {F.}~\bibnamefont {Lutton}}, \bibinfo {author}
  {\bibfnamefont {I.~D.}\ \bibnamefont {Moore}}, \bibinfo {author}
  {\bibfnamefont {Y.}~\bibnamefont {Mart{\'{i}}nez}}, \bibinfo {author}
  {\bibfnamefont {E.}~\bibnamefont {Mogilevskiy}}, \bibinfo {author}
  {\bibfnamefont {P.}~\bibnamefont {Naubereit}}, \bibinfo {author}
  {\bibfnamefont {J.}~\bibnamefont {Piot}}, \bibinfo {author} {\bibfnamefont
  {S.}~\bibnamefont {Raeder}}, \bibinfo {author} {\bibfnamefont
  {S.}~\bibnamefont {Rothe}}, \bibinfo {author} {\bibfnamefont
  {H.}~\bibnamefont {Savajols}}, \bibinfo {author} {\bibfnamefont
  {S.}~\bibnamefont {Sels}}, \bibinfo {author} {\bibfnamefont {V.}~\bibnamefont
  {Sonnenschein}}, \bibinfo {author} {\bibfnamefont {J.-C.}\ \bibnamefont
  {Thomas}}, \bibinfo {author} {\bibfnamefont {E.}~\bibnamefont {Traykov}},
  \bibinfo {author} {\bibfnamefont {C.}~\bibnamefont {{Van Beveren}}}, \bibinfo
  {author} {\bibfnamefont {P.}~\bibnamefont {{Van den Bergh}}}, \bibinfo
  {author} {\bibfnamefont {P.}~\bibnamefont {{Van Duppen}}}, \bibinfo {author}
  {\bibfnamefont {K.}~\bibnamefont {Wendt}},\ and\ \bibinfo {author}
  {\bibfnamefont {A.}~\bibnamefont {Zadvornaya}},\ }\bibfield  {title}
  {\bibinfo {title} {{Towards high-resolution laser ionization spectroscopy of
  the heaviest elements in supersonic gas jet expansion}},\ }\href
  {https://doi.org/10.1038/ncomms14520} {\bibfield  {journal} {\bibinfo
  {journal} {Nature Communications}\ }\textbf {\bibinfo {volume} {8}},\
  \bibinfo {pages} {14520} (\bibinfo {year} {2017})}\BibitemShut {NoStop}%
\bibitem [{\citenamefont {Ferrer}\ \emph {et~al.}(2021)\citenamefont {Ferrer},
  \citenamefont {Verlinde}, \citenamefont {Verstraelen}, \citenamefont
  {Claessens}, \citenamefont {Huyse}, \citenamefont {Kraemer}, \citenamefont
  {Kudryavtsev}, \citenamefont {Romans}, \citenamefont {{Van den Bergh}},
  \citenamefont {{Van Duppen}}, \citenamefont {Zadvornaya}, \citenamefont
  {Chazot}, \citenamefont {Grossir}, \citenamefont {Kalikmanov}, \citenamefont
  {Nabuurs},\ and\ \citenamefont {Reynaerts}}]{Ferrer2021}%
  \BibitemOpen
  \bibfield  {author} {\bibinfo {author} {\bibfnamefont {R.}~\bibnamefont
  {Ferrer}}, \bibinfo {author} {\bibfnamefont {M.}~\bibnamefont {Verlinde}},
  \bibinfo {author} {\bibfnamefont {E.}~\bibnamefont {Verstraelen}}, \bibinfo
  {author} {\bibfnamefont {A.}~\bibnamefont {Claessens}}, \bibinfo {author}
  {\bibfnamefont {M.}~\bibnamefont {Huyse}}, \bibinfo {author} {\bibfnamefont
  {S.}~\bibnamefont {Kraemer}}, \bibinfo {author} {\bibfnamefont
  {Y.}~\bibnamefont {Kudryavtsev}}, \bibinfo {author} {\bibfnamefont
  {J.}~\bibnamefont {Romans}}, \bibinfo {author} {\bibfnamefont
  {P.}~\bibnamefont {{Van den Bergh}}}, \bibinfo {author} {\bibfnamefont
  {P.}~\bibnamefont {{Van Duppen}}}, \bibinfo {author} {\bibfnamefont
  {A.}~\bibnamefont {Zadvornaya}}, \bibinfo {author} {\bibfnamefont
  {O.}~\bibnamefont {Chazot}}, \bibinfo {author} {\bibfnamefont
  {G.}~\bibnamefont {Grossir}}, \bibinfo {author} {\bibfnamefont {V.~I.}\
  \bibnamefont {Kalikmanov}}, \bibinfo {author} {\bibfnamefont
  {M.}~\bibnamefont {Nabuurs}},\ and\ \bibinfo {author} {\bibfnamefont
  {D.}~\bibnamefont {Reynaerts}},\ }\bibfield  {title} {\bibinfo {title}
  {{Hypersonic nozzle for laser-spectroscopy studies at 17 K characterized by
  resonance-ionization-spectroscopy-based flow mapping}},\ }\href
  {https://doi.org/10.1103/PhysRevResearch.3.043041} {\bibfield  {journal}
  {\bibinfo  {journal} {Physical Review Research}\ }\textbf {\bibinfo {volume}
  {3}},\ \bibinfo {pages} {043041} (\bibinfo {year} {2021})}\BibitemShut
  {NoStop}%
\bibitem [{\citenamefont {Lantis}\ \emph {et~al.}(2024)\citenamefont {Lantis},
  \citenamefont {Claessens}, \citenamefont {M{\"{u}}nzberg}, \citenamefont
  {Auler}, \citenamefont {Block}, \citenamefont {Chhetri}, \citenamefont
  {D{\"{u}}llmann}, \citenamefont {Ferrer}, \citenamefont {Giacoppo},
  \citenamefont {Guti{\'{e}}rrez}, \citenamefont {Ivandikov}, \citenamefont
  {Kaleja}, \citenamefont {Kieck}, \citenamefont {Kim}, \citenamefont
  {Laatiaoui}, \citenamefont {Lecesne}, \citenamefont {Manea}, \citenamefont
  {Nothhelfer}, \citenamefont {Raeder}, \citenamefont {Romans}, \citenamefont
  {Romero-Romero}, \citenamefont {{De Roubin}}, \citenamefont {Savajols},
  \citenamefont {Sels}, \citenamefont {Stemmler}, \citenamefont {{Van Duppen}},
  \citenamefont {Walther}, \citenamefont {Warbinek}, \citenamefont {Wendt},
  \citenamefont {Yakushev},\ and\ \citenamefont {Zadvornaya}}]{Lantis2024}%
  \BibitemOpen
  \bibfield  {author} {\bibinfo {author} {\bibfnamefont {J.}~\bibnamefont
  {Lantis}}, \bibinfo {author} {\bibfnamefont {A.}~\bibnamefont {Claessens}},
  \bibinfo {author} {\bibfnamefont {D.}~\bibnamefont {M{\"{u}}nzberg}},
  \bibinfo {author} {\bibfnamefont {J.}~\bibnamefont {Auler}}, \bibinfo
  {author} {\bibfnamefont {M.}~\bibnamefont {Block}}, \bibinfo {author}
  {\bibfnamefont {P.}~\bibnamefont {Chhetri}}, \bibinfo {author} {\bibfnamefont
  {{\relax Ch}.~E.}\ \bibnamefont {D{\"{u}}llmann}}, \bibinfo {author}
  {\bibfnamefont {R.}~\bibnamefont {Ferrer}}, \bibinfo {author} {\bibfnamefont
  {F.}~\bibnamefont {Giacoppo}}, \bibinfo {author} {\bibfnamefont {M.~J.}\
  \bibnamefont {Guti{\'{e}}rrez}}, \bibinfo {author} {\bibfnamefont
  {F.}~\bibnamefont {Ivandikov}}, \bibinfo {author} {\bibfnamefont
  {O.}~\bibnamefont {Kaleja}}, \bibinfo {author} {\bibfnamefont
  {T.}~\bibnamefont {Kieck}}, \bibinfo {author} {\bibfnamefont
  {E.}~\bibnamefont {Kim}}, \bibinfo {author} {\bibfnamefont {M.}~\bibnamefont
  {Laatiaoui}}, \bibinfo {author} {\bibfnamefont {N.}~\bibnamefont {Lecesne}},
  \bibinfo {author} {\bibfnamefont {V.}~\bibnamefont {Manea}}, \bibinfo
  {author} {\bibfnamefont {S.}~\bibnamefont {Nothhelfer}}, \bibinfo {author}
  {\bibfnamefont {S.}~\bibnamefont {Raeder}}, \bibinfo {author} {\bibfnamefont
  {J.}~\bibnamefont {Romans}}, \bibinfo {author} {\bibfnamefont
  {E.}~\bibnamefont {Romero-Romero}}, \bibinfo {author} {\bibfnamefont
  {A.}~\bibnamefont {{De Roubin}}}, \bibinfo {author} {\bibfnamefont
  {H.}~\bibnamefont {Savajols}}, \bibinfo {author} {\bibfnamefont
  {S.}~\bibnamefont {Sels}}, \bibinfo {author} {\bibfnamefont {M.}~\bibnamefont
  {Stemmler}}, \bibinfo {author} {\bibfnamefont {P.}~\bibnamefont {{Van
  Duppen}}}, \bibinfo {author} {\bibfnamefont {T.}~\bibnamefont {Walther}},
  \bibinfo {author} {\bibfnamefont {J.}~\bibnamefont {Warbinek}}, \bibinfo
  {author} {\bibfnamefont {K.}~\bibnamefont {Wendt}}, \bibinfo {author}
  {\bibfnamefont {A.}~\bibnamefont {Yakushev}},\ and\ \bibinfo {author}
  {\bibfnamefont {A.}~\bibnamefont {Zadvornaya}},\ }\bibfield  {title}
  {\bibinfo {title} {{In-gas-jet laser spectroscopy of $^{254}$No with
  JetRIS}},\ }\href {https://doi.org/10.1103/PhysRevResearch.6.023318}
  {\bibfield  {journal} {\bibinfo  {journal} {Physical Review Research}\
  }\textbf {\bibinfo {volume} {6}},\ \bibinfo {pages} {1} (\bibinfo {year}
  {2024})}\BibitemShut {NoStop}%
\bibitem [{\citenamefont {Claessens}(2024)}]{thesisClaessens}%
  \BibitemOpen
  \bibfield  {author} {\bibinfo {author} {\bibfnamefont {A.}~\bibnamefont
  {Claessens}},\ }\emph {\bibinfo {title} {Laser ionization spectroscopy of
  $^{254}$No and $^{229}$Th in hypersonic gas jets}},\ \href@noop {} {Ph.D.
  thesis},\ \bibinfo  {school} {KU Leuven} (\bibinfo {year} {2024})\BibitemShut
  {NoStop}%
\bibitem [{\citenamefont {Verlinde}\ \emph {et~al.}(2020)\citenamefont
  {Verlinde}, \citenamefont {Ferrer}, \citenamefont {Claessens}, \citenamefont
  {Granados}, \citenamefont {Kraemer}, \citenamefont {Kudryavtsev},
  \citenamefont {Li}, \citenamefont {{Van den Bergh}}, \citenamefont {{Van
  Duppen}},\ and\ \citenamefont {Verstraelen}}]{Verlinde2020}%
  \BibitemOpen
  \bibfield  {author} {\bibinfo {author} {\bibfnamefont {M.}~\bibnamefont
  {Verlinde}}, \bibinfo {author} {\bibfnamefont {R.}~\bibnamefont {Ferrer}},
  \bibinfo {author} {\bibfnamefont {A.}~\bibnamefont {Claessens}}, \bibinfo
  {author} {\bibfnamefont {C.~A.}\ \bibnamefont {Granados}}, \bibinfo {author}
  {\bibfnamefont {S.}~\bibnamefont {Kraemer}}, \bibinfo {author} {\bibfnamefont
  {Y.}~\bibnamefont {Kudryavtsev}}, \bibinfo {author} {\bibfnamefont
  {D.}~\bibnamefont {Li}}, \bibinfo {author} {\bibfnamefont {P.}~\bibnamefont
  {{Van den Bergh}}}, \bibinfo {author} {\bibfnamefont {P.}~\bibnamefont {{Van
  Duppen}}},\ and\ \bibinfo {author} {\bibfnamefont {E.}~\bibnamefont
  {Verstraelen}},\ }\bibfield  {title} {\bibinfo {title}
  {{Single-longitudinal-mode pumped pulsed-dye amplifier for high-resolution
  laser spectroscopy}},\ }\href {https://doi.org/10.1063/5.0017985} {\bibfield
  {journal} {\bibinfo  {journal} {Review of Scientific Instruments}\ }\textbf
  {\bibinfo {volume} {91}},\ \bibinfo {pages} {103002} (\bibinfo {year}
  {2020})}\BibitemShut {NoStop}%
\bibitem [{\citenamefont {Herrera-Sancho}\ \emph {et~al.}(2012)\citenamefont
  {Herrera-Sancho}, \citenamefont {Okhapkin}, \citenamefont {Zimmermann},
  \citenamefont {Tamm}, \citenamefont {Peik}, \citenamefont {Taichenachev},
  \citenamefont {Yudin},\ and\ \citenamefont
  {G{\l}owacki}}]{Herrera-Sancho2012}%
  \BibitemOpen
  \bibfield  {author} {\bibinfo {author} {\bibfnamefont {O.~A.}\ \bibnamefont
  {Herrera-Sancho}}, \bibinfo {author} {\bibfnamefont {M.~V.}\ \bibnamefont
  {Okhapkin}}, \bibinfo {author} {\bibfnamefont {K.}~\bibnamefont
  {Zimmermann}}, \bibinfo {author} {\bibfnamefont {C.}~\bibnamefont {Tamm}},
  \bibinfo {author} {\bibfnamefont {E.}~\bibnamefont {Peik}}, \bibinfo {author}
  {\bibfnamefont {A.~V.}\ \bibnamefont {Taichenachev}}, \bibinfo {author}
  {\bibfnamefont {V.~I.}\ \bibnamefont {Yudin}},\ and\ \bibinfo {author}
  {\bibfnamefont {P.}~\bibnamefont {G{\l}owacki}},\ }\bibfield  {title}
  {\bibinfo {title} {{Two-photon laser excitation of trapped \ch{^{232}Th^+}
  ions via the 402-nm resonance line}},\ }\href
  {https://doi.org/10.1103/PhysRevA.85.033402} {\bibfield  {journal} {\bibinfo
  {journal} {Physical Review A - Atomic, Molecular, and Optical Physics}\
  }\textbf {\bibinfo {volume} {85}},\ \bibinfo {pages} {1} (\bibinfo {year}
  {2012})}\BibitemShut {NoStop}%
\bibitem [{\citenamefont {Claessens}\ \emph {et~al.}(2023)\citenamefont
  {Claessens}, \citenamefont {Ivandikov}, \citenamefont {Bara}, \citenamefont
  {Chhetri}, \citenamefont {Dragoun}, \citenamefont {D{\"{u}}llmann},
  \citenamefont {Elskens}, \citenamefont {Ferrer}, \citenamefont {Kraemer},
  \citenamefont {Kudryavtsev}, \citenamefont {Renisch}, \citenamefont {Romans},
  \citenamefont {Rosecker}, \citenamefont {de~Roubin}, \citenamefont {Schumm},
  \citenamefont {{Van den Bergh}},\ and\ \citenamefont {{Van
  Duppen}}}]{Claessens2023}%
  \BibitemOpen
  \bibfield  {author} {\bibinfo {author} {\bibfnamefont {A.}~\bibnamefont
  {Claessens}}, \bibinfo {author} {\bibfnamefont {F.}~\bibnamefont
  {Ivandikov}}, \bibinfo {author} {\bibfnamefont {S.}~\bibnamefont {Bara}},
  \bibinfo {author} {\bibfnamefont {P.}~\bibnamefont {Chhetri}}, \bibinfo
  {author} {\bibfnamefont {A.}~\bibnamefont {Dragoun}}, \bibinfo {author}
  {\bibfnamefont {{\relax Ch}.~E.}\ \bibnamefont {D{\"{u}}llmann}}, \bibinfo
  {author} {\bibfnamefont {Y.}~\bibnamefont {Elskens}}, \bibinfo {author}
  {\bibfnamefont {R.}~\bibnamefont {Ferrer}}, \bibinfo {author} {\bibfnamefont
  {S.}~\bibnamefont {Kraemer}}, \bibinfo {author} {\bibfnamefont
  {Y.}~\bibnamefont {Kudryavtsev}}, \bibinfo {author} {\bibfnamefont
  {D.}~\bibnamefont {Renisch}}, \bibinfo {author} {\bibfnamefont
  {J.}~\bibnamefont {Romans}}, \bibinfo {author} {\bibfnamefont
  {V.}~\bibnamefont {Rosecker}}, \bibinfo {author} {\bibfnamefont
  {A.}~\bibnamefont {de~Roubin}}, \bibinfo {author} {\bibfnamefont
  {T.}~\bibnamefont {Schumm}}, \bibinfo {author} {\bibfnamefont
  {P.}~\bibnamefont {{Van den Bergh}}},\ and\ \bibinfo {author} {\bibfnamefont
  {P.}~\bibnamefont {{Van Duppen}}},\ }\bibfield  {title} {\bibinfo {title}
  {{Laser ionization scheme development for in-gas-jet spectroscopy studies of
  Th+}},\ }\href {https://doi.org/10.1016/j.nimb.2023.04.019} {\bibfield
  {journal} {\bibinfo  {journal} {Nuclear Instruments and Methods in Physics
  Research Section B: Beam Interactions with Materials and Atoms}\ }\textbf
  {\bibinfo {volume} {540}},\ \bibinfo {pages} {224} (\bibinfo {year}
  {2023})}\BibitemShut {NoStop}%
\bibitem [{\citenamefont {Redman}\ and\ \citenamefont
  {Nave}(2014)}]{Redman2014}%
  \BibitemOpen
  \bibfield  {author} {\bibinfo {author} {\bibfnamefont {S.~L.}\ \bibnamefont
  {Redman}}\ and\ \bibinfo {author} {\bibfnamefont {C.~J.}\ \bibnamefont
  {Nave}, \bibfnamefont {.and~Sansonetti}},\ }\bibfield  {title} {\bibinfo
  {title} {{The spectrum of thorium from 250 nm to 5500 nm: Ritz wavelengths
  and optimized energy levels}},\ }\bibfield  {journal} {\bibinfo  {journal}
  {Astrophysical Journal, Supplement Series}\ }\textbf {\bibinfo {volume}
  {211}},\ \href {https://doi.org/10.1088/0067-0049/211/1/4}
  {10.1088/0067-0049/211/1/4} (\bibinfo {year} {2014}),\ \Eprint
  {https://arxiv.org/abs/1308.5229} {arXiv:1308.5229} \BibitemShut {NoStop}%
\bibitem [{\citenamefont {Kramida}\ \emph {et~al.}(2023)\citenamefont
  {Kramida}, \citenamefont {{Yu.~Ralchenko}}, \citenamefont {Reader},\ and\
  \citenamefont {{and NIST ASD Team}}}]{NIST_ASD_ThI}%
  \BibitemOpen
  \bibfield  {author} {\bibinfo {author} {\bibfnamefont {A.}~\bibnamefont
  {Kramida}}, \bibinfo {author} {\bibnamefont {{Yu.~Ralchenko}}}, \bibinfo
  {author} {\bibfnamefont {J.}~\bibnamefont {Reader}},\ and\ \bibinfo {author}
  {\bibnamefont {{and NIST ASD Team}}},\ }\href@noop {} {}\bibinfo
  {howpublished} {{NIST Atomic Spectra Database (ver. 5.11), [Online]. \\
  Available: {\tt{https://physics.nist.gov/asd}} [2024, January 20]. National
  Institute of Standards and Technology, Gaithersburg, MD.}} (\bibinfo {year}
  {2023})\BibitemShut {NoStop}%
\bibitem [{\citenamefont {Rothe}\ \emph {et~al.}(2013)\citenamefont {Rothe},
  \citenamefont {Andreyev}, \citenamefont {Antalic}, \citenamefont
  {Borschevsky}, \citenamefont {Capponi}, \citenamefont {Cocolios},
  \citenamefont {{De Witte}}, \citenamefont {Eliav}, \citenamefont {Fedorov},
  \citenamefont {Fedosseev}, \citenamefont {Fink}, \citenamefont {Fritzsche},
  \citenamefont {Ghys}, \citenamefont {Huyse}, \citenamefont {Imai},
  \citenamefont {Kaldor}, \citenamefont {Kudryavtsev}, \citenamefont
  {K{\"{o}}ster}, \citenamefont {Lane}, \citenamefont {Lassen}, \citenamefont
  {Liberati}, \citenamefont {Lynch}, \citenamefont {Marsh}, \citenamefont
  {Nishio}, \citenamefont {Pauwels}, \citenamefont {Pershina}, \citenamefont
  {Popescu}, \citenamefont {Procter}, \citenamefont {Radulov}, \citenamefont
  {Raeder}, \citenamefont {Rajabali}, \citenamefont {Rapisarda}, \citenamefont
  {Rossel}, \citenamefont {Sandhu}, \citenamefont {Seliverstov}, \citenamefont
  {Sj{\"{o}}din}, \citenamefont {{Van Den Bergh}}, \citenamefont {{Van
  Duppen}}, \citenamefont {Venhart}, \citenamefont {Wakabayashi},\ and\
  \citenamefont {Wendt}}]{Rothe2013}%
  \BibitemOpen
  \bibfield  {author} {\bibinfo {author} {\bibfnamefont {S.}~\bibnamefont
  {Rothe}}, \bibinfo {author} {\bibfnamefont {A.~N.}\ \bibnamefont {Andreyev}},
  \bibinfo {author} {\bibfnamefont {S.}~\bibnamefont {Antalic}}, \bibinfo
  {author} {\bibfnamefont {A.}~\bibnamefont {Borschevsky}}, \bibinfo {author}
  {\bibfnamefont {L.}~\bibnamefont {Capponi}}, \bibinfo {author} {\bibfnamefont
  {T.~E.}\ \bibnamefont {Cocolios}}, \bibinfo {author} {\bibfnamefont
  {H.}~\bibnamefont {{De Witte}}}, \bibinfo {author} {\bibfnamefont
  {E.}~\bibnamefont {Eliav}}, \bibinfo {author} {\bibfnamefont {D.~V.}\
  \bibnamefont {Fedorov}}, \bibinfo {author} {\bibfnamefont {V.~N.}\
  \bibnamefont {Fedosseev}}, \bibinfo {author} {\bibfnamefont {D.~A.}\
  \bibnamefont {Fink}}, \bibinfo {author} {\bibfnamefont {S.}~\bibnamefont
  {Fritzsche}}, \bibinfo {author} {\bibfnamefont {L.}~\bibnamefont {Ghys}},
  \bibinfo {author} {\bibfnamefont {M.}~\bibnamefont {Huyse}}, \bibinfo
  {author} {\bibfnamefont {N.}~\bibnamefont {Imai}}, \bibinfo {author}
  {\bibfnamefont {U.}~\bibnamefont {Kaldor}}, \bibinfo {author} {\bibfnamefont
  {Y.}~\bibnamefont {Kudryavtsev}}, \bibinfo {author} {\bibfnamefont
  {U.}~\bibnamefont {K{\"{o}}ster}}, \bibinfo {author} {\bibfnamefont {J.~F.}\
  \bibnamefont {Lane}}, \bibinfo {author} {\bibfnamefont {J.}~\bibnamefont
  {Lassen}}, \bibinfo {author} {\bibfnamefont {V.}~\bibnamefont {Liberati}},
  \bibinfo {author} {\bibfnamefont {K.~M.}\ \bibnamefont {Lynch}}, \bibinfo
  {author} {\bibfnamefont {B.~A.}\ \bibnamefont {Marsh}}, \bibinfo {author}
  {\bibfnamefont {K.}~\bibnamefont {Nishio}}, \bibinfo {author} {\bibfnamefont
  {D.}~\bibnamefont {Pauwels}}, \bibinfo {author} {\bibfnamefont
  {V.}~\bibnamefont {Pershina}}, \bibinfo {author} {\bibfnamefont
  {L.}~\bibnamefont {Popescu}}, \bibinfo {author} {\bibfnamefont {T.~J.}\
  \bibnamefont {Procter}}, \bibinfo {author} {\bibfnamefont {D.}~\bibnamefont
  {Radulov}}, \bibinfo {author} {\bibfnamefont {S.}~\bibnamefont {Raeder}},
  \bibinfo {author} {\bibfnamefont {M.~M.}\ \bibnamefont {Rajabali}}, \bibinfo
  {author} {\bibfnamefont {E.}~\bibnamefont {Rapisarda}}, \bibinfo {author}
  {\bibfnamefont {R.~E.}\ \bibnamefont {Rossel}}, \bibinfo {author}
  {\bibfnamefont {K.}~\bibnamefont {Sandhu}}, \bibinfo {author} {\bibfnamefont
  {M.~D.}\ \bibnamefont {Seliverstov}}, \bibinfo {author} {\bibfnamefont
  {A.~M.}\ \bibnamefont {Sj{\"{o}}din}}, \bibinfo {author} {\bibfnamefont
  {P.}~\bibnamefont {{Van Den Bergh}}}, \bibinfo {author} {\bibfnamefont
  {P.}~\bibnamefont {{Van Duppen}}}, \bibinfo {author} {\bibfnamefont
  {M.}~\bibnamefont {Venhart}}, \bibinfo {author} {\bibfnamefont
  {Y.}~\bibnamefont {Wakabayashi}},\ and\ \bibinfo {author} {\bibfnamefont
  {K.~D.}\ \bibnamefont {Wendt}},\ }\bibfield  {title} {\bibinfo {title}
  {Measurement of the first ionization potential of astatine by laser
  ionization spectroscopy},\ }\href {https://doi.org/10.1038/ncomms2819}
  {\bibfield  {journal} {\bibinfo  {journal} {Nature Communications}\ }\textbf
  {\bibinfo {volume} {4}},\ \bibinfo {pages} {1} (\bibinfo {year}
  {2013})}\BibitemShut {NoStop}%
\bibitem [{\citenamefont {Rothe}(2012)}]{thesisRothe}%
  \BibitemOpen
  \bibfield  {author} {\bibinfo {author} {\bibfnamefont {S.}~\bibnamefont
  {Rothe}},\ }\emph {\bibinfo {title} {An all-solid state laser system for the
  laser ion source RILIS and in-source laser spectroscopy of astatine at
  ISOLDE/CERN}},\ \href@noop {} {Ph.D. thesis},\ \bibinfo  {school} {JGU Mainz}
  (\bibinfo {year} {2012})\BibitemShut {NoStop}%
\bibitem [{\citenamefont {Vascon}\ \emph {et~al.}(2012)\citenamefont {Vascon},
  \citenamefont {Santi}, \citenamefont {Isse}, \citenamefont {Reich},
  \citenamefont {Drebert}, \citenamefont {Christ}, \citenamefont
  {D{\"{u}}llmann},\ and\ \citenamefont {Eberhardt}}]{Vascon2012}%
  \BibitemOpen
  \bibfield  {author} {\bibinfo {author} {\bibfnamefont {A.}~\bibnamefont
  {Vascon}}, \bibinfo {author} {\bibfnamefont {S.}~\bibnamefont {Santi}},
  \bibinfo {author} {\bibfnamefont {A.~A.}\ \bibnamefont {Isse}}, \bibinfo
  {author} {\bibfnamefont {T.}~\bibnamefont {Reich}}, \bibinfo {author}
  {\bibfnamefont {J.}~\bibnamefont {Drebert}}, \bibinfo {author} {\bibfnamefont
  {H.}~\bibnamefont {Christ}}, \bibinfo {author} {\bibfnamefont {{\relax
  Ch}.~E.}\ \bibnamefont {D{\"{u}}llmann}},\ and\ \bibinfo {author}
  {\bibfnamefont {K.}~\bibnamefont {Eberhardt}},\ }\bibfield  {title} {\bibinfo
  {title} {Elucidation of constant current density molecular plating},\ }\href
  {https://doi.org/10.1016/j.nima.2012.08.072} {\bibfield  {journal} {\bibinfo
  {journal} {Nuclear Instruments and Methods in Physics Research, Section A:
  Accelerators, Spectrometers, Detectors and Associated Equipment}\ }\textbf
  {\bibinfo {volume} {696}},\ \bibinfo {pages} {180} (\bibinfo {year}
  {2012})}\BibitemShut {NoStop}%
\bibitem [{\citenamefont {Haas}\ \emph {et~al.}(2020)\citenamefont {Haas},
  \citenamefont {Hufnagel}, \citenamefont {Abrosimov}, \citenamefont
  {D{\"{u}}llmann}, \citenamefont {Krupp}, \citenamefont {Mokry}, \citenamefont
  {Renisch}, \citenamefont {Runke},\ and\ \citenamefont {Scherer}}]{Haas2020}%
  \BibitemOpen
  \bibfield  {author} {\bibinfo {author} {\bibfnamefont {R.}~\bibnamefont
  {Haas}}, \bibinfo {author} {\bibfnamefont {M.}~\bibnamefont {Hufnagel}},
  \bibinfo {author} {\bibfnamefont {R.}~\bibnamefont {Abrosimov}}, \bibinfo
  {author} {\bibfnamefont {{\relax Ch}.~E.}\ \bibnamefont {D{\"{u}}llmann}},
  \bibinfo {author} {\bibfnamefont {D.}~\bibnamefont {Krupp}}, \bibinfo
  {author} {\bibfnamefont {{\relax Ch}.}~\bibnamefont {Mokry}}, \bibinfo
  {author} {\bibfnamefont {D.}~\bibnamefont {Renisch}}, \bibinfo {author}
  {\bibfnamefont {J.}~\bibnamefont {Runke}},\ and\ \bibinfo {author}
  {\bibfnamefont {U.~W.}\ \bibnamefont {Scherer}},\ }\bibfield  {title}
  {\bibinfo {title} {{Alpha spectrometric characterization of thin \ch{^{233}U}
  sources for \ch{^{229(m)}Th} production}},\ }\href
  {https://doi.org/10.1515/ract-2020-0032} {\bibfield  {journal} {\bibinfo
  {journal} {Radiochimica Acta}\ }\textbf {\bibinfo {volume} {108}},\ \bibinfo
  {pages} {923} (\bibinfo {year} {2020})}\BibitemShut {NoStop}%
\bibitem [{rep(2021)}]{reportSourcesMainz}%
  \BibitemOpen
  \href@noop {} {\emph {\bibinfo {title} {Uranium-233 recoil ion sources from
  JGU Mainz for KU Leuven}}},\ \bibinfo {type} {Progress report}\ (\bibinfo
  {institution} {JGU Mainz},\ \bibinfo {year} {2021})\BibitemShut {NoStop}%
\bibitem [{\citenamefont {K{\"{o}}hler}\ \emph {et~al.}(1997)\citenamefont
  {K{\"{o}}hler}, \citenamefont {Dei{\ss}enberger}, \citenamefont {Eberhardt},
  \citenamefont {Erdmann}, \citenamefont {Herrmann}, \citenamefont {Huber},
  \citenamefont {Kratz}, \citenamefont {Nunnemann}, \citenamefont {Passler},
  \citenamefont {Rao}, \citenamefont {Riegel}, \citenamefont {Trautmann},\ and\
  \citenamefont {Wendt}}]{Kohler1997}%
  \BibitemOpen
  \bibfield  {author} {\bibinfo {author} {\bibfnamefont {S.}~\bibnamefont
  {K{\"{o}}hler}}, \bibinfo {author} {\bibfnamefont {R.}~\bibnamefont
  {Dei{\ss}enberger}}, \bibinfo {author} {\bibfnamefont {K.}~\bibnamefont
  {Eberhardt}}, \bibinfo {author} {\bibfnamefont {N.}~\bibnamefont {Erdmann}},
  \bibinfo {author} {\bibfnamefont {G.}~\bibnamefont {Herrmann}}, \bibinfo
  {author} {\bibfnamefont {G.}~\bibnamefont {Huber}}, \bibinfo {author}
  {\bibfnamefont {J.~V.}\ \bibnamefont {Kratz}}, \bibinfo {author}
  {\bibfnamefont {M.}~\bibnamefont {Nunnemann}}, \bibinfo {author}
  {\bibfnamefont {G.}~\bibnamefont {Passler}}, \bibinfo {author} {\bibfnamefont
  {P.~M.}\ \bibnamefont {Rao}}, \bibinfo {author} {\bibfnamefont
  {J.}~\bibnamefont {Riegel}}, \bibinfo {author} {\bibfnamefont
  {N.}~\bibnamefont {Trautmann}},\ and\ \bibinfo {author} {\bibfnamefont
  {K.}~\bibnamefont {Wendt}},\ }\bibfield  {title} {\bibinfo {title}
  {{Determination of the first ionization potential of actinide elements by
  resonance ionization mass spectroscopy}},\ }\href
  {https://doi.org/10.1016/s0584-8547(96)01670-9} {\bibfield  {journal}
  {\bibinfo  {journal} {Spectrochimica Acta - Part B Atomic Spectroscopy}\
  }\textbf {\bibinfo {volume} {52}},\ \bibinfo {pages} {717} (\bibinfo {year}
  {1997})}\BibitemShut {NoStop}%
\bibitem [{\citenamefont {Liu}\ and\ \citenamefont
  {Stracener}(2020)}]{Liu2020}%
  \BibitemOpen
  \bibfield  {author} {\bibinfo {author} {\bibfnamefont {Y.}~\bibnamefont
  {Liu}}\ and\ \bibinfo {author} {\bibfnamefont {D.}~\bibnamefont
  {Stracener}},\ }\bibfield  {title} {\bibinfo {title} {{High efficiency
  resonance ionization of thorium}},\ }\href
  {https://doi.org/10.1016/j.nimb.2019.11.006} {\bibfield  {journal} {\bibinfo
  {journal} {Nuclear Instruments and Methods in Physics Research Section B:
  Beam Interactions with Materials and Atoms}\ }\textbf {\bibinfo {volume}
  {462}},\ \bibinfo {pages} {95} (\bibinfo {year} {2020})}\BibitemShut
  {NoStop}%
\bibitem [{\citenamefont {Raeder}\ \emph {et~al.}(2011)\citenamefont {Raeder},
  \citenamefont {Sonnenschein}, \citenamefont {Gottwald}, \citenamefont
  {Moore}, \citenamefont {Reponen}, \citenamefont {Rothe}, \citenamefont
  {Trautmann},\ and\ \citenamefont {Wendt}}]{Raeder2011}%
  \BibitemOpen
  \bibfield  {author} {\bibinfo {author} {\bibfnamefont {S.}~\bibnamefont
  {Raeder}}, \bibinfo {author} {\bibfnamefont {V.}~\bibnamefont
  {Sonnenschein}}, \bibinfo {author} {\bibfnamefont {T.}~\bibnamefont
  {Gottwald}}, \bibinfo {author} {\bibfnamefont {I.~D.}\ \bibnamefont {Moore}},
  \bibinfo {author} {\bibfnamefont {M.}~\bibnamefont {Reponen}}, \bibinfo
  {author} {\bibfnamefont {S.}~\bibnamefont {Rothe}}, \bibinfo {author}
  {\bibfnamefont {N.}~\bibnamefont {Trautmann}},\ and\ \bibinfo {author}
  {\bibfnamefont {K.}~\bibnamefont {Wendt}},\ }\bibfield  {title} {\bibinfo
  {title} {{Resonance ionization spectroscopy of thorium isotopes–towards a
  laser spectroscopic identification of the low-lying 7.6 eV isomer of
  \ch{^{229}Th} }},\ }\href {https://doi.org/10.1088/0953-4075/44/16/165005}
  {\bibfield  {journal} {\bibinfo  {journal} {Journal of Physics B: Atomic,
  Molecular and Optical Physics}\ }\textbf {\bibinfo {volume} {44}},\ \bibinfo
  {pages} {165005} (\bibinfo {year} {2011})},\ \Eprint
  {https://arxiv.org/abs/1105.4646} {arXiv:1105.4646} \BibitemShut {NoStop}%
\bibitem [{\citenamefont {Tomita}\ \emph {et~al.}(2018)\citenamefont {Tomita},
  \citenamefont {Nakamura}, \citenamefont {Matsui}, \citenamefont {Ohtake},
  \citenamefont {Sonnenschein}, \citenamefont {Saito}, \citenamefont {Kato},
  \citenamefont {Ohashi}, \citenamefont {Degner}, \citenamefont {Wendt},
  \citenamefont {Morita}, \citenamefont {Sakamoto}, \citenamefont {Kawai},
  \citenamefont {Okumura}, \citenamefont {Moore},\ and\ \citenamefont
  {Iguchi}}]{Tomita2018}%
  \BibitemOpen
  \bibfield  {author} {\bibinfo {author} {\bibfnamefont {H.}~\bibnamefont
  {Tomita}}, \bibinfo {author} {\bibfnamefont {A.}~\bibnamefont {Nakamura}},
  \bibinfo {author} {\bibfnamefont {D.}~\bibnamefont {Matsui}}, \bibinfo
  {author} {\bibfnamefont {R.}~\bibnamefont {Ohtake}}, \bibinfo {author}
  {\bibfnamefont {V.}~\bibnamefont {Sonnenschein}}, \bibinfo {author}
  {\bibfnamefont {K.}~\bibnamefont {Saito}}, \bibinfo {author} {\bibfnamefont
  {K.}~\bibnamefont {Kato}}, \bibinfo {author} {\bibfnamefont {M.}~\bibnamefont
  {Ohashi}}, \bibinfo {author} {\bibfnamefont {V.}~\bibnamefont {Degner}},
  \bibinfo {author} {\bibfnamefont {K.}~\bibnamefont {Wendt}}, \bibinfo
  {author} {\bibfnamefont {M.}~\bibnamefont {Morita}}, \bibinfo {author}
  {\bibfnamefont {T.}~\bibnamefont {Sakamoto}}, \bibinfo {author}
  {\bibfnamefont {T.}~\bibnamefont {Kawai}}, \bibinfo {author} {\bibfnamefont
  {T.}~\bibnamefont {Okumura}}, \bibinfo {author} {\bibfnamefont
  {I.}~\bibnamefont {Moore}},\ and\ \bibinfo {author} {\bibfnamefont
  {T.}~\bibnamefont {Iguchi}},\ }\bibfield  {title} {\bibinfo {title}
  {{Development of two-color resonance ionization scheme for Th using an
  automated wide-range tunable Ti:sapphire laser system}},\ }\href
  {https://doi.org/10.15669/pnst.5.97} {\bibfield  {journal} {\bibinfo
  {journal} {Progress in Nuclear Science and Technology}\ }\textbf {\bibinfo
  {volume} {5}},\ \bibinfo {pages} {97} (\bibinfo {year} {2018})}\BibitemShut
  {NoStop}%
\bibitem [{\citenamefont {Tiesinga}\ \emph {et~al.}(2021)\citenamefont
  {Tiesinga}, \citenamefont {Mohr}, \citenamefont {Newell},\ and\ \citenamefont
  {Taylor}}]{CODATA2018}%
  \BibitemOpen
  \bibfield  {author} {\bibinfo {author} {\bibfnamefont {E.}~\bibnamefont
  {Tiesinga}}, \bibinfo {author} {\bibfnamefont {P.}~\bibnamefont {Mohr}},
  \bibinfo {author} {\bibfnamefont {D.}~\bibnamefont {Newell}},\ and\ \bibinfo
  {author} {\bibfnamefont {B.}~\bibnamefont {Taylor}},\ }\bibfield  {title}
  {\bibinfo {title} {Codata recommended values of the fundamental physical
  constants: 2018}\ }\href
  {https://doi.org/https://doi.org/10.1103/RevModPhys.93.025010}
  {https://doi.org/10.1103/RevModPhys.93.025010} (\bibinfo {year}
  {2021})\BibitemShut {NoStop}%
\bibitem [{\citenamefont {Lethokhov}(1987)}]{Lethokhov}%
  \BibitemOpen
  \bibfield  {author} {\bibinfo {author} {\bibfnamefont {V.~S.}\ \bibnamefont
  {Lethokhov}},\ }\href@noop {} {\emph {\bibinfo {title} {Laser Photoionization
  Spectroscopy}}}\ (\bibinfo  {publisher} {Academic press},\ \bibinfo {year}
  {1987})\BibitemShut {NoStop}%
\bibitem [{\citenamefont {Zalubas}(1976)}]{Zalubas1976}%
  \BibitemOpen
  \bibfield  {author} {\bibinfo {author} {\bibfnamefont {R.}~\bibnamefont
  {Zalubas}},\ }\bibfield  {title} {\bibinfo {title} {{Energy levels,
  classified lines, and Zeeman effect of neutral thorium}},\ }\href
  {https://doi.org/10.6028/jres.080A.023} {\bibfield  {journal} {\bibinfo
  {journal} {Journal of Research of the National Bureau of Standards Section A:
  Physics and Chemistry}\ }\textbf {\bibinfo {volume} {80A}},\ \bibinfo {pages}
  {221} (\bibinfo {year} {1976})}\BibitemShut {NoStop}%
\bibitem [{\citenamefont {Naubereit}\ \emph {et~al.}(2018)\citenamefont
  {Naubereit}, \citenamefont {Gottwald}, \citenamefont {Studer},\ and\
  \citenamefont {Wendt}}]{Naubereit2018}%
  \BibitemOpen
  \bibfield  {author} {\bibinfo {author} {\bibfnamefont {P.}~\bibnamefont
  {Naubereit}}, \bibinfo {author} {\bibfnamefont {T.}~\bibnamefont {Gottwald}},
  \bibinfo {author} {\bibfnamefont {D.}~\bibnamefont {Studer}},\ and\ \bibinfo
  {author} {\bibfnamefont {K.}~\bibnamefont {Wendt}},\ }\bibfield  {title}
  {\bibinfo {title} {{Excited atomic energy levels in protactinium by resonance
  ionization spectroscopy}},\ }\href
  {https://doi.org/10.1103/PhysRevA.98.022505} {\bibfield  {journal} {\bibinfo
  {journal} {Physical Review A}\ }\textbf {\bibinfo {volume} {98}},\ \bibinfo
  {pages} {1} (\bibinfo {year} {2018})}\BibitemShut {NoStop}%
\bibitem [{\citenamefont {Sonnenschein}\ \emph {et~al.}(2012)\citenamefont
  {Sonnenschein}, \citenamefont {Raeder}, \citenamefont {Hakimi}, \citenamefont
  {Moore},\ and\ \citenamefont {Wendt}}]{Sonnenschein2012}%
  \BibitemOpen
  \bibfield  {author} {\bibinfo {author} {\bibfnamefont {V.}~\bibnamefont
  {Sonnenschein}}, \bibinfo {author} {\bibfnamefont {S.}~\bibnamefont
  {Raeder}}, \bibinfo {author} {\bibfnamefont {A.}~\bibnamefont {Hakimi}},
  \bibinfo {author} {\bibfnamefont {D.}~\bibnamefont {Moore}},\ and\ \bibinfo
  {author} {\bibfnamefont {K.}~\bibnamefont {Wendt}},\ }\bibfield  {title}
  {\bibinfo {title} {{Determination of the ground-state hyperfine structure in
  neutral \ch{^{229}Th}}},\ }\bibfield  {journal} {\bibinfo  {journal} {Journal
  of Physics B: Atomic, Molecular and Optical Physics}\ }\textbf {\bibinfo
  {volume} {45}},\ \href {https://doi.org/10.1088/0953-4075/45/16/165005}
  {10.1088/0953-4075/45/16/165005} (\bibinfo {year} {2012})\BibitemShut
  {NoStop}%
\bibitem [{\citenamefont {Cowan}(1981)}]{Cowan1981}%
  \BibitemOpen
  \bibfield  {author} {\bibinfo {author} {\bibfnamefont {R.~D.}\ \bibnamefont
  {Cowan}},\ }\href@noop {} {\emph {\bibinfo {title} {The Theory of Atomic
  Structure and Spectra}}}\ (\bibinfo  {publisher} {University of California
  Press},\ \bibinfo {address} {Berkeley},\ \bibinfo {year} {1981})\BibitemShut
  {NoStop}%
\bibitem [{\citenamefont {Quinet}\ \emph {et~al.}(1999)\citenamefont {Quinet},
  \citenamefont {Palmeri}, \citenamefont {Bi\'emont}, \citenamefont {McCurdy},
  \citenamefont {Rieger}, \citenamefont {Pinnington}, \citenamefont
  {Wickliffe},\ and\ \citenamefont {Lawler}}]{Quinet1999}%
  \BibitemOpen
  \bibfield  {author} {\bibinfo {author} {\bibfnamefont {P.}~\bibnamefont
  {Quinet}}, \bibinfo {author} {\bibfnamefont {P.}~\bibnamefont {Palmeri}},
  \bibinfo {author} {\bibfnamefont {E.}~\bibnamefont {Bi\'emont}}, \bibinfo
  {author} {\bibfnamefont {M.~M.}\ \bibnamefont {McCurdy}}, \bibinfo {author}
  {\bibfnamefont {G.}~\bibnamefont {Rieger}}, \bibinfo {author} {\bibfnamefont
  {E.~H.}\ \bibnamefont {Pinnington}}, \bibinfo {author} {\bibfnamefont
  {M.~E.}\ \bibnamefont {Wickliffe}},\ and\ \bibinfo {author} {\bibfnamefont
  {J.~E.}\ \bibnamefont {Lawler}},\ }\bibfield  {title} {\bibinfo {title}
  {{Experimental and Theoretical Radiative Lifetimes, Branching Fractions and
  Oscillator Strengths in Lu~II }},\ }\href
  {https://doi.org/10.1046/j.1365-8711.1999.02689.x} {\bibfield  {journal}
  {\bibinfo  {journal} {Monthly Notices of the Royal Astronomical Society}\
  }\textbf {\bibinfo {volume} {307}},\ \bibinfo {pages} {934} (\bibinfo {year}
  {1999})}\BibitemShut {NoStop}%
\bibitem [{\citenamefont {Quinet}\ \emph {et~al.}(2002)\citenamefont {Quinet},
  \citenamefont {Palmeri}, \citenamefont {Bi\'emont}, \citenamefont {Li},\ and\
  \citenamefont {Svanberg}}]{Quinet2002}%
  \BibitemOpen
  \bibfield  {author} {\bibinfo {author} {\bibfnamefont {P.}~\bibnamefont
  {Quinet}}, \bibinfo {author} {\bibfnamefont {P.}~\bibnamefont {Palmeri}},
  \bibinfo {author} {\bibfnamefont {E.}~\bibnamefont {Bi\'emont}}, \bibinfo
  {author} {\bibfnamefont {Z.~S.}\ \bibnamefont {Li}},\ and\ \bibinfo {author}
  {\bibfnamefont {S.}~\bibnamefont {Svanberg}},\ }\bibfield  {title} {\bibinfo
  {title} {{Radiative Lifetime Measurements and Transition probability
  Calculations in Lanthanide Ions}},\ }\href
  {https://doi.org/https://doi.org/10.1016/S0925-8388(02)00363-8} {\bibfield
  {journal} {\bibinfo  {journal} {J. Alloys Compounds}\ }\textbf {\bibinfo
  {volume} {344}},\ \bibinfo {pages} {255} (\bibinfo {year}
  {2002})}\BibitemShut {NoStop}%
\bibitem [{\citenamefont {Fraga}\ \emph {et~al.}(1976)\citenamefont {Fraga},
  \citenamefont {Karwowski},\ and\ \citenamefont {Saxena}}]{Fraga1976}%
  \BibitemOpen
  \bibfield  {author} {\bibinfo {author} {\bibfnamefont {S.}~\bibnamefont
  {Fraga}}, \bibinfo {author} {\bibfnamefont {J.}~\bibnamefont {Karwowski}},\
  and\ \bibinfo {author} {\bibfnamefont {K.~M.~S.}\ \bibnamefont {Saxena}},\
  }\href@noop {} {\emph {\bibinfo {title} {Handbook of Atomic Data}}}\
  (\bibinfo  {publisher} {Elsevier},\ \bibinfo {address} {Amsterdam},\ \bibinfo
  {year} {1976})\BibitemShut {NoStop}%
\bibitem [{\citenamefont {Grant}(2007)}]{Grant2007}%
  \BibitemOpen
  \bibfield  {author} {\bibinfo {author} {\bibfnamefont {I.~P.}\ \bibnamefont
  {Grant}},\ }\href@noop {} {\emph {\bibinfo {title} {Relativistic quantum
  theory of atoms and molecules. Theory and computation}}}\ (\bibinfo
  {publisher} {Springer},\ \bibinfo {address} {New York},\ \bibinfo {year}
  {2007})\BibitemShut {NoStop}%
\bibitem [{\citenamefont {Froese~Fischer}\ \emph {et~al.}(2019)\citenamefont
  {Froese~Fischer}, \citenamefont {Gaigalas}, \citenamefont {J\"onsson},\ and\
  \citenamefont {Biero\'n}}]{Froese2019}%
  \BibitemOpen
  \bibfield  {author} {\bibinfo {author} {\bibfnamefont {C.}~\bibnamefont
  {Froese~Fischer}}, \bibinfo {author} {\bibfnamefont {G.}~\bibnamefont
  {Gaigalas}}, \bibinfo {author} {\bibfnamefont {P.}~\bibnamefont
  {J\"onsson}},\ and\ \bibinfo {author} {\bibfnamefont {J.}~\bibnamefont
  {Biero\'n}},\ }\bibfield  {title} {\bibinfo {title} {{GRASP2018-A Fortran 95
  version of the General Relativistic Atomic Structure Package }},\ }\href
  {https://doi.org/https://doi.org/10.1016/j.cpc.2018.10.032} {\bibfield
  {journal} {\bibinfo  {journal} {Computer Physics Communication}\ }\textbf
  {\bibinfo {volume} {237}},\ \bibinfo {pages} {184} (\bibinfo {year}
  {2019})}\BibitemShut {NoStop}%
\bibitem [{\citenamefont {Weigand}\ \emph {et~al.}(2014)\citenamefont
  {Weigand}, \citenamefont {Cao}, \citenamefont {Hangele},\ and\ \citenamefont
  {Dolg}}]{Weigand2014}%
  \BibitemOpen
  \bibfield  {author} {\bibinfo {author} {\bibfnamefont {A.}~\bibnamefont
  {Weigand}}, \bibinfo {author} {\bibfnamefont {X.}~\bibnamefont {Cao}},
  \bibinfo {author} {\bibfnamefont {T.}~\bibnamefont {Hangele}},\ and\ \bibinfo
  {author} {\bibfnamefont {M.}~\bibnamefont {Dolg}},\ }\bibfield  {title}
  {\bibinfo {title} {{Relativistic small-core pseudopotentials for actinium,
  thorium, and protactinium}},\ }\href {https://doi.org/10.1021/jp500215z}
  {\bibfield  {journal} {\bibinfo  {journal} {Journal of Physical Chemistry A}\
  }\textbf {\bibinfo {volume} {118}},\ \bibinfo {pages} {2519} (\bibinfo {year}
  {2014})}\BibitemShut {NoStop}%
\end{thebibliography}%

\end{document}